\documentclass[11pt]{article}
\pdfoutput=1

\usepackage{amsmath}
\usepackage{amssymb}
\usepackage{amsfonts}
\usepackage{mathrsfs}

\usepackage{mathrsfs}
\usepackage{fullpage}
\usepackage{setspace}
\usepackage{bm}
\usepackage{bbm}
\usepackage{relsize}
\usepackage{adjustbox}
\usepackage{multirow}
\usepackage{cite}
\usepackage{filecontents}
\usepackage[T1]{fontenc}

\usepackage[normalem]{ulem}
\usepackage{enumerate}
\usepackage{yfonts}

\usepackage{psfrag}
\usepackage{array}
\usepackage{booktabs}
\newcolumntype{N}{>{\centering\arraybackslash}m{.5in}}
\newcolumntype{G}{>{\centering\arraybackslash}m{2in}}

\usepackage[latin1]{inputenc}
\usepackage{graphicx}
\usepackage{cancel}
\usepackage{slashed}
\usepackage{mathtools}

\usepackage[font=small,labelfont=bf]{caption}
\usepackage{subcaption}

\usepackage[colorlinks=true]{hyperref} 
\hypersetup{
    unicode=false,          % non-Latin characters 
    pdftoolbar=true,        % show Acrobat
    pdfmenubar=true,        % show Acrobat 
    pdffitwindow=false,     % window fit to page when opened
    pdfstartview={FitH},    % fits the width of the page to the window
    pdftitle={Probing astrophysical environment with eccentric extreme mass-ratio inspirals},    % title
    pdfauthor={Mostafizur Rahman, Shailesh Kumar and Arpan Bhattacharyya},     % author
    pdfnewwindow=true,      % links in new window
    colorlinks=true,       % false: boxed links; true: colored links
    linkcolor=blue,          % color of internal links (change box color with linkbordercolor)
    citecolor=red,        % color of links to bibliography
    filecolor=magenta,      % color of file links
    urlcolor=blue           % color of external links
    }
\def\In{\textrm{in}}
\def\up{\textrm{up}}
\usepackage{hyperref}
\hypersetup{colorlinks=true}

\def\equationautorefname~#1\null{%
	Eq.~(#1)\null
}
\def\figureautorefname~#1\null{%
	Fig.~#1\null
}
\def\tableautorefname~#1\null{%
	Table.~#1\null
}
\def\sectionautorefname~#1\null{%
	Section #1\null
}
\def\appendixautorefname~#1\null{%
	Appendix #1\null
}

\newcommand{\ca}[1]{\mathcal{#1}}

\begin{document}
%%%%%%%%%%%%%%%%%%%%%%%%%%%%%%%%%%%%%%%%%%%%%%%%%%%%%%%%%%%%%%%%%%%%%%%%%%%%%%%%

%%%%%%%%%%%%%%%%%%%%%%%%%%%%%%%%%%%%%%%%%%%%%%%%%%%%%%%%%%%%%%%%%%%%%%%%%%%%%%%%
% Title page
%%%%%%%%%%%%%%%%%%%%%%%%%%%%%%%%%%%%%%%%%%%%%%%%%%%%%%%%%%%%%%%%%%%%%%%%%%%%%%%%
\numberwithin{equation}{section}
{
\begin{titlepage}
\begin{center}

\hfill \\
\hfill \\
\vskip 0.75in
{\Large \bf Probing astrophysical environment with eccentric extreme mass-ratio inspirals 
}\\

\vskip 0.3in

{\large
Mostafizur Rahman${}$\footnote{\href{mailto: mostafizur.r@iitgn.ac.in}{mostafizur.r@iitgn.ac.in}}, Shailesh Kumar${}$\footnote{\href{mailto: shailesh.k@iitgn.ac.in}{shailesh.k@iitgn.ac.in}} and Arpan Bhattacharyya${}$\footnote{\href{mailto: abhattacharyya@iitgn.ac.in}{abhattacharyya@iitgn.ac.in}}}

\vskip 0.3in

{\it ${}$Indian Institute of Technology, Gandhinagar, Gujarat-382355, India}

\vskip.5mm

\end{center}

\vskip 0.35in

\begin{center} 
{\bf ABSTRACT }
\end{center}

The discovery of gravitational waves and black holes has started a new era of gravitational wave astronomy that allows us to probe the underpinning features of gravity and astrophysics in extreme environments of the universe. In this article, we investigate one such study with an extreme mass-ratio inspiral system where the primary object is a spherically symmetric static black hole immersed in a dark matter halo governed by the Hernquist density distribution. We consider the eccentric equatorial orbital motion of the steller-mass object orbiting around the primary and compute measurable effects. We examine the behaviour of dark matter mass and halo radius in generated gravitational wave fluxes and the evolution of eccentric orbital parameters- eccentricity and semi-latus rectum. We further provide an estimate of gravitational wave dephasing and find the seminal role of low-frequency detectors in the observational prospects of such an astrophysical environment.
%that could fundamentally put forward a new understanding of the universe.

\vfill

%\noindent \today

\end{titlepage}
}

%%%%%%%%%%%%%%%%%%%%%%%%%%%%%%%%%%%%%%%%%%%%%%%%%%%%%%%%%%%%%%%%%%%%%%%%%%%%%%%%
% Table of contents
%%%%%%%%%%%%%%%%%%%%%%%%%%%%%%%%%%%%%%%%%%%%%%%%%%%%%%%%%%%%%%%%%%%%%%%%%%%%%%%%
\newpage
%\tableofcontents

\section{Introduction}
Astrophysical environments surrounding black hole systems may offer insights into the longstanding questions that could impact how we perceive the cosmos. 
One such entity is dark matter which could potentially add a new chapter to the fundamental understanding of how it fits into the Standard Model of particle physics. It is claimed to be some form of elementary particle that plays an essential role in various astrophysical processes and remains one of the most enigmatic conundrums. There are indirect pieces of evidence that support the existence of dark matter in the universe \cite{1990ApJ...356..359H, Navarro:1995iw, Freese:2008cz, Clowe:2006eq, Bertone:2004pz, vandenBergh:1999sa, Persic:1995ru, Corbelli:1999af, Massey:2010hh, Ellis:2010kf}. The interaction between dark matter and massive compact sources can provide meaningful insights into the composition and properties of this mystifying matter. It is known that galaxies are surrounded by a dark matter halo that is substantially larger than the visible galaxy, and almost every big galaxy has a supermassive black hole in its centre \cite{Kormendy:2013dxa, Harris:2015vxa}. So it is natural to investigate the effects of dark matter on the dynamics of black hole spacetimes and gravitational waves (GWs) that could infer deep insights into galactic properties \cite{Macedo:2013qea,Coogan:2021uqv,Baryakhtar:2022hbu,Singh:2022wvw,Bhattacharya:2023stq, Bhattacharyya:2023kbh,AbhishekChowdhuri:2023cle}.

Typically, an anisotropic fluid with a certain density distribution is used to model galactic matter that further indicates a halo dominated by dark matter \cite{BENSON201033}. One can attribute a general relativistic metric to such a distribution of galactic matter that encompasses a black hole spacetime. There are various spacetimes reported in the literature that deal with an isolated black hole spacetime which is basically matched to matter distribution via some mass function \cite{Xu_2018, PhysRevD.104.124082, PhysRevD.104.104042, Jusufi2020, Hou_2018, KONOPLYA20191}. With an overcome to such a cut-paste approach, recently an interesting solution was provided by Cardoso et al. \cite{PhysRevD.105.L061501}, where the general relativistic description of a black hole immersed in dark matter halo has been considered self-consistently. It is the solution of Einstein field equations with corresponding stress-energy tensor constituting matter distribution with the Hernquist density profile \cite{1990ApJ...356..359H}. However, various exact solutions of the spacetime with a supermassive black hole immersed in dark matter halo were then derived in \cite{Konoplya:2022hbl}. Numerous measurable effects have been studied with the spacetime provided in \cite{PhysRevD.105.L061501}, starting from the orbital trajectory and black hole shadow analysis to GW signatures with extreme mass-ratio inspiral (EMRI) systems \cite{Konoplya:2022hbl, Dai:2023cft, Xavier:2023exm, Stuchlik:2021gwg, Figueiredo:2023gas, Cardoso:2022whc, Destounis:2022obl}.

The development of GW astronomy may fundamentally alter how we understand the invisible universe \cite{LIGOScientific:2016aoc, LIGOScientific:2017vox}. The dynamics of black holes and generated GWs can get affected by the presence of the dark matter environment at the centre of a galaxy and close to black holes. In this direction, the interplay between dark matter and EMRI systems can provide meaningful insights into the properties of galactic matter. An EMRI is a binary system where a steller-mass object (secondary object with mass $\mu$) inspirals a supermassive black hole (primary object with mass $M_{\textrm{BH}}$). As the mass-ratio for EMRI lies in the range ($q \equiv \mu/M_{\textrm{BH}} = 10^{-7} - 10^{-4}$), the secondary can be treated as a background perturbation to the primary supermassive black hole enabling us to analyze the system with black hole perturbation techniques. EMRIs have gained considerable attention from several directions of gravitational physics, which directly have observational consequences \cite{Babak:2017tow, Amaro-Seoane:2007osp, Barack:2009ux, Hinderer:2008dm, Drasco:2005kz, Gair:2004iv, Sopuerta:2009iy,Rahman:2021eay, Rahman:2022fay,Fan:2022wio,Liang:2022gdk,Maselli:2021men,Drummond:2023loz, PhysRevD.102.024041}. Since EMRIs are anticipated to be located in star clusters, and the galactic centre, dark matter may impact both its dynamics and GWs, providing an understanding of the astrophysical environment. Current developments focusing on such studies infer the possible detectability of these sources through space-borne detectors like Laser Interferometer Space Antenna Experiments (LISA) \cite{Gair:2017ynp, Babak:2017tow, LISA:2022kgy}.

In \cite{Cardoso:2022whc}, authors considered an EMRI system with spherically symmetric, static, non-vacuum black hole spacetime derived in \cite{PhysRevD.105.L061501} where the secondary object exhibits the circular motion and further discuss measurable aspects of galactic parameters from GW measurements. On the other hand, environmental effects are capable of increasing the eccentricity of a binary system \cite{Cardoso:2020iji}. In particular, the addition of eccentricity is pertinent for the orbits in EMRI systems where the secondary object can exhibit a large initial eccentricity \cite{Barsanti:2022ana}. %before settling into a circular inspiral motion\ 
Apart from that, the study of the evolution of eccentricity is important to constrain the presence of new degrees of freedom. This motivates us to examine the detectability of dark matter distribution with EMRIs where the secondary object shows the eccentric orbital motion. Therefore, in this article, we study an EMRI system where the central massive object is a black hole immersed in a dark matter medium \cite{PhysRevD.105.L061501}, and the secondary object moves on eccentric equatorial orbits. We provide a detailed analysis of the effect of the dark matter medium on energy and angular momentum fluxes as well as study the orbital evolution of the secondary and the orbital phase. We further explicitly discuss the possible detectability of dark matter environment by estimating the GW dephasing and notice the essential role of LISA observations with exceptionally high accuracy. We find significant deviations from Schwarzschild in computed quantities with different dark matter parameters (dark matter mass $M$ and typical length scale of a galaxy or halo radius $a_{0}$) and eccentricities.  

Let us briefly review how we organize the draft. Section (\ref{BH}) touches upon the introduction of black hole spacetime immersed in dark matter halo that we consider in the analysis and its eccentric orbital motion. In Section (\ref{pert1}), we describe the basic setup of metric perturbation equations, including the stress-energy tensor of the environment and the secondary object. Next, Section (\ref{axpt}) elaborates on the axial and polar perturbation equations together with the details and numerical results on estimating fluxes, orbital evolution and further detectability of such EMRI systems in dark matter environments through low-frequency GW detectors in Section (\ref{nmr}). Lastly, we conclude the article with Section (\ref{dscn}) by highlighting the outcomes of the study and future outlooks.

\par 
\textit{Notation and Convention: } We set the fundamental constants $G$ and $c$ to unity and  adopt positive sign convention $(-1,1,1,1)$. Greek letters are used to represent four-dimensional indices.

%%%%%%%%%%%%%%%%%%%%%%%%%%%%%%%%%%%%%%%%%%%%%%%%%%%%%%%%%%%%%%%%%%%%%%%%%%%%%%%%%%%%%%%%%%%%%%%%%%%%%%
\section{Black hole immersed in dark matter halo and orbital motion}\label{BH}

To comprehend how the surrounding environment of a black hole affects the evolution of generated GWs, one requires a spacetime geometry that corresponds to a density profile of the matter distribution. Here, we consider Hernquist-type density distribution which offers an accurate description of the profiles found in bulges and elliptical galaxies \cite{1990ApJ...356..359H, PhysRevD.105.L061501}. If the total mass of the \textit{halo} is $M$ and $a_{0}$ denotes the typical length-scale of a galaxy, the Hernquist-type density distribution is given by \cite{1990ApJ...356..359H, PhysRevD.105.L061501},
\begin{align} \label{dmtt}
\rho_{\textrm{HD}} = \frac{M a_{0}}{2\pi r (r+a_{0})^{3}},
\end{align}
where $\rho_{\textrm{HD}}$ is the \textit{Hernquist density} profile. 
Following \cite{PhysRevD.105.L061501}, we consider a spherically symmetric static black hole spacetime described by the line element 
\begin{align}\label{met1}
    ds^{2} = g_{\mu\nu}^{(0)}dx^{\mu}dx^{\nu}=-f(r) dt^{2} + b(r)^{-1} dr^{2} + r^{2} \left[d\theta^{2}+\sin^{2}\theta d\phi^{2}\right],
\end{align}
which is immersed in the dark matter halo. The energy-momentum tensor of the dark  matter halo distribution containing a supermassive black hole with the mass $M_{\textrm{BH}}$ is an anisotropic fluid, %that has vanishing radial pressure,
\begin{align}\label{stt1}
%T^{\mu}_{\nu} = \text{diag} (-\rho_{\textrm{DM}}, 0, p_{t}, p_{t}),
T^{\textrm{DM}(0)}_{\mu\nu} = \rho_{\textrm{DM}} u_{\mu}^{(0)}u_{\nu}^{(0)}+p_{r}k_{\mu}^{(0)}k_{\nu}^{(0)}+p_{t}\Pi_{\mu\nu}^{(0)}, 
\end{align}
where  $\rho_{\textrm{DM}}$ is the density of the dark matter halo in the presence of the massive black hole, $p_{r}$ is the radial pressure, $p_{t}$ is the tangential pressure, $u^{\mu}_{(0)}=\frac{1}{\sqrt{f(r)}}\delta^{\mu}_{t}$ is the 4-velocity of the fluid and $k_{\mu}^{(0)}=\frac{1}{\sqrt{b(r)}}\delta^{r}_{\mu}$ is the unit spacelike vector orthogonal to $u^{\mu}_{(0)}$,  satisfying the normalization conditions  $u_{(0)}^{\mu}k_{\mu}^{(0)}=0$ and $k_{(0)}^{\mu}k_{\mu}^{(0)}=1$. Further, the projection operator $\Pi_{\mu\nu}^{(0)}=g_{\mu\nu}^{(0)}+u_{\mu}^{(0)}u_{\nu}^{(0)}-k_{\mu}^{(0)}k_{\nu}^{(0)}$ is orthogonal to ($u_{(0)}^{\mu},k_{(0)}^{\mu}$). Throughout the paper, a superscript (or a subscript) ``${(0)}$'' denotes background quantities. It is noted that we consider our system where the fluid's energy-momentum tensor has vanishing radial pressure; however, we have a non-vanishing contribution of the same when we take the perturbed quantities which we mention in the subsequent section. 

Now from the Einstein field equations, we have the following solutions
%and $p_t$ is the tangential pressure the expression of which is given by %the massive black hole is located in the centre of the distribution (\ref{dmtt}) that takes the following form of the density profile \cite{PhysRevD.105.L061501},
\begin{align}\label{dm22}
    4\pi\rho_{\textrm{DM}} = \frac{m'(r)}{r^{2}}
  %  = \frac{2M (a_{0}+2M_{\textrm{BH}})}{r (r+a_{0})^{3}} \Big(1-\frac{2M_{\textrm{BH}}}{r}\Big) 
    \hspace{3mm} ; \hspace{3mm}  p_{t} = \frac{1}{2} \frac{m(r) \rho_{\textrm{DM}}}{r-2m(r)}
\end{align}
% with the tangential pressure,
% \begin{align}\label{tan1}
%     p_{t} = \frac{1}{2} \frac{m(r) \rho_{\textrm{DM}}}{r-2m(r)},
% \end{align}
where  $m(r)$ is the mass function and the prime denotes the derivative of $m(r)$ with respect to its argument. Following \cite{PhysRevD.105.L061501}, we consider the following form of the mass function
\begin{align}\label{mass_function}
    m(r) = M_{\textrm{BH}} + \frac{M r^{2}}{(a_{0}+r)^{2}} \Big(1-\frac{2M_{\textrm{BH}}}{r}\Big)^{2}\,.
\end{align}
Utilizing Eq. (\ref{dm22}) and Eq.~(\ref{mass_function}), we can easily show that if there is no black hole present, the density profile reduces to Eq. (\ref{dmtt}). With the expression of mass function in Eq.~(\ref{mass_function}), we finally have the following metric components of the Eq. (\ref{met1}),

\begin{align}\label{df}
     f(r) = \Big(1-\frac{2M_{\textrm{BH}}}{r}\Big) e^{\Gamma} \hspace{5mm} ; \hspace{5mm}
     b(r) = \Big(1-\frac{2m(r)}{r}\Big),
 \end{align}
 with
\begin{align} \label{cmp1}
    \Gamma = \sqrt{\frac{M}{\zeta}}\Big(-\pi+2\arctan\Big(\frac{r+a_{0}-M}{\sqrt{M\zeta}}\Big)\Big) \hspace{5mm} ; \hspace{5mm}
    \zeta = 2a_{0}-M+4M_{\textrm{BH}}. 
\end{align}
The spacetime Eq. (\ref{met1}) with components Eq. (\ref{df}) is a black hole immersed in the dark matter profile Eq. (\ref{dm22}) that has a horizon at $r=r_+\equiv 2M_{\textrm{BH}}$ with a curvature singularity  at $r=0$. At large distances, the halo mass will dominate over the black hole mass; hence the ADM mass of the spacetime becomes ($M_{\textrm{BH}}+M$). It is to note that the matter density vanishes at the horizon; however, the tangential pressure Eq. (\ref{dm22}) remains regular. %At the galactic length scales, it is logical to presume $M_{\textrm{BH}}<<M<<a$. 
One can further investigate the effects of a specific parameter termed as \textit{halo compactness} ($M/a_{0}$) that ascertains the gravitational features or properties of a galaxy. This parameter is usually small and for galaxies, compactness can be as large as $10^{-4}$ \cite{Navarro:1995iw}.

%To our interest, we re-write metric coefficients (\ref{df}) in terms of halo compactness, $\epsilon_{hc} \equiv \frac{M}{a} << 1$. With this assumption, we can structure the functions $\Gamma$, $m_{fun}(r)$, $P(r)$ and $Q(r)$ in terms of $\epsilon_{hc}$. Here, we consider only the leading order contribution in $\epsilon_{hc}$. Thus the metric components take the following form,
%\begin{align}
%    P(r) =& \Big(1-\frac{2 M_{\textrm{BH}}}{r}\Big) \Big(1-\frac{2a}{a+r}\epsilon_{hc}\Big) + \mathcal{O}(\epsilon_{hc}^{2}), \label{p1} \\
%    Q(r) =& \Big(1-\frac{2 M_{\textrm{BH}}}{r}\Big)^{-1}+\frac{2 a r}{(a+r)^2}\epsilon_{hc} + \mathcal{O}(\epsilon_{hc}^{2}). \label{q1}
%\end{align}
%The corresponding density profile becomes,
%\begin{align}\label{dm22n}
%    \rho_{\textrm{DM}} = \frac{2\epsilon_{hc}}{ra^{2}(1+r/a)^{3}} \Big(1+\frac{2M_{\textrm{BH}}}{a}\Big) \Big(1-\frac{2M_{\textrm{BH}}}{r}\Big)
%\end{align}
%Since the spacetime is asymptotically flat, therefore,  $\lim_{r\to +\infty} P(r) = 1 $ and $\lim_{r\to +\infty} Q(r) = 1$. Our further calculations and results are based on the consideration of the functions $P(r)$ and $Q(r)$ represented in (\ref{p1}) and (\ref{q1}) respectively along with the matter density distribution (\ref{dm22n}). The Schwarzschild results can be recovered by setting $\epsilon_{hc}=0$ and $a=0$.

\subsection{Geodesic equations and orbital parameters}

In this section, we study the orbital motion of a test particle of mass $\mu$ in the spacetime described by the line element Eq. (\ref{met1}). In particular, we focus on eccentric orbitals, which would further assist us in examining the evolution of the secondary object. For the sake of simplicity, we set  $M_{\textrm{BH}}=1$, i.e., we consider that the time and distances are measured in the unit of black hole mass. Furthermore, since the spacetime is spherically symmetric, we set $\theta=\pi/2$ with the loss of generality.  
%It is noted that we scale everything in terms of $M_{\textrm{BH}}$ throughout the article by setting this to be unity. %For the computational purpose, we find useful introducing dimensionless quantities $\hat{E}=\frac{E}{\mu}, \hat{J}=\frac{J}{M_{\textrm{BH}}\mu}$ and $\hat{r}=\frac{r}{M_{\textrm{BH}}}$. 
Note that the spacetime exhibits two constants of motion ($\overline{E}, \overline{J}_{z}$), namely energy and angular momentum, corresponding to two Killing vectors ($\frac{\partial}{\partial t}, \frac{\partial}{\partial\phi}$). We can express the equation of motion of the particle in terms of these constants of motion in the following manner
%In order to study the orbital dynamics, we start with timeline geodesics which are expressed in the following form
\begin{align} \label{gd1}
    U^{t}=\frac{dt}{d\tau} = \frac{E}{f(r)} \hspace{7mm} ; \hspace{7mm} U^{\phi}=\frac{d\phi}{d\tau} = \frac{J_{z}}{r^{2}} \hspace{7mm} ; \hspace{7mm}
    \left(U^{r}\right)^2=\Big(\frac{dr}{d\tau}\Big)^{2} =- V_{\textrm{eff}}(r)~,
\end{align}
where $E=\overline{E}/\mu$ and $J_z=\overline{J}_z/\mu$ and $V_{\textrm{eff}}(r)$ denotes the effective potential that determines the orbital motion, which can be expressed as follows
\begin{align}\label{vef}
    V_{\textrm{eff}}(r) = \Big(- \frac{E^{2}}{f(r)} +\frac{J_{z}^{2}}{r^{2}}+1\Big) b(r) ~.
\end{align}
%$V_{\textrm{eff}}(r)$ denotes the effective potential that determines the orbital motion. 

\begin{figure}[t!]
	%%%%%%%%%%%%%%%%%%%%%%%%
	\centering
	\minipage{0.48\textwidth}
	\includegraphics[width=\linewidth]{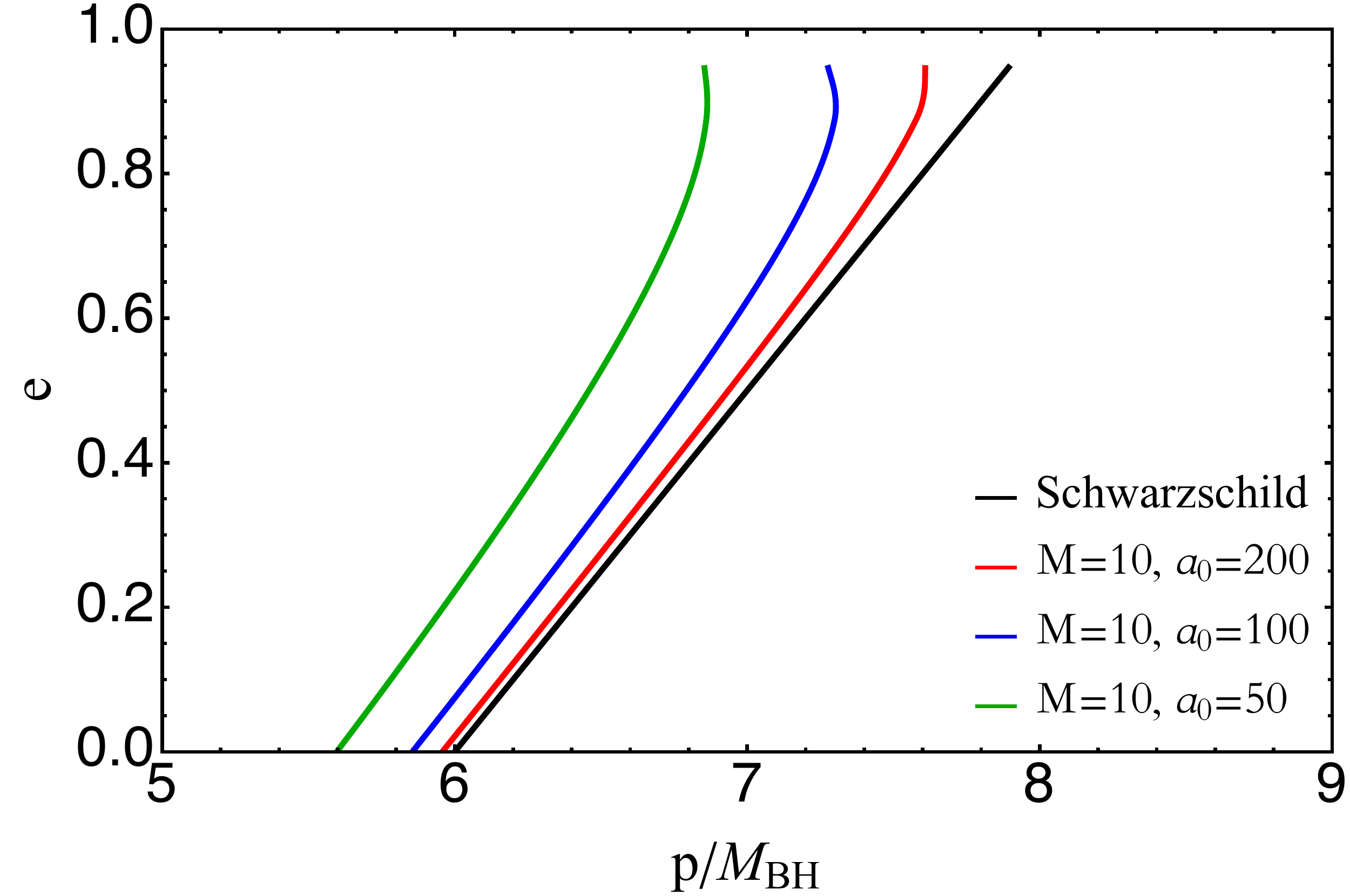}
% \caption{Wormholes for $\Lambda=0$}
	\endminipage
	\caption{The plot of separatrices in the ($p, e$) plane for different values of dark matter parameters is presented. The black straight line is the separatrix for a Schwarzschild black hole which follows the relation $p_{\textrm{min}}=6+2e$.  
 The parameter space to the right of the separatrix represents the set of all points that allows bounded orbits, with the points on the separatrix depicting the position of the last stable orbit. The parameter space that allows bounded orbits expands with the increase of halo compactness parameter  $M/a_0$.  }\label{fig_Separatrix}
\end{figure}

Since we are interested in bounded orbits, we consider the two turning points $r_{p}$ and $r_{a}$, representing periastron and apastron, respectively. These are the points where the radial velocity vanishes, i.e., $V_{\textrm{eff}}(r) = 0$. To have a bounded orbit between these turning points, we require $V_{\textrm{eff}}(r)<0$ in the range $r_p<r<r_a$. This condition is satisfied only when \cite{PhysRevD.103.104045}
\begin{equation}
    \begin{aligned}
        V_{\textrm{eff}}'(r_p)\leq 0\,,\qquad  V_{\textrm{eff}}'(r_a)>0~,
    \end{aligned}
\end{equation}
where ``prime'' denotes the derivative of the function with respect to the $r$. The aforementioned conditions lead to the expressions for the energy and angular momentum of a bounded orbit, which is given by
\begin{equation}\label{Ep_Jz}
  \begin{aligned}
    E^{2} = f(r_{p}) f(r_{a}) \frac{(r_{a}^{2}-r_{p}^{2})}{r_{a}^{2}f(r_{p})-r_{p}^{2}f(r_{a})} \hspace{3mm} ; \hspace{3mm} J_{z}^{2} = r_{p}^{2} r_{a}^{2} \frac{f(r_{p})-f(r_{a})}{r_{p}^{2}f(r_{a})-r_{a}^{2}f(r_{p})}\,.
\end{aligned}  
\end{equation}
%The condition for having a bounded orbit between the turning points then translates to the following 
Note that it is more convenient to express the bounded orbits in terms of the eccentricity ($e$) and semi-latus rectum ($p$) instead of the energy and angular momentum \cite{PhysRevD.50.3816}. The semi-latus rectum $p$ infers the size of the orbits, whereas the eccentricity $e$ indicates the degree of non-circularity. These parameters are related to the turning points in the following manner
%The bound orbits of spacetime can be obtained with $0 \leq e < 1$.
%This provides a separatrix and a minimum condition on elliptic orbital parameters, eccentricity ($e$), and semi-latus rectum ($p$) to have bound orbits. 
\begin{align}\label{rp_ra}
    r_{p} = \frac{p}{1+e} \hspace{5mm} ; \hspace{5mm} r_{a} = \frac{p}{1-e}~\,.
\end{align}
As mentioned earlier, we are rescaling parameters in the unit $M_{\textrm{BH}}$. The relationship between ($p, e$) and ($E, J_{z}$) can be obtained by replacing Eq. (\ref{rp_ra}) in Eq. (\ref{Ep_Jz}). We can obtain the last stable orbit (LSO) or marginally stable orbits by imposing the condition \cite{PhysRevD.103.104045, PhysRevD.77.124050}
\begin{equation}\label{LSO}
  \begin{aligned}
V_{\textrm{eff}}(r_p)&=0\,,\qquad V_{\textrm{eff}}(r_a)=0, \\ 
V_{\textrm{eff}}'(r_p)&=0\,,\qquad V_{\textrm{eff}}'(r_a)>0~.
\end{aligned}  
\end{equation}
By replacing Eq. (\ref{rp_ra}) in Eq. (\ref{LSO}), we can find the set of all points in the ($p, e$) plane, termed as the \textit{separatrix}, which separates a bounded orbit from an unbounded (plunging) one \cite{PhysRevD.103.104045, PhysRevD.77.124050, PhysRevD.50.3816}. In other words, for a given eccentricity value, the separatrix defines the minimum value of the semi-latus rectum $p_{\textrm{min}}$ for which spacetime allows bound orbits. It is worth noting that the separatrix is called the innermost stable circular orbit (ISCO) for the circular orbits ($e=0$). In Fig. (\ref{fig_Separatrix}), we plot the separatrix for different values of dark matter parameters $M$ and $a_0$. In the plot, the black curve represents the separatrix for a Schwarzschild black hole which takes the value $p_{\textrm{min}}=6+2e$ \cite{PhysRevD.50.3816}. As can be seen from the plot, for a given value of eccentricity, $p_{\textrm{min}}$ decreases as we increase the value of the halo compactness parameter $M/a_0$. The parameter space that allows bounded orbits expands with the increase of halo compactness parameter. \par
%%%%%%%%%%%%%%%%%%%%%%%%%%%%%%%%%%%%%%%%%%%%%%%%%%%%%%%%%%%%%%%%%%%%%%%%%%%%%%%%%%%%%%%%%%%%%%%%%%%
%%%%%%%%%%%%%%%%%%%%%%%%%%%%%%%%%%%%%%%%%%%%%%%%%%%%%%%%%%%%%%%%%
%A \textit{separatrix} is the locus of all points in the ($p,~e$) plane which separates a bounded orbit from an unbounded (plunging one) . 
 %Quantities ($p, e$) are dimensionless. $p$ infers the size of the orbits and $e$ indicates the degree of non-circularity. The bound orbits of spacetime can be obtained with $0 \leq e < 1$ and points at least greater than the separatrix; which for Schwarzschild is: $p = 6+2e$ \cite{PhysRevD.50.3816}. In our case, the metric (\ref{met1}) is complicated, we can not have similar algebraic relation. Therefore, we would consider numerical values of $p$ where the secondary object truncates its trajectory; however, it will be close to Schwarzschild values. We would point this out in subsequent sections wherever it is required.
%Further, the relationship between ($p, e$) and ($E, J_{z}$) is obtained by setting $V_{\textrm{eff}}(r) = 0$ at ($r_{p}, r_{a}$),
%Here, capitalized suffix $P$ represents the quantities associated with the secondary object. 
One can integrate equations in Eq. (\ref{gd1}) by eliminating proper time $\tau$ and considering $r$ to be the orbit parameter. The motion happens in two branches, from $r_{p}$ to $r_{a}$ and back $r_{a}$ to $r_{p}$. In order to overcome the divergences at the turning points ($r_{p},r_{a}$ where $V_{\textrm{eff}}=0$), we parametrize the radial coordinate as
\begin{align}
    r = \frac{p}{1+e\cos\chi},
\end{align}
This parametrization helps in removing the singularity in differential equations at ($r_{p}, r_{a}$) corresponding to ($\chi=0, \chi=\pi$). Further, the $V_{\textrm{eff}}$ takes the following form
\begin{align}
    V_{\textrm{eff}} = \Bigg[\frac{-4ef(r_{p})f(r_{a})}{f(r) \Big(f(r_{p})(1+e)^{2}-f(r_{a})(1-e)^{2}\Big)}+ \frac{(1+e\cos\chi)^{2} (f(r_{p})-f(r_{a}))}{\Big(f(r_{a})(1-e)^{2}-f(r_{p})(1+e)^{2}\Big)}+1\Bigg]b(r)\,.
\end{align}
The orbital period from $r_{p}$ to $r_{p}$ is
\begin{align}
    T_{P} = \int_{0}^{2\pi}d\chi \frac{dt}{d\chi} \hspace{3mm} ; \hspace{7mm} \textup{where}, \hspace{7mm} \frac{dt}{d\chi} = \frac{E}{f(r)\sqrt{-V_{\textrm{eff}}}}\frac{dr}{d\chi}\,.
\end{align}
 For a given $\chi$, $e$ and halo parameters, one can always find the minimum value of $p$ for which $dt/d\chi$ is real; this makes sure the square-root quantity is real in $dt/d\chi$. Also, the minimum possible value of $p$ implies the location where the object ends its trajectory. Unlike the Schwarzschild metric, the spacetime under consideration is not simple, we take only numerical values of minimum $p$. \par
We know that eccentric orbits exhibit two fundamental frequencies- azimuthal ($\Omega_{\phi}$) and radial ($\Omega_{r}$). As in the case of the Schwarzschild \cite{PhysRevD.50.3816}, the radial motion shows the periodicity, not the azimuthal. Therefore, the corresponding expressions can be written in the following manner, %Further, one can determine the average rate at which the azimuthal angle changes and can be calculated by averaging the angular frequency ($d\phi/dt$),
\begin{align}
  \Omega_{r} = \frac{2\pi}{T_{P}} \hspace{3mm} ; \hspace{3mm}  \Omega_{\phi} \equiv \frac{d\phi}{dt} = \frac{J_{z}^{2}f(r)}{r^{2}E} \,. %<\Omega_{\phi}> \equiv \frac{1}{T_{P}}\int_{0}^{2\pi}\frac{d\phi_{P}}{dt}dt \,.
\end{align}
These quantities ultimately help us to estimate GW flux and orbital phase wherever required in subsequent sections. Let us now turn the discussion to perturbation equations.
%%%%%%%%%%%%%%%%%%%%%%%%%%%%%%%%%%%%%%%%%%%%%%%%%%%%%%%%%%%%%%%%%%%%%%%%%%%%%%%%%%%%%%%%%%%%%%%%%%%%%%%%%%%%%%%

\section{Perturbation equations}\label{pert1}

In this section, we provide the setup for perturbation equations, including the energy-momentum tensor of the dark matter profile and the secondary object \cite{PhysRevD.67.104017, Cardoso:2022whc}. Due to the extremely small mass-ratio of the EMRI system, we can consider the secondary object as a perturber to the background spacetime described by Eq.~(\ref{met1}). 
%As we are considering an EMRI system, the secondary object hovering around the primary causes perturbations in the spacetime and energy-momentum tensor. 
The resultant geometry and dark matter energy momentum tensor have the following forms
\begin{align}\label{perturbation}
    g_{\mu\nu} = g_{\mu\nu}^{(0)} + g_{\mu\nu}^{(1)} \hspace{5mm} ; \hspace{5mm} T^{DM}_{\mu\nu} = T^{\textrm{DM}(0)}_{\mu\nu} + T^{\textrm{DM}(1)}_{\mu\nu},
\end{align}
where superscript (1) denotes the perturbations and `DM' denotes the dark matter. In the Regge-Wheeler-Zerilli gauge  \cite{PhysRevLett.24.737, PhysRevD.2.2141}, the metric perturbation can be decomposed into tensor spherical harmonics as $g_{\mu\nu}^{(1)} = g_{\mu\nu}^{(1) \textrm{axial}}+g_{\mu\nu}^{(1) \textrm{polar}}$, where $g_{\mu\nu}^{(1) \textrm{axial}}$ and  $g_{\mu\nu}^{(1) \textrm{polar}}$ represent the axial and polar perturbations respectively, whose expressions are provided in Eq.~(\ref{prt1}).
 The perturbed density and pressure of the dark matter halo can also be decomposed into tensor spherical harmonics as 
\begin{equation}\label{delta_rho_pt}
\mathcal{C}^{(1)} = \sum_{l=2}^{\infty}\sum_{m=-l}^{l} \delta \mathcal{C}^{lm}(t,r) Y_{lm}(\theta, \phi)\,,\qquad \mathcal{C}\in\left\{\rho, p_{t}, p_{r}\right\} .
\end{equation}
where $Y_{lm}(\theta, \phi)$ denotes the spherical harmonics on 2-sphere.
We can define the radial and transverse sound speeds ($c_{sr}, c_{st}$) through the following relations $\delta p^{lm}_{t} = c^{2}_{st} \delta\rho^{lm}$ and $\delta p^{lm}_{r} = c^{2}_{sr} \delta\rho^{lm}$. To construct the perturbed energy-momentum tensor of the halo, we first perturb the four-velocity of the fluid and the normal vector as  
\begin{equation}
\begin{aligned}
u^{\mu}=u^{\mu}_{(0)}+u^{\mu}_{(1)}\,,\qquad k^{\mu}=k^{\mu}_{(0)}+k^{\mu}_{(1)}
\end{aligned}
\end{equation}
We can describe $u^{\mu}_{(1)}$ and $k^{\mu}_{(1)}$ in terms of three functions $\{U_{l m}(t,r), V_{l m}(t,r), W_{l m}(t,r)\}$ (see  Eq.~(\ref{fluid_4_vel}) and Eq.~(\ref{fluid_normal_vec})) \cite{PhysRevD.67.104017}.
By substituting the values of the perturbed 4-velocity $u^{\mu}$ and the normal vector $k^{\mu}$ along with Eq.~(\ref{delta_rho_pt}) in Eq.~(\ref{stt1}), we can find the expression for perturbed energy-momentum tensor $T_{\mu\nu}^{DM (1)}$. The non-vanishing components of the tensor are provided in  Eq.~(\ref{fluid_pert_em_tensor}).\par
%where, , defined with the radial and transverse sound speeds ($c_{sr}(r), c_{st}(r)$). %It is to add that we ($l, m$) indices in the suffix now for improving the readability. 
%Further, the perturbations to the fluid environment are given by
So far we have the expressions for the perturbed energy-momentum tensor of the dark matter halo. We now model the source of perturbation, the secondary object, as a pointlike whose energy-momentum tensor can be written as \cite{Poisson:2011nh}

\begin{align}\label{src}
    T^{\mu\nu}_{P} = \mu \int d\tau \frac{\delta^{(4)}(x^{\mu}-z^{\mu}_{P}(\tau))}{\sqrt{-g}}U^{\mu}_{P}U^{\nu}_{P},
\end{align}
where $\mu$ is the mass of the secondary object and $U^{\mu}_{P}$ denotes the four-velocity. $\tau$ is the proper time along the worldline $z^{\mu}_{P}$ and $U^{\mu}_{P} = dz^{\mu}_{P}/d\tau$ represents the tangent to the line. Since, we are considering the equatorial eccentric orbits, the four-velocity of the secondary object, exhibiting eccentric motion, can be written as $U^{\mu}_{P} = \Big(U^{t}, U^{r}, 0, U^{\phi}\Big)$, where the expressions for $U^{t},~ U^{r}$ and $U^{\phi}$ are given in Eq.~(\ref{gd1}).  

%%%%%%%%%%%%%%%%%%%%%%%%%%%%%%%%%%%%%%%%%%%%%%%%%%%%%%%%%%%%%%%

\section{Axial and Polar perturbations}\label{axpt}
In this section, we obtain the equations for the axial and polar perturbations where we use the relevant expressions of the perturbed metric and energy-momentum tensor given in Eq. (\ref{perturbation}). We recall again that we are considering the vanishing background radial pressure. Let us start with the axial perturbation setup.

\subsection{Axial sector}\label{ax1}
We define the perturbed Einstein field equation by $\mathcal{E}_{\mu\nu}$ that carries the perturbed geometry of the spacetime as well as the perturbed energy-momentum tensor of the dark matter environment and the secondary. 
%In order to deal with the axial perturbations, it involves finding the solutions for $(h_{0}^{lm}, h_{1}^{lm}, U^{lm})$. 
The perturbed field equation is given by
\begin{align}
    \mathcal{E}_{\mu\nu} = G^{(1)}_{\mu\nu} - 8\pi (T_{\mu\nu}^{DM (1)}+T^{P}_{\mu\nu})=0,
\end{align}
where $G^{(1)}_{\mu\nu}$ is the perturbed Einstein tensor. The perturbation in the axial sector can be described in terms of a master function $\mathcal{R}_{lm\omega}(r)$ which satisfies the following differential equation 
%%%
\begin{align}
    \Big(\frac{d^{2}}{dr_{*}^{2}}+\omega^{2}-V^{ax}\Big)\mathcal{R}_{lm\omega} = S^{ax}_{lm\omega}, \label{eq:master_axial}
\end{align}
where the tortoise coordinate follows the relation $dr_{*}=dr/\sqrt{fb}$, and the potential is given by
\begin{equation}
V^{ax} = \dfrac{f}{r^2}\Big(l(l+1) - \dfrac{6m(r)}{r} + m'(r)  \Big),
\end{equation}
%%%
and the expression of the source term is given in Eq.~(\ref{eq:Source}). The details about the computation of the axial perturbation equation along with boundary conditions are given in Appendix \ref{App_axial}.
\subsection{Polar sector}
 In this section, we mention the details regarding the polar perturbation equation. It turns out that the perturbation in the polar sector can be described in terms of a set of five coupled inhomogeneous first-order ordinary differential equations %that in general can be written as
%In this section, we derive a set of first-order differential equations for the metric polar perturbations using components of $\mathcal{E}_{\mu\nu}$. The components $\mathcal{E}_{tr}$, $\mathcal{E}_{t\theta}$ and $\mathcal{E}_{r\theta}$ three inhomogeneous differential equations for $\frac{dK}{dr}$, $\frac{dH_{1}}{dr}$ and $\frac{dH_{0}}{dr}$ respectively. We further use the conservation of energy-momentum tensor ($\nabla_{\mu}T^{\mu\nu}=0$) in order to find two other differential equations for variables $W'(r)$ and $\delta\rho'(r)$ together with an algebraic relation for fluid velocity component $V$. We further simplify the equations by replacing $H_{2}$ and its derivatives with the help of $\mathcal{E}_{\theta\phi}$. With this setup, we get five coupled inhomogeneous first-order ordinary differential equations (ODEs) for $\vec{\psi}=(H_1, H_0, K, W,\delta\rho)$ that can be ultimately written in the matrix form in the following way,
\begin{align}\label{ppte1}
  \frac{d\vec{\psi}_{lm\omega}}{dr}-\boldsymbol{\alpha}\vec{\psi}_{lm\omega}=\vec S^{pol}_{lm\omega} ,
\end{align}
where $\vec{\psi}_{lm\omega}=(H_1^{lm}, H_0^{lm}, K^{lm}, W^{lm},\delta\rho^{lm})$ and $\vec S^{pol}_{lm\omega}$ denotes the source term and $\boldsymbol{\alpha}$ is a matrix. The components of $\vec S^{pol}_{lm\omega}$ and $\boldsymbol{\alpha}$ is given in Eq.~(\ref{source_polar}) and 
Eq.~(\ref{alpha_vals}), respectively. The solution of the inhomogeneous equation mentioned above can be written in terms of the \textit{fundamental matrix} solution of the corresponding homogeneous equation. The fundamental matrix of the homogeneous equation is a matrix function $\boldsymbol{\Psi_{lm\omega}}(r)$ which satisfies the differential equation  $d\boldsymbol{\Psi_{lm\omega}}/dr=\boldsymbol{\alpha}\boldsymbol{\Psi_{lm\omega}} $ and every column of the matrix represents a linearly independent solution of the homogeneous equation \cite{wasow1965asymptotic}. The general solution of Eq. (\ref{ppte1}) can be written in terms of the fundamental matrix solution as follows \cite{wasow1965asymptotic}
%%%%%%%%%%%%%%%%%%%%%%%%%%%%%%%%%%%%%%%%%%%%%%%%%%%%%%%%%%%%%%%%
\begin{equation}\label{fundamental_matrix}
    \begin{aligned}
        \vec{\psi}_{lm\omega}(r)=\boldsymbol{\Psi}_{lm\omega}(r)\boldsymbol{\Psi}_{lm\omega}^{-1}(r_B)\vec{\psi}_{lm\omega}(r_B)+\boldsymbol{\Psi}_{lm\omega}(r)\int_{r_B}^{r}dx~\boldsymbol{\Psi}^{-1}_{lm\omega}(x) \vec{S}^{pol}_{lm\omega}(x),
    \end{aligned}
\end{equation}
%%%%%%%%%%%%%%%%%%%%%%%%%%%%%%%%%%%%%%%%%%%%%%%%%%%%%%%%%%%%%%%%
where $r_B$ represents the position of the boundary. We use the above equation to determine the boundary conditions for Eq. (\ref{ppte1}) by series expanding $\vec{\psi}_{lm\omega}(r)$ about the horizon and the infinity. Note that, in this paper, we are interested in the inspiral phase of the EMRI system; thus, only consider the dynamics of secondary in the range $p\in (p_{\textrm{ini}},p_{\textrm{min}})$, where $p_{ini}$ is the starting point of the inspiral which we set $p_{ini}=10$ and $p_{\textrm{min}}$ is the position of the separatrix. Given that the term $\vec{S}^{pol}_{lm\omega}(x)$ contains a Dirac delta function $\delta(r-r_P)$ with $r_P$ representing the position of the secondary, the second term in the right-hand side of Eq. (\ref{fundamental_matrix}) vanishes identically if we expand $\vec{\psi}_{lm\omega}(r)$ about either the event horizon or infinity. Thus, the boundary conditions for the inhomogeneous equation would be the same as the homogeneous equation.\par
In order to determine the boundary condition about the event horizon, we note that Eq. (\ref{ppte1}) has a regular singular point at $r=r_+\equiv 2$. Thus, it is more appropriate to write the homogeneous equation in the following manner
%%%%%%%%%%%%%%%%%%%%%%%%%%%%%%%%%%%%%%%%%%%%%%%%%%%%%%%%%%%%%%%%
\begin{align}\label{ppte2}
 (r-r_+) \frac{d\vec{\psi}_{lm\omega}}{dr}=\tilde{\boldsymbol{\alpha}}\vec{\psi}_{lm\omega} ,
\end{align}
%%%%%%%%%%%%%%%%%%%%%%%%%%%%%%%%%%%%%%%%%%%%%%%%%%%%%%%%%%%%%%%%
where $\tilde{\boldsymbol{\alpha}}=(r-r_+)\boldsymbol{\alpha}$ and $\vec{\tilde{S}}^{pol}_{lm\omega}=(r-r_+)\vec{S}^{pol}_{lm\omega}$. We further, in this particular section, avoid the suffix ($lm\omega$) for writing convenience. Provided that all the eigenvalues $\lambda_j,~(j=1,2,..,5)$ of $\tilde{\boldsymbol{\alpha}}(r=r_+)$ are distinct, the solution of the above equation has the following form \cite{wasow1965asymptotic}
%%%%%%%%%%%%%%%%%%%%%%%%%%%%%%%%%%%%%%%%%%%%%%%%%%%%%%%%%%%%%%%%
\begin{equation}\label{fundamental_matrix_regular}
    \begin{aligned}
       \vec{\psi}_j(r)=\vec q_j(r) (r-r_+)^{\lambda_j}=\left(\sum_{i=0}^{N_H}q_{ji} (r-r_+)^{i}\right)(r-r_+)^{\lambda_j},
    \end{aligned}
\end{equation}
%%%%%%%%%%%%%%%%%%%%%%%%%%%%%%%%%%%%%%%%%%%%%%%%%%%%%%%%%%%%%%%%
where $\vec q_j(r)$ is holomorphic at $r=r_+$, $\vec{\psi}_j$ is the $j$th column of the fundamental matrix $\boldsymbol{\Psi}_{lm\omega}$ and $N_H$ is a finite integer. We have checked that  all the eigenvalues of $\tilde{\boldsymbol{\alpha}}(r=r_+)$ are distinct; however, the solution corresponding to the eigenvalue $\lambda_H\equiv-1+i\omega/f'(r_+)$ represents an incoming solution at the event horizon. We have calculated the vectors $q_{j0},~q_{j1},...,$ by inserting the eigenvalue $\lambda_H$  in Eq.~(\ref{ppte2}) and using the following Taylor expansion $\tilde{\boldsymbol{\alpha}}=\sum_{i=0}^{N_H}\tilde{\boldsymbol{\alpha}}_i(r-r_+)^{i}$. Here, we have taken the expansion upto $N_H=4$ order.\footnote{The \texttt{Mathematica} \cite{Mathematica} function \texttt{AsymptoticDSolveValue} provides an alternative way to calculate the boundary conditions at the horizon and we have checked that the numerical values of the boundary conditions obtained from the fundamental matrix method agree quite well with those obtained using \texttt{AsymptoticDSolveValue}.} Finally, to obtain the boundary conditions at infinity, we use the following expansion, 
%%%%%%%%%%%%%%%%%%%%%%%%%%%%%%%%%%%%%%%%%%%%%%%%%%%%%%%%%%%%%%%%
\begin{equation}\label{infinity_expansion}
    \begin{aligned}
       \psi^{nv}(r)&=e^{i\omega r_{*}}\sum_{i=0}^{N_\infty}\frac{\psi^{nv}_i}{r^i} \,,\qquad \psi^{nv}\in (H_1^{lm}, H_0^{lm}, K^{lm})\\
       \psi^{v}(r)&=0\,,\qquad \psi^{v}\in ( W^{lm},\delta\rho^{lm})
    \end{aligned}
\end{equation}
%%%%%%%%%%%%%%%%%%%%%%%%%%%%%%%%%%%%%%%%%%%%%%%%%%%%%%%%%%%%%%%%
where, $r_{*}$ is the tortoise coordinate. Replacing  Eq. (\ref{infinity_expansion}) in Eq. (\ref{ppte1}), we can write the coefficients  $\psi^{nv}_i$ in terms of $K_0^{lm}$. Here, we have taken the expansion up to $N_{\infty}=4$ order. This gives the behaviour of $\vec\psi$ at large distance of $r=r_{\textrm{inf}}$.\par
%Given that all the eigenvalues $\lambda_j,~(j=1,2,..,5)$ of $\tilde{\boldsymbol{\alpha}}(r=2)$ are distinct, the fundamental matrix solution of the above equation has the following form
%%%%%%%%%%%%%%%%%%%%%%%%%%%%%%%%%%%%%%%%%%%%%%%%%%%%%%%%%%%%%%%%
%\begin{equation}\label{fundamental_matrix_regular}
 %   \begin{aligned}
%        \boldsymbol{\Psi}(r)=\boldsymbol{Q}(r) (r-2)^{\textrm{diag}(\lambda_1,~\lambda_2,...,~\lambda_5)}
%    \end{aligned}
%\end{equation}
%%%%%%%%%%%%%%%%%%%%%%%%%%%%%%%%%%%%%%%%%%%%%%%%%%%%%%%%%%%%%%%%
%where, $\boldsymbol{Q}(r)$ is holomorphic at $r=2$ and $\det \boldsymbol{Q}\neq 0$.
%This equation has a fundamental matrix solution of the form

\section{Numerical Methods and Results}\label{nmr}
In this section, we briefly describe our method to calculate the gravitational wave flux emitted by the secondary and study how the emitted flux influences the dynamics of the secondary. As discussed earlier, we consider that the secondary object revolves around the black hole in an equatorial eccentric orbit. 
%%%%%%%%%%%%%%%%%%%%%%%%%%%%%%%%%%%%%%%%%%%%%%%%%%%%%%%%%%%%%%%%
\subsection{Gravitational wave flux}

Having obtained the boundary conditions for the axial and polar perturbation, we can solve the respective equations. We employ the standard Green function method to solve the axial perturbation equation Eq. (\ref{eq:master_axial}). Consider that $\ca{R}_{lm\omega}^{\In}$ and $\ca{R}_{lm\omega}^{\up}$ representing the solution of homogeneous axial perturbation equation where $\ca{R}_{lm\omega}^{\In}$ satisfies purely incoming boundary condition at the event horizon whereas $\ca{R}_{lm\omega}^{\up}$ satisfies purely outgoing boundary condition at infinity. The Green function method dictates that the solution of Eq. (\ref{eq:master_axial}) can be written as 
%%%%%%%%%%%%%%%%%%%%%%%%%%%%%%%%%%%%%%%%%%%%%%%%%%%%%%%%%%%%%%%%
\begin{equation}\label{axial_sol}
\begin{aligned}
\ca{R}_{lm\omega}(r)=\frac{1}{\ca{W}}\left[\ca{R}_{lm\omega}^{\up}\int_{r_+}^{r}dr_{*}\ca{R}_{lm\omega}^{\In} \ca{S}_{lm\omega}^{ax}+\ca{R}_{lm\omega}^{\In}\int_{r}^{\infty}dr_{*}\ca{R}_{lm\omega}^{\up} \ca{S}_{lm\omega}^{ax}\right]~,
\end{aligned}
\end{equation}
%%%%%%%%%%%%%%%%%%%%%%%%%%%%%%%%%%%%%%%%%%%%%%%%%%%%%%%%%%%%%%%%
where $\ca{W}$ is the constant Wronskian which is given by the following expression
%%%%%%%%%%%%%%%%%%%%%%%%%%%%%%%%%%%%%%%%%%%%%%%%%%%%%%%%%%%%%%%%
\begin{equation}\label{Wronskian}
\begin{aligned}
\ca{W}\equiv \left[\ca{R}_{lm\omega}^{\In}\frac{d\ca{R}_{lm\omega}^{\up}}{dr_*}-\ca{R}_{lm\omega}^{\up}\frac{d\ca{R}_{lm\omega}^{\In}}{dr_*}\right]\,.
\end{aligned}
\end{equation}
%%%%%%%%%%%%%%%%%%%%%%%%%%%%%%%%%%%%%%%%%%%%%%%%%%%%%%%%%%%%%%%%
%%%%%%%%%%%%%%%%%%%%%%%%%%%%%%%%%%%%%%%%%%%%%%%%%%%%%%%%%%%%%%%%%%%

%%%%%%%%%%%%%%%%%%%%%%%%%%%%%%%%%%%%%%%%%%%%%%%%%%%%%%%%%%%%%%%%%%%%%%%%%%%%%%%%%%%%%%%%%%%%%%%%%%%
To obtain $\ca{R}_{lm\omega}^{\In}$ and $\ca{R}_{lm\omega}^{\up}$, we solve the homogeneous part of Eq. (\ref{eq:master_axial}) with boundary conditions mentioned in Eq.~(\ref{bc_axial_horizon}) and Eq.~(\ref{bc_axial_inf}) in the domain $r\in (r_0,r_{\textrm{inf}})$, where $r_0=2(1+\epsilon)$ ($\epsilon\ll 1$). The upper limit $r_{\textrm{inf}}$ is determined as $r_{\textrm{inf}}=\textrm{max}[10^3/\Omega_\phi,2a_0]$, where $\Omega_\phi$ represents the azimuthal frequency. Furthermore, following \cite{Cardoso:2022whc}, we approximate the Dirac delta function appearing in the expression of $\ca{S}_{lm\omega}^{ax}$ (see Eq.~(\ref{eq:Source}) and Eq.~(\ref{dfvcxv})) by a Gaussian distribution function 
%%%%%%%%%%%%%%%%%%%%%%%%%%%%%%%%%%%%%%%%%%%%%%%%%%%%%%%%%%%%%%%%
\begin{equation}\label{dirac_delta}
\begin{aligned}
\delta(r-r_P)=\frac{1}{\sqrt{2\pi}\sigma}\exp{\left[-\frac{(r-r_P)^2}{2\sigma^2}\right]},
\end{aligned}
\end{equation}
%%%%%%%%%%%%%%%%%%%%%%%%%%%%%%%%%%%%%%%%%%%%%%%%%%%%%%%%%%%%%%%%
where the variance $\sigma$ is adjusted iteratively to get better convergences.\par
%solve the governing axial and polar 
%\begin{center}
  
%\end{center}
To solve the polar perturbation equation Eq. (\ref{ppte1}), we utilize the packages \cite{cardoso, maseli} with slight modifications for our eccentric case. These packages employ the ``shooting method'' to obtain the solution of the polar perturbation equation \cite{Berti:2009kk, Cardoso:2017njb, Cardoso:2008bp}. The basic idea behind this method is as follows: given the boundary condition Eq. (\ref{fundamental_matrix_regular}) at the event horizon, we integrate equation Eq. (\ref{ppte1}) from a point $r_0=2(1+\epsilon)$ (where $\epsilon\ll 1$) in close proximity to the horizon up to $r_{\textrm{inf}}$. We then obtain  $\vec\psi_{lm\omega}$ by comparing this solution with the boundary condition at $r_{\textrm{inf}}$.\par
With the solution of the axial and polar perturbation equation, we can calculate the energy and angular momentum flux at infinity using the following relation \cite{RevModPhys.52.299, PhysRevD.70.084044, PhysRevD.69.044025}
%%%%%%%%%%%%%%%%%%%%%%%%%%%%%%%%%%%%%%%%%%%%%%%%%%%%%%%%%%%%%%%%
\begin{equation}\label{Flux_eqn}
\begin{aligned}
\frac{dE}{dt}=\sum_{lm}\frac{dE_{lm}}{dt}&=\frac{1}{32\pi}\sum_{lm}\frac{(l+2)!}{(l-2)!}\left[|\dot Z_{lm\omega}^{\textrm{polar}}|^2+4|Z_{lm\omega}^{\textrm{axial}}|^2\right]\,,\\
\frac{dJ_z}{dt}=\sum_{lm}\frac{dJ_{lm}}{dt}=\frac{1}{32\pi}&\sum_{lm}im\frac{(l+2)!}{(l-2)!}\left[\dot Z_{lm\omega}^{\textrm{polar}}\tilde Z_{lm\omega}^{\textrm{polar}}+4 Z_{lm\omega}^{\textrm{axial}}\int dt~\tilde Z_{lm\omega}^{\textrm{axial}}\right]+\textrm{c.c},
\end{aligned}
\end{equation}
%%%%%%%%%%%%%%%%%%%%%%%%%%%%%%%%%%%%%%%%%%%%%%%%%%%%%%%%%%%%%%%%
where ``tilde'' represents complex conjugation, $\textrm{c.c}$ represents complex conjugate, $E_{lm}$ and $J_{lm}$ represents the energy and angular momentum flux for each multipole mode $\{l,m\}$ respectively, $Z_{lm\omega}^{\textrm{axial}}=\mathcal{R}_{lm\omega}$ and following \cite{Cardoso:2022whc}, we take $Z_{lm\omega}^{\textrm{polar}}$ as %\footnote{write footnote here}
%%%%%%%%%%%%%%%%%%%%%%%%%%%%%%%%%%%%%%%%%%%%%%%%%%%%%%%%%%%%%%%%
\begin{equation}\label{polar_master}
\begin{aligned}
Z_{lm\omega}^{\textrm{polar}}=\frac{r}{n+1}\left[K^{lm}+\frac{f}{n}\left(H^{lm}_2-r\frac{\partial K^{lm}}{\partial r}\right)\right],
\end{aligned}
\end{equation}
%%%%%%%%%%%%%%%%%%%%%%%%%%%%%%%%%%%%%%%%%%%%%%%%%%%%%%%%%%%%%%%%
where $n=l(l+1)/2-1$. The average flux for each multipole mode $\{l,m\}$ over an orbital cycle is given by the following relation \cite{PhysRevD.53.3064}
%%%%%%%%%%%%%%%%%%%%%%%%%%%%%%%%%%%%%%%%%%%%%%%%%%%%%%%%%%%%%%%%
\begin{equation}\label{ave_flux}
\begin{aligned}
\left\langle\frac{dE_{lm}}{dt}\right\rangle_{\textrm{GW}}=\frac{1}{T_P}\int_{0}^{2\pi}d\chi \frac{dt}{d\chi}\frac{dE_{lm}}{dt}\,,\qquad\left\langle\frac{dJ_{lm}}{dt}\right\rangle_{\textrm{GW}}=\frac{1}{T_P}\int_{0}^{2\pi}d\chi \frac{dt}{d\chi}\frac{dJ_{lm}}{dt}\,.
\end{aligned}
\end{equation}
%%%%%%%%%%%%%%%%%%%%%%%%%%%%%%%%%%%%%%%%%%%%%%%%%%%%%%%%%%%%%%%%

\begin{figure}[t!]
	%%%%%%%%%%%%%%%%%%%%%%%%
	\centering
	\minipage{0.48\textwidth}
	\includegraphics[width=\linewidth]{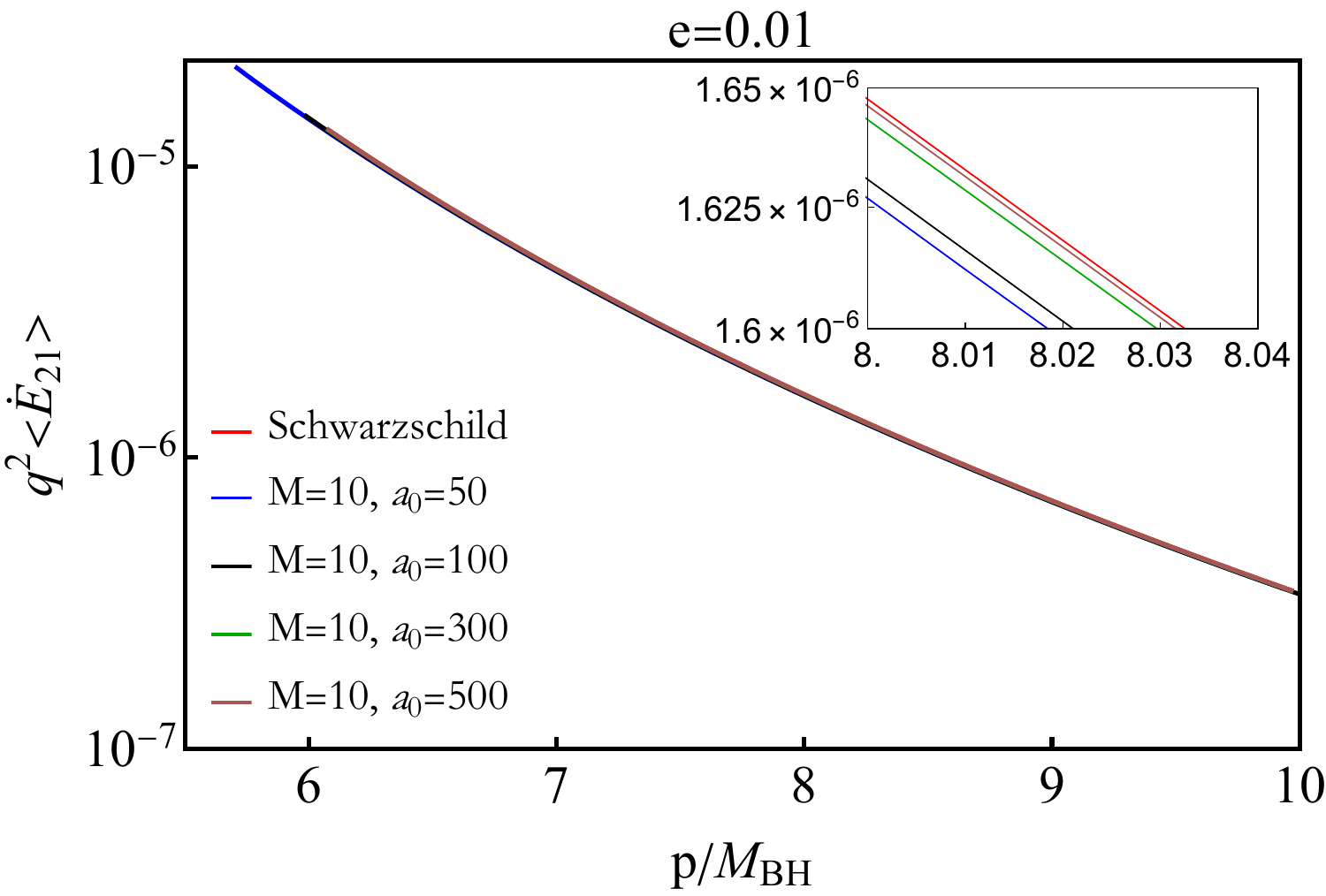}
% \caption{Wormholes for $\Lambda=0$}
	\endminipage\hfill
	%%%%%%%%%%%%%%%%%%%%%%%%
	\minipage{0.48\textwidth}
	\includegraphics[width=\linewidth]{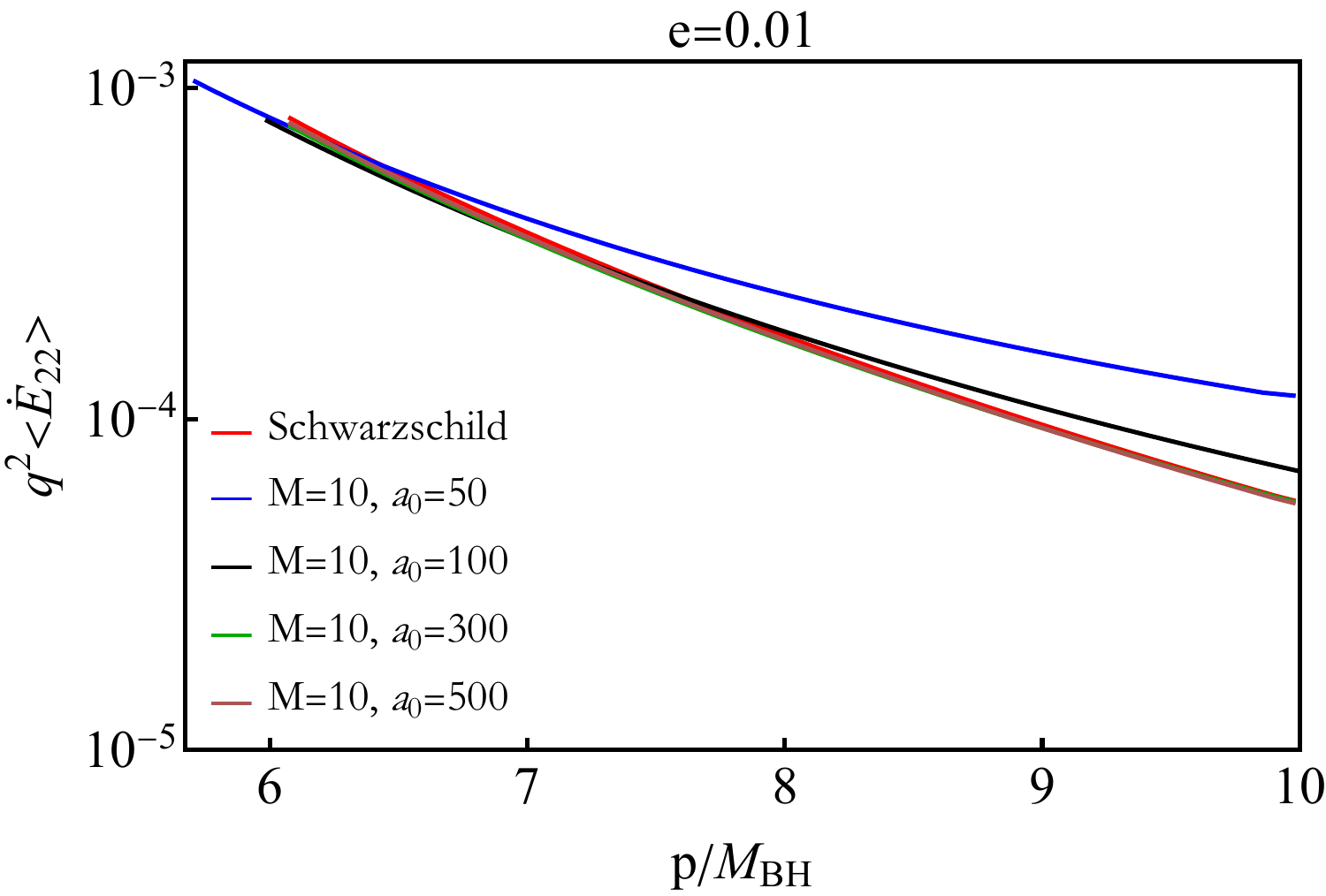}
	\endminipage\hfill
 \minipage{0.48\textwidth}
	\includegraphics[width=\linewidth]{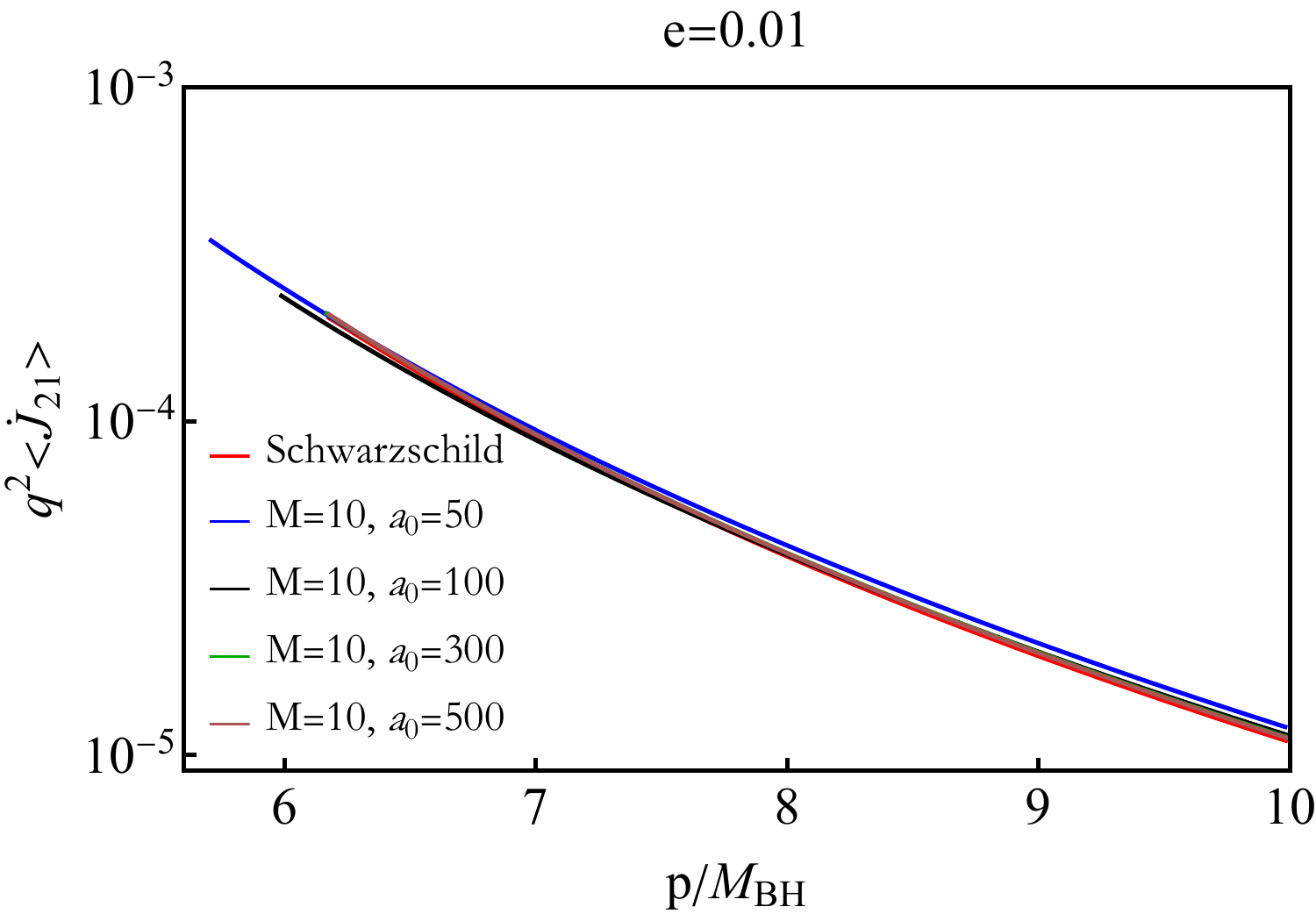}
	\endminipage\hfill
 \centering
	\minipage{0.48\textwidth}
	\includegraphics[width=\linewidth]{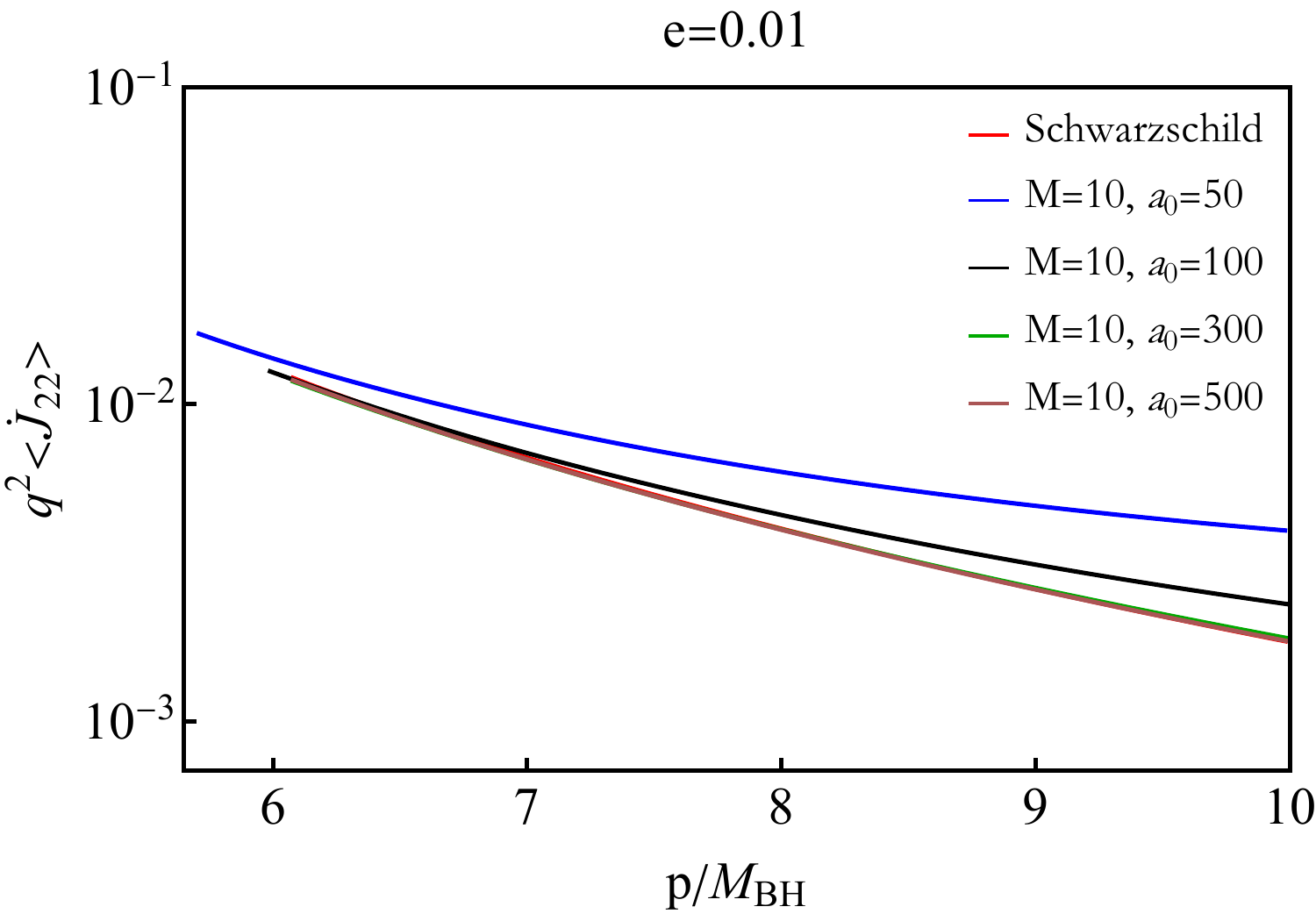}
% \caption{Wormholes for $\Lambda=0$}
	\endminipage
	\caption{In the upper panel of Fig. (\ref{fig_enr_flux_same_e}), we show the average energy flux for $\{2,1\}$ (left panel) and $\{2,2\}$ (right panel) mode in the range $p\in(p_{\textrm{min}}(e),~p_{\textrm{ini}})$ for a fixed value of orbital eccentricity and different values of dark matter parameters $M$ and $a_0$. The lower panel depicts the same for angular momentum flux. As stated before, we set $p_{\textrm{ini}}=10$. In the upper panel, the plot of the average energy flux $\langle\dot{E}_{lm}\rangle$ for the axial $\{2,1\}$ mode (left panel) and the polar $\{2,2\}$ mode (right panel) as a function of the semi-latus rectum $p$, considering different values of dark matter parameters, is presented. The lower panel displays the corresponding plots for the average angular momentum flux $\langle\dot{J}_{lm}\rangle$ of the $\{2,1\}$ mode (left panel) and the $\{2,2\}$ mode (right panel). We consider the eccentricity value as $e=0.01$. The red line in each of these plots represents the flux for a Schwarzschild black hole.
 }\label{fig_enr_flux_same_e}
\end{figure}

In the upper panel of Fig. (\ref{fig_enr_flux_same_e}), we show the average energy flux for $\{2,1\}$ (left panel) and $\{2,2\}$ (right panel) mode in the range $p\in(p_{\textrm{min}}(e),~p_{\textrm{ini}})$ for a fixed value of orbital eccentricity and different values of dark matter parameters $M$ and $a_0$. The lower panel depicts the same for angular momentum flux. As stated before, we set $p_{\textrm{ini}}=10$. To mimic a real astrophysical scenario, we consider $M_{\textrm{BH}}\ll M\ll a_0$. Furthermore, to make a comparison with \cite{Cardoso:2022whc}, we consider $c_{sr}=0.9$ and $c_{st}=0$ throughout the paper. In each of these plots, the red curve represents the value of energy (upper panel) and angular momentum  (lower panel) flux for a Schwarzschild black hole. The energy and angular momentum flux is almost identical to that of a Schwarzschild black hole when the halo compactness parameter $M/a_0$ is small. However, as we increase the value of the halo compactness parameter, the value of the energy and angular momentum flux shift away from the respective Schwarzschild value. 

  \begin{table}[t!]\centering
	\begin{tabular}{c c c c c c} 
		\hline\hline
		$l$ & \hspace{3mm} $m$ & \hspace{3mm} $p$ & \hspace{3mm} $\langle\dot{E}_{lm}\rangle$ & \hspace{3mm} $\langle\dot{J}_{lm}\rangle$ \\
		\hline\hline
		\multirow{3}*{2} & \hspace{3mm}\multirow{3}*{2} & \hspace{3mm} 10 & \hspace{3mm} 9.67803e-5 & \hspace{3mm} 3.27568e-3 \\ 
						 &	  			    & \hspace{3mm} 8 & \hspace{3mm} 2.32578e-4 &	\hspace{3mm} 5.71542e-3	    \\
       				 &	  			    & \hspace{3mm} 6.1 & \hspace{3mm} 7.88156e-4 &		\hspace{3mm} 1.25629e-2    \\
		\hline
		\multirow{3}*{2} & \hspace{3mm} \multirow{3}*{1} & \hspace{3mm} 10 & \hspace{3mm} 3.96527e-7 & \hspace{3mm} 1.34211e-5 \\ 
						 &	  			    & \hspace{3mm} 8 & \hspace{3mm} 1.85898e-6 &	\hspace{3mm} 4.56830e-5	    \\
       				 &	  			    & \hspace{3mm} 6.1 & \hspace{3mm} 1.68476e-5 &		\hspace{3mm} 2.68545e-4  			  \\
		\hline  
		%				%
		\multirow{3}*{3} & \hspace{3mm} \multirow{3}*{3} & \hspace{3mm} 10 & \hspace{3mm} 4.38703e-5 & \hspace{3mm} 1.48486e-3 \\ 
						 &	  			    & \hspace{3mm} 8 & \hspace{3mm} 8.05536e-5 &	\hspace{3mm} 1.97954e-3	    \\
       				 &	  			    & \hspace{3mm} 6.1 & \hspace{3mm} 1.01981e-4 &		\hspace{3mm} 1.62554e-3  			\\
		\hline
		\multirow{3}*{3} & \hspace{3mm} \multirow{3}*{2} & \hspace{3mm} 10 & \hspace{3mm} 1.00542e-7 & \hspace{3mm} 3.40300e-6 \\ 
						 &	  			    & \hspace{3mm} 8 & \hspace{3mm} 5.69611e-7 &	\hspace{3mm} 1.39977e-5	    \\
       				 &	  			    & \hspace{3mm} 6.1 & \hspace{3mm} 6.55838e-6 &		\hspace{3mm} 1.04538e-4  		\\
		\hline\hline
	\end{tabular}
\caption{
%$e=0.1, M=10, a_{0}=100$. 
Average energy $\langle\dot{E}_{lm}\rangle$ and angular momentum flux $\langle\dot{J}_{lm}\rangle$ for different modes at different semi-latus rectum ($p$) points. We provide the data considering  $e=0.1$, $M=10M_{\textrm{BH}}$ and $a_0=10M$. Polar modes are represented with $l=m$ modes, whereas $l=m+1$ correspond to axial modes.
}\label{tab:Fluxes_Comparison}
\end{table}

In Table~\ref{tab:Fluxes_Comparison}, we present the energy and angular momentum flux of different modes. Given that  $l=m+1$ and $l=m$ modes correspond to axial and polar mode excitation, \textit{we can see that flux contribution from the leading order polar mode $\{2,2\}$ is the order of magnitude higher than the axial and higher order polar modes. Thus, the orbital dynamics are mainly dictated by leading order polar flux.}
%With the boundary conditions at our hand, we solve \ref{ppte1} using ``shooting method''. The basis idea is to integrate the equation \ref{ppte1} from a point $r_0=2(1+\epsilon)$ ($\epsilon\ll 1$) very close to the event horizon to $r_{\textrm{inf}}$. We obtain $\vec\psi$ by comparing the solution with boundary condition at $r_{\textrm{inf}}$. Here, we have set $r_{\textrm{inf}}=\textrm{max}\{10^3/\Omega_P,2a_0\}$. Furthermore, following \cite{}, we have approximated the Dirac delta function by a Gaussian distribution function $\delta(r-r_P)=\exp{-(r-r_P)^2/2\sigma^2}/\sqrt{2\pi}\sigma$.
%%%%%%%%%%%%%%%%%%%%%%%%%%%%%%%%%%%%%%%%%%%%%%%%%%%%%%%%%%%%%%%%
%%%%%%%%%%%%%%%%%%%%%%%%%%%%%%%%%%%%%%%%%%%%%%%%%%%%%%%%%%%%%%%%
\subsection{Orbital Evolution}
%%%%%%%%%%%%%%%%%%%%%%%%%%%%%%%%%%%%%%%%%%%%%%%%%%%%%%%%%%%%%%%%

%%%%%%%%%%%%%%%%%%%%%%%%%%%%%%%%%%%%%%%%%%%%%%%%%%%%%%%%%%%%%%%%%
%%%%%%%%%%%%%%%%%%%%%%%%%%%%%%%%%%%%%%%%%%%%%%%%%%%
We make use of the \textit{adiabatic approximation} to study the evolution of the object \cite{PhysRevD.78.064028, PhysRevD.103.104014,Isoyama:2021jjd}. This approximation makes use of the fact that the timescale associated with the orbital evolution ($T_P\sim M_{\textrm{BH}}=1$) is much shorter than inspiral timescale ($T_P\sim M_{\textrm{BH}}/q \gg 1$). Thus, over short timescales ($\sim M_{\textrm{BH}}$), we can approximate the particle's trajectory as geodesic, characterized by its energy and angular momentum. However, over a longer timescale ($\sim M_{\textrm{BH}}/q$), the radiation backreaction comes into play. The gravitational wave emission causes the system to lose energy and angular momentum at the rate \cite{PhysRevD.103.104014} 
%%%%%%%%%%%%%%%%%%%%%%%%%%%%%%%%%%%%%%%%%%%%%%%%%%%%%%%%%%%%%%%%
\begin{equation}\label{balance_law}
    \begin{aligned}
        \frac{dE}{dt}=-\left\langle\frac{dE}{dt}\right\rangle_{\textrm{GW}}\,,\qquad \frac{dJ_z}{dt}=-\left\langle\frac{dJ_z}{dt}\right\rangle_{\textrm{GW}}\,.
    \end{aligned}
\end{equation}
%%%%%%%%%%%%%%%%%%%%%%%%%%%%%%%%%%%%%%%%%%%%%%%%%%%%%%%%%%%%%%%%
As a result, the orbital separation between the objects decreases. Thus, the adiabatic approximation can be thought of as a flow between a sequence of geodesic with the flow rate dictated by the balance law given by Eq. (\ref{balance_law}). Note that the approximation breaks down when the object crosses the last stable orbit and  the object started to plunge into the horizon, as the radiation backreaction no longer drives the evolution of the system in the plunging phase \cite{PhysRevD.78.064028}.  Thus, we study the evolution of the system in the domain $p\in (p_{\textrm{ini}}, p_{\textrm{min}})$ where we have set $p_{\textrm{ini}}=10$.\par%The gravitational wave emission cause the system to lose energy and angular momentum
%Thus, over a short duration ($T_P$), we   
From Eq. (\ref{balance_law}), we can calculate the rate of change of semi-latus rectum and eccentricity by inverting the relation 
%%%%%%%%%%%%%%%%%%%%%%%%%%%%%%%%%%%%%%%%%%%%%%%%%%%%%%%%%%%%%%%%
\begin{equation}\label{EJtope}
    \begin{aligned}
      \frac{dE}{dt}&=\frac{\partial E}{\partial p}\frac{dp}{dt}+\frac{\partial E}{\partial e}\frac{de}{dt}\,,\qquad\frac{dJ_z}{dt}&=\frac{\partial J_z}{\partial p}\frac{dp}{dt}+\frac{\partial J_z}{\partial e}\frac{de}{dt},
    \end{aligned}
\end{equation}
%%%%%%%%%%%%%%%%%%%%%%%%%%%%%%%%%%%%%%%%%%%%%%%%%%%%%%%%%%%%%%%%
in the following manner \cite{Barsanti:2022ana, PhysRevD.50.3816}
%%%%%%%%%%%%%%%%%%%%%%%%%%%%%%%%%%%%%%%%%%%%%%%%%%%%%%%%%%%%%%%%
\begin{equation}\label{petoEJ}
    \begin{aligned}
      \frac{dp}{dt}&=\frac{1}{H}\left[\frac{\partial J_z}{\partial e}\frac{dE}{dt}-\frac{\partial E}{\partial e}\frac{dJ_z}{dt}\right]\,,\qquad\frac{de}{dt}&=\frac{1}{H}\left[\frac{\partial E}{\partial p}\frac{dJ_z}{dt}-\frac{\partial J_z}{\partial p}\frac{dE}{dt}\right],
    \end{aligned}
\end{equation}
%%%%%%%%%%%%%%%%%%%%%%%%%%%%%%%%%%%%%%%%%%%%%%%%%%%%%%%%%%%%%%%%
where, $H=\partial_p E~\partial_e J_z-\partial_p J_z~\partial_e E$. %In the absence of gravitational wave radiation, the semi-latus rectum $p$ and the eccentricity $e$ are constants of motion. However, the radiation backreaction effect causes these parameters to evolve adiabatically in accordance to the relation \ref{petoEJ}. Thus, the evolution generates curves in the $p-e$ plane where the tangent vector $de/dp$ can be obatined from the relation \ref{petoEJ}.
%%%%%%%%%%%%%%%%%%%%%%%%%%%%%%%%%%%%%%%%%%%%%%%%%%%%%%%%%%%%%%%%%
\begin{figure}[b!]
	%%%%%%%%%%%%%%%%%%%%%%%%
	\centering
	\minipage{0.48\textwidth}
	\includegraphics[width=\linewidth]{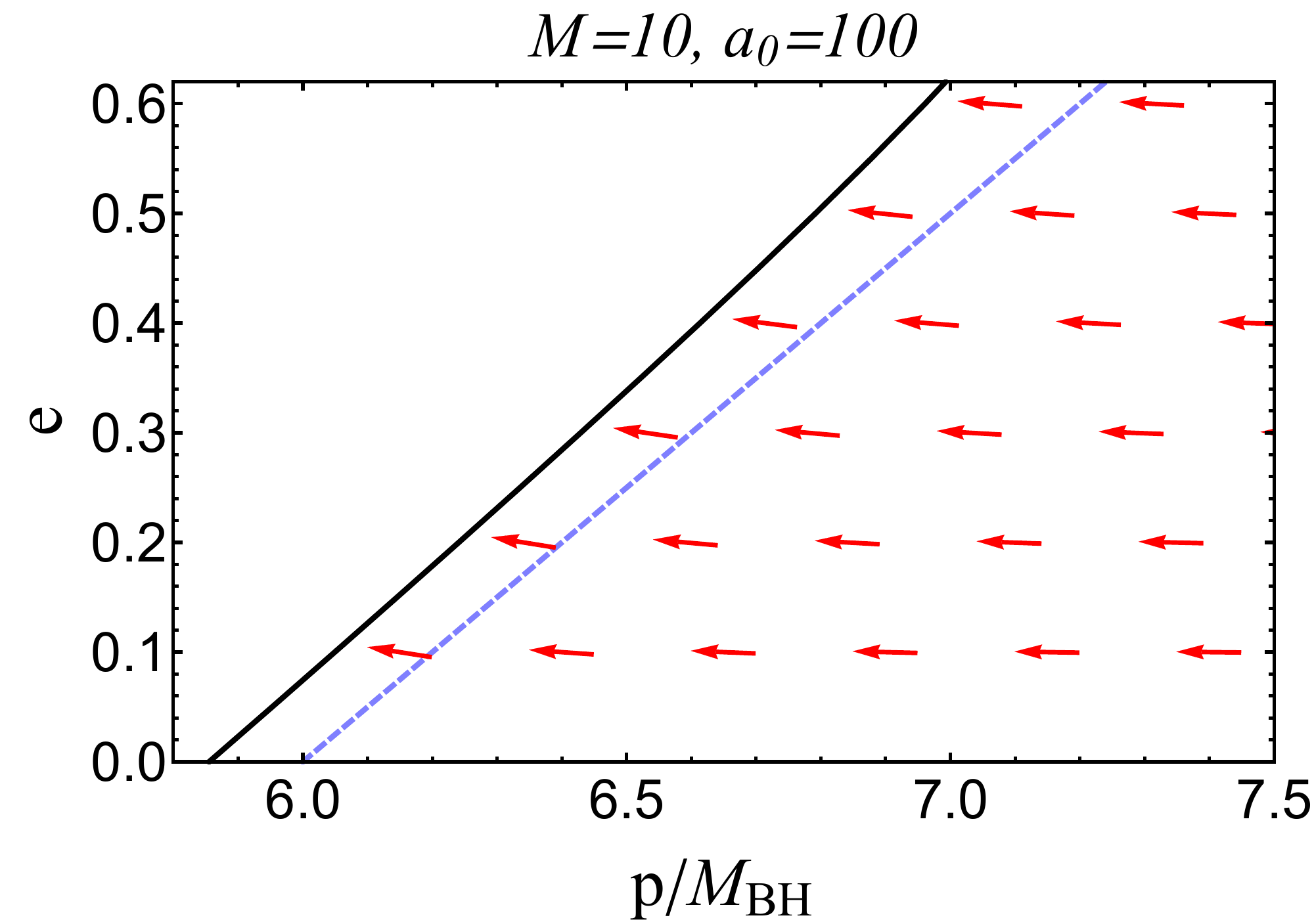}
% \caption{Wormholes for $\Lambda=0$}
	\endminipage
	\caption{The tangent vector field $\vec{v}_{\textrm{tan}}(p,e)=(1,de/dp)$ depicted as red arrows in the $p$-$e$ plane. 
 The dark matter parameters are taken as $M=10$ and $a_0=100$. The solid black line represents the separatrix for the aforementioned values of the dark matter parameters, while the blue dashed line represents the separatrix for Schwarzschild black hole. The downward-pointing arrows indicate a decrease in eccentricity while upward-pointing arrows indicate an increase in eccentricity caused by radiation backreaction. Note that the radiation backreaction effects increase the eccentricity value near the separatrix.
 }\label{fig_pevector_plot}
\end{figure}	
%%%%%%%%%%%%%%%%%%%%%%%%%%%%%%%%%%%%%%%%%%%%%%%%%%%

When gravitational wave radiation is absent, the semi-latus rectum $p$ and the eccentricity $e$ remain constant. However, due to the effect of radiation backreaction, these parameters undergo adiabatic evolution described by the relation Eq. (\ref{petoEJ}). Consequently, this evolution gives rise to curves in the $p$-$e$ plane, where the tangent vector $\vec{v}_{\textrm{tan}}(p,e)=(1,de/dp)$ at the point $(p,e)$ can be determined from Eq. (\ref{petoEJ}). In Fig. (\ref{fig_pevector_plot}), we show tangent vector fields as red arrows in the $p$-$e$ plane. We take the dark matter parameters as $M=10$ and $a_0=100$. The black solid line represents the separatrix for the above-mentioned values of the dark matter parameters while the blue dashed line is the separatrix for Schwarzschild black hole.  In this plot, downward-pointing arrows indicate a decrease in eccentricity due to the effect of radiation backreaction, while upward-pointing arrows indicate an increase in eccentricity caused by radiation backreaction. It is evident from the plot that the radiation backreaction effect increases the eccentricity value near the separatrix. This phenomenon is also observed in the Schwarzschild black hole. When the gravitational field is extremely strong i.e., if we consider points that are close to the separatrix of a Schwarzschild black hole $p_{\textrm{min}}=6+2e$ in the $p$-$e$ plane, then the rate of change of orbital parameters $(p,e)$ due to radiation backreaction effects follows the relation \cite{PhysRevD.50.3816}
%%%%%%%%%%%%%%%%%%%%%%%%%%%%%%%%%%%%%%%%%%%%%%%%%%%%%%%%%%%%%%%%
\begin{equation}
    \begin{aligned}
        \frac{d\ln e}{d\ln p}\bigg|_{p\to  6+2e,e\gg ~\varepsilon/4 }\sim -\frac{1-e}{e}.
    \end{aligned}
\end{equation}
%%%%%%%%%%%%%%%%%%%%%%%%%%%%%%%%%%%%%%%%%%%%%%%%%%%%%%%%%%%%%%%%
with $dp/dt<0$, $de/dt>0$ and $\varepsilon=p-6-2e$. Thus, the eccentricity value increases near the separatrix due to the radiation backreaction even for the Schwarzschild black hole.\par
%%%%%%%%%%%%%%%%%%%%%%%%%%%%%%%%%%%%%%%%%%%%%%%%%%%%%%%%%%%%%%%%%%
\begin{figure}[b!]
	%%%%%%%%%%%%%%%%%%%%%%%%
	\centering
	\minipage{0.48\textwidth}
	\includegraphics[width=\linewidth]{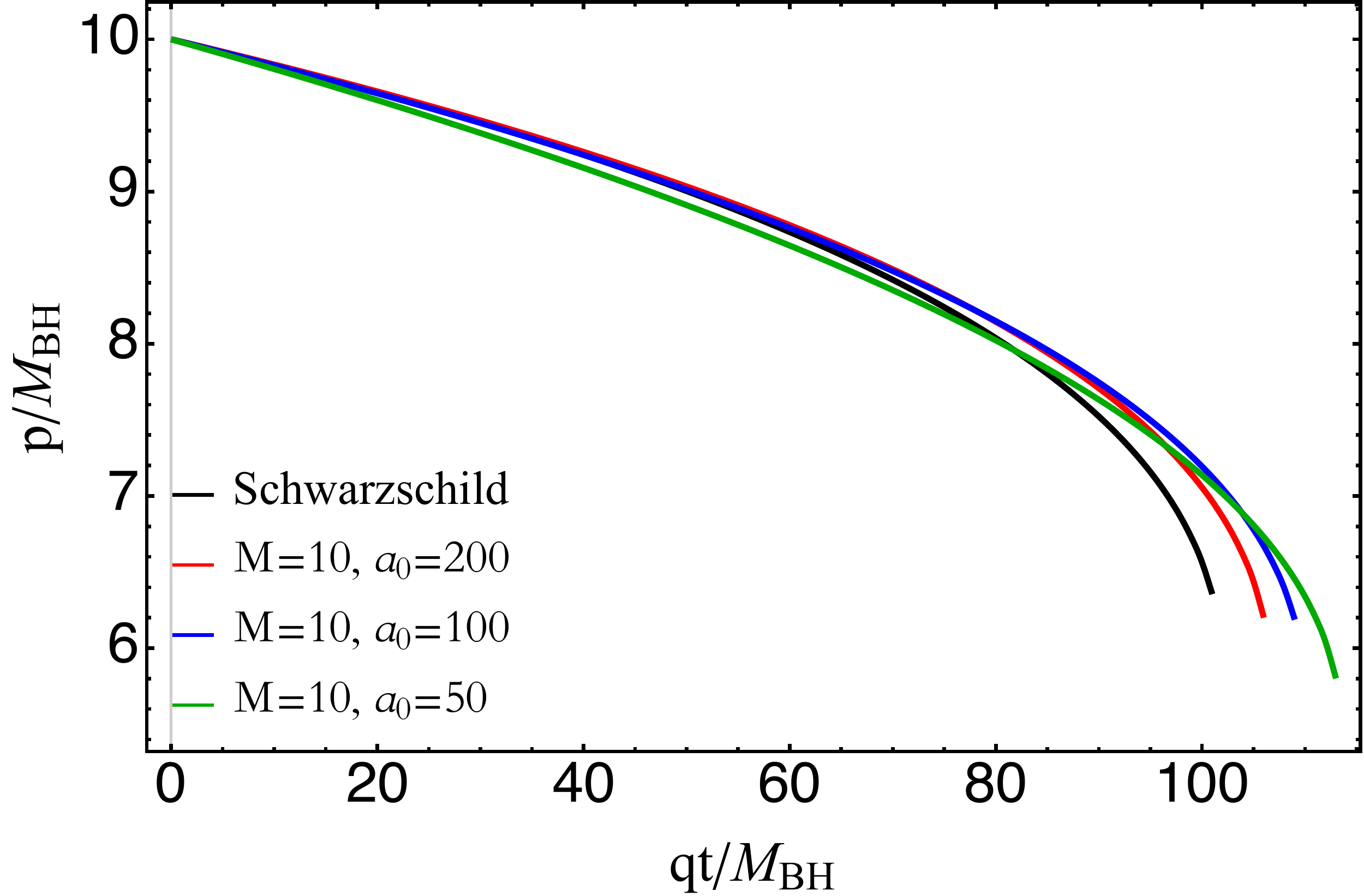}
% \caption{Wormholes for $\Lambda=0$}
	\endminipage\hfill
	%%%%%%%%%%%%%%%%%%%%%%%%
	\minipage{0.48\textwidth}
	\includegraphics[width=\linewidth]{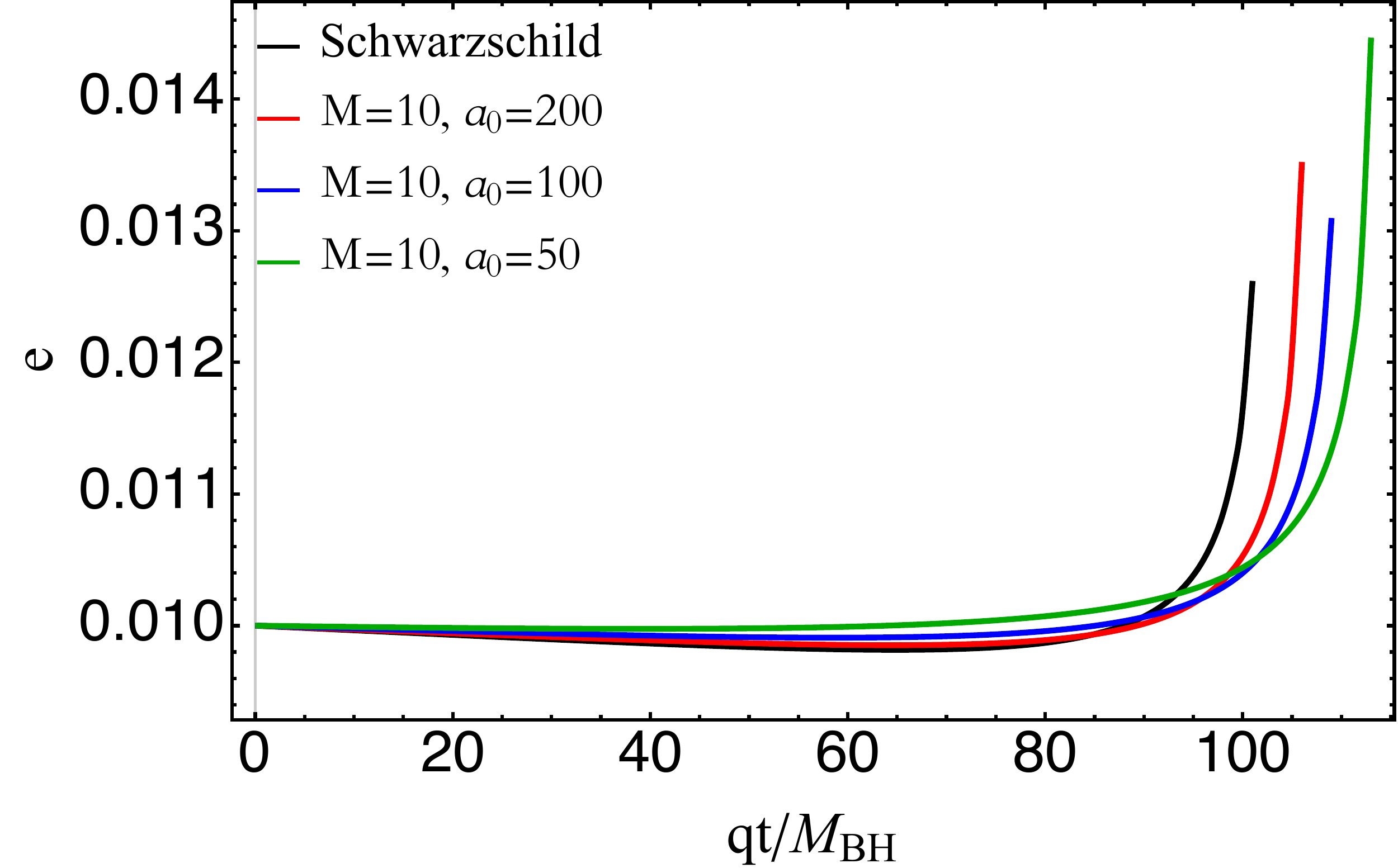}
	\endminipage
	\caption{The time evolution of the semi-latus rectum p (left panel) and the orbital eccentricity $e$ (right panel) is shown, considering different values of dark matter parameters. We consider that the evolution starts at $p_{\textrm{ini}}=10$ and $e_{\textrm{ini}}=0.01$ and ends when the secondary object reaches the last stable orbit $p_{\textrm{end}}$. In each of these plots, the black curve represents the evolution of the orbital parameters for the Schwarzschild black hole. The presence of dark matter lengthens the inspiral phase; the inspiral time increases with the increase of the halo compactness parameter $M/a_0$. }\label{fig_orbital_evolution}
\end{figure}
\begin{figure}[htb!]
	%%%%%%%%%%%%%%%%%%%%%%%%
	\centering
	\minipage{0.48\textwidth}
	\includegraphics[width=\linewidth]{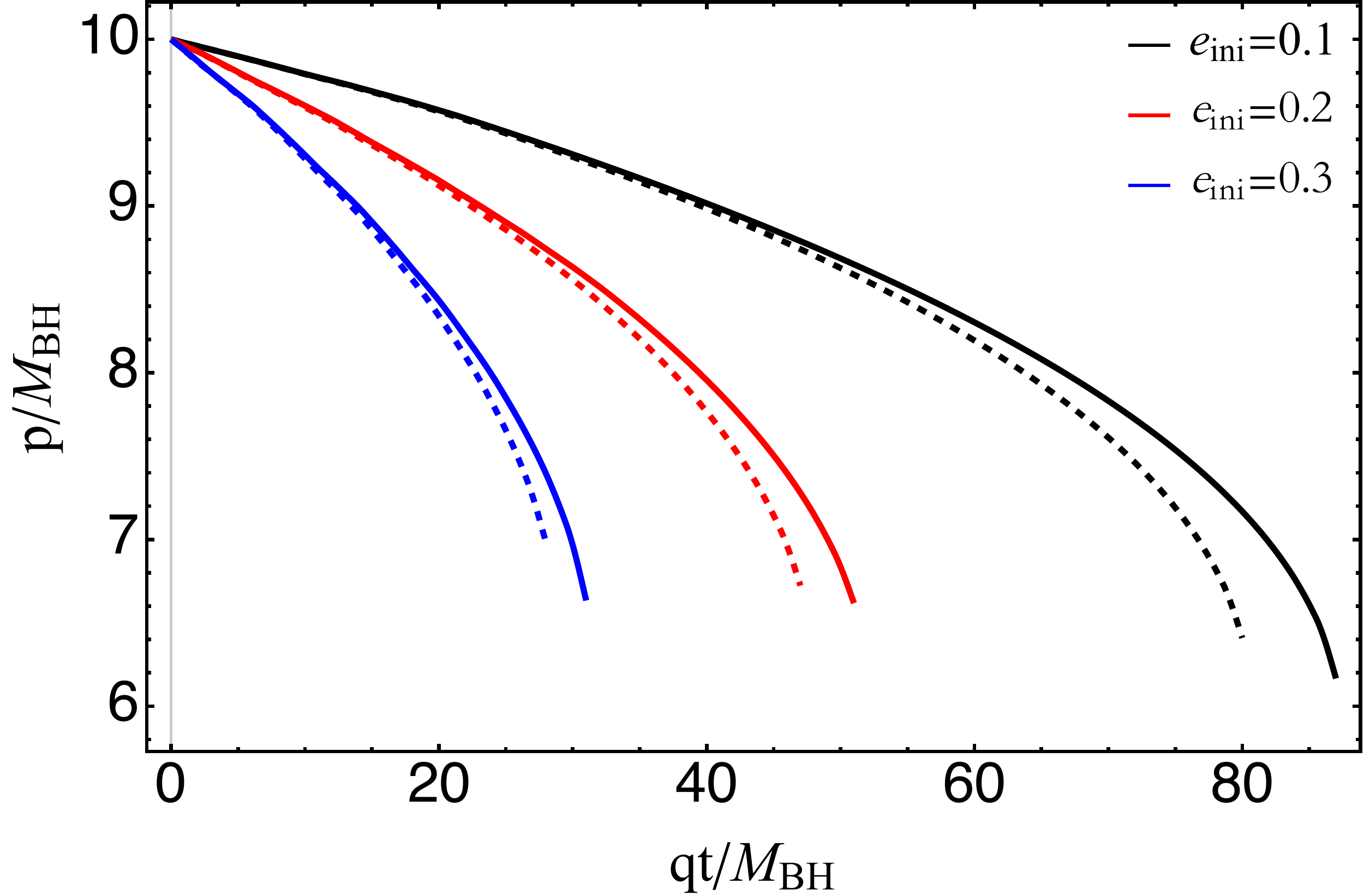}
% \caption{Wormholes for $\Lambda=0$}
	\endminipage\hfill
	%%%%%%%%%%%%%%%%%%%%%%%%
	\minipage{0.48\textwidth}
	\includegraphics[width=\linewidth]{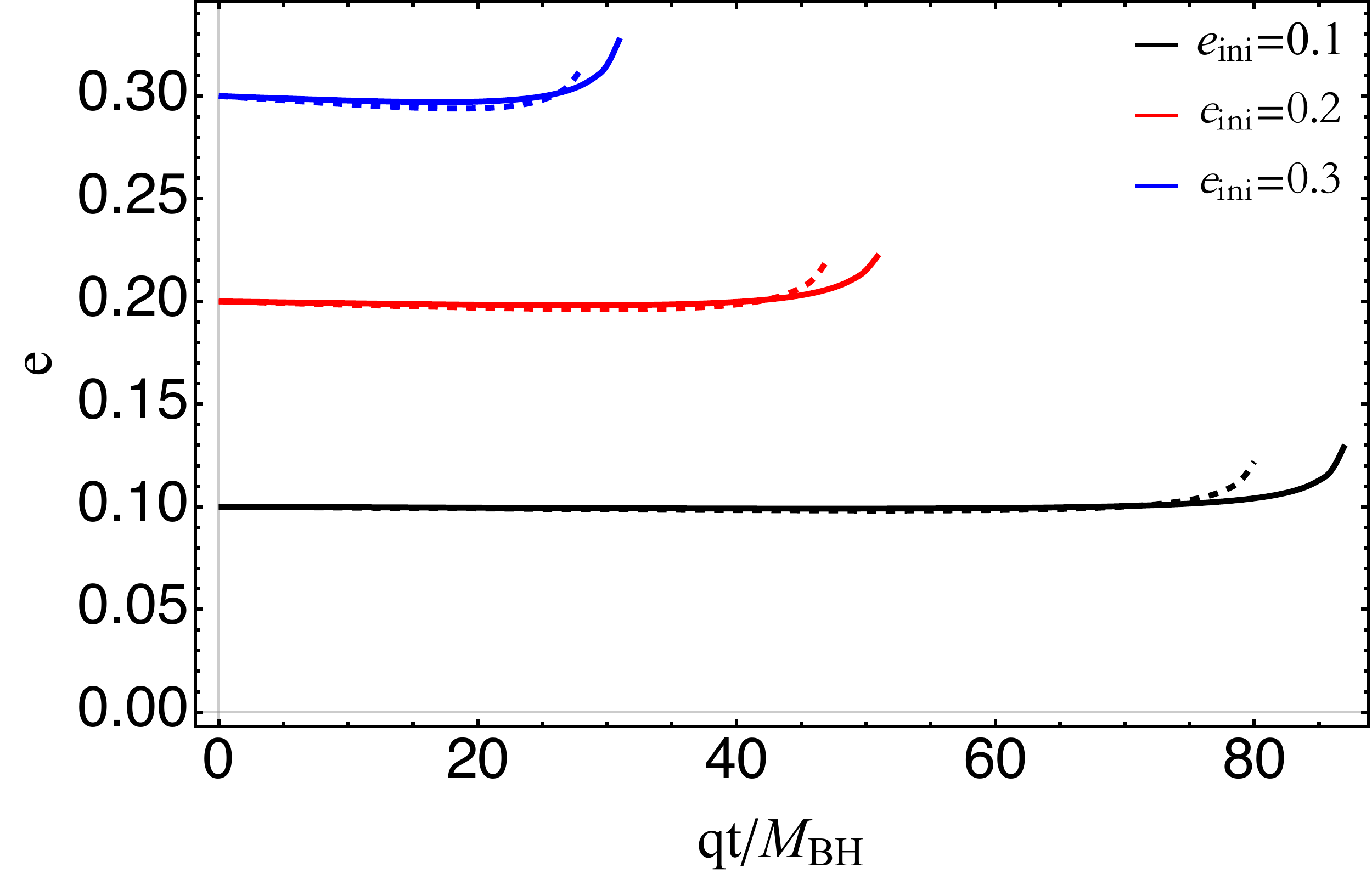}
	\endminipage
	\caption{The time evolution of the semi-latus rectum p (left panel) and the orbital eccentricity $e$ (right panel) is shown, considering different values of initial eccentricity $e_{\textrm{ini}}$. As before, we consider that the evolution starts at $p_{\textrm{ini}}=10$ and ends when the secondary object reaches the last stable orbit $p_{\textrm{end}}$. In each of these plots, the solid line represents the evolution of an EMRI system immersed in the dark matter halo with dark matter parameters $M=10$ and $a_0=100$. The dashed line shows the same for a vacuum EMRI system. As can be seen, the inspiral time is significantly less for highly eccentric system. }\label{fig_orbital_evolution_diff_e}
\end{figure}

	
%%%%%%%%%%%%%%%%%%%%%%%%%%%%%%%%%%%%%%%%%%%%%%%%%%%%%%%%%%%%%%%%%
We can study the orbital evolution by integrating the equations in Eq. (\ref{petoEJ}) simultaneously in the domain $p\in (p_{\textrm{ini}}, p_{\textrm{min}})$. Here, we use explicit Euler's method to solve the equation \cite{holmes2006introduction}. Since the integration process is computationally very expensive (mostly due to the computation of orbital average value of energy and angular momentum flux, given in Eq. (\ref{ave_flux}), we consider the contribution of $\{2,2\}$ and $\{3,3\}$ mode only to compute the orbit-averaged flux. The contribution of these two polar modes are the order of magnitude higher than the axial and the higher-order polar modes (see Table
%\ref{fig_enr_flux_same_e}, \ref{fig_flux_higher}, \ref{fig_AngFx2} and 
\ref{tab:Fluxes_Comparison}); thus, gives a good estimation of the orbital evolution. In Fig. (\ref{fig_orbital_evolution}), we show the evolution of semi-latus rectum and eccentricity for different values of dark matter parameters. Here, we consider the inspiral starts at $p_{\textrm{ini}}=10$ and set the initial value of eccentricity as $e_{\textrm{ini}}=0.01$. To make a comparison with the vacuum scenario, we also show the evolution of the orbital parameters for the Schwarzschild black hole, represented by the black curve in the plot. \textit{As can be seen, the secondary object takes more time to reach the last stable orbit $p_{\textrm{min}}$ in the presence of dark matter. Moreover, as we increase the value of the halo compactness parameter, inspiral time increases. The semi-latus decreases monotonically due to the radiation backreaction effect. However, the orbital eccentricity decreases at first when the orbital separation is large; but as the object reaches the last stable orbit, the eccentricity increases due to the radiation reaction effect.} In Fig. (\ref{fig_orbital_evolution_diff_e}), we show the orbital evolution for $M=10$ and $a_0=100$ and for different values of $e_{\textrm{ini}}$. In each of these plots, the dashed line represents the same for a Schwarzschild black hole. \textit{As can be seen, highly eccentric orbits take less time to reach the last stable orbit.}

%%%%%%%%%%%%%%%%%%%%%%%%%%%%%%%%%%%%%%%%%%%%%%%%%%%%%%%%%%%%%%%%%
%%%%%%%%%%%%%%%%%%%%%%%%%%%%%%%%%%%%%%%%%%%%%%%%%%%%%%%%%%%%%%%%%%

%\par%the inspiral time decreases drastically with the increase of initial eccentricity value. \par
%Thus, in the adiabatic approximation, the 

\subsection{Gravitational wave phase and Detectability}
\begin{figure}[t!]
	%%%%%%%%%%%%%%%%%%%%%%%%
	\centering
	\minipage{0.48\textwidth}
	\includegraphics[width=\linewidth]{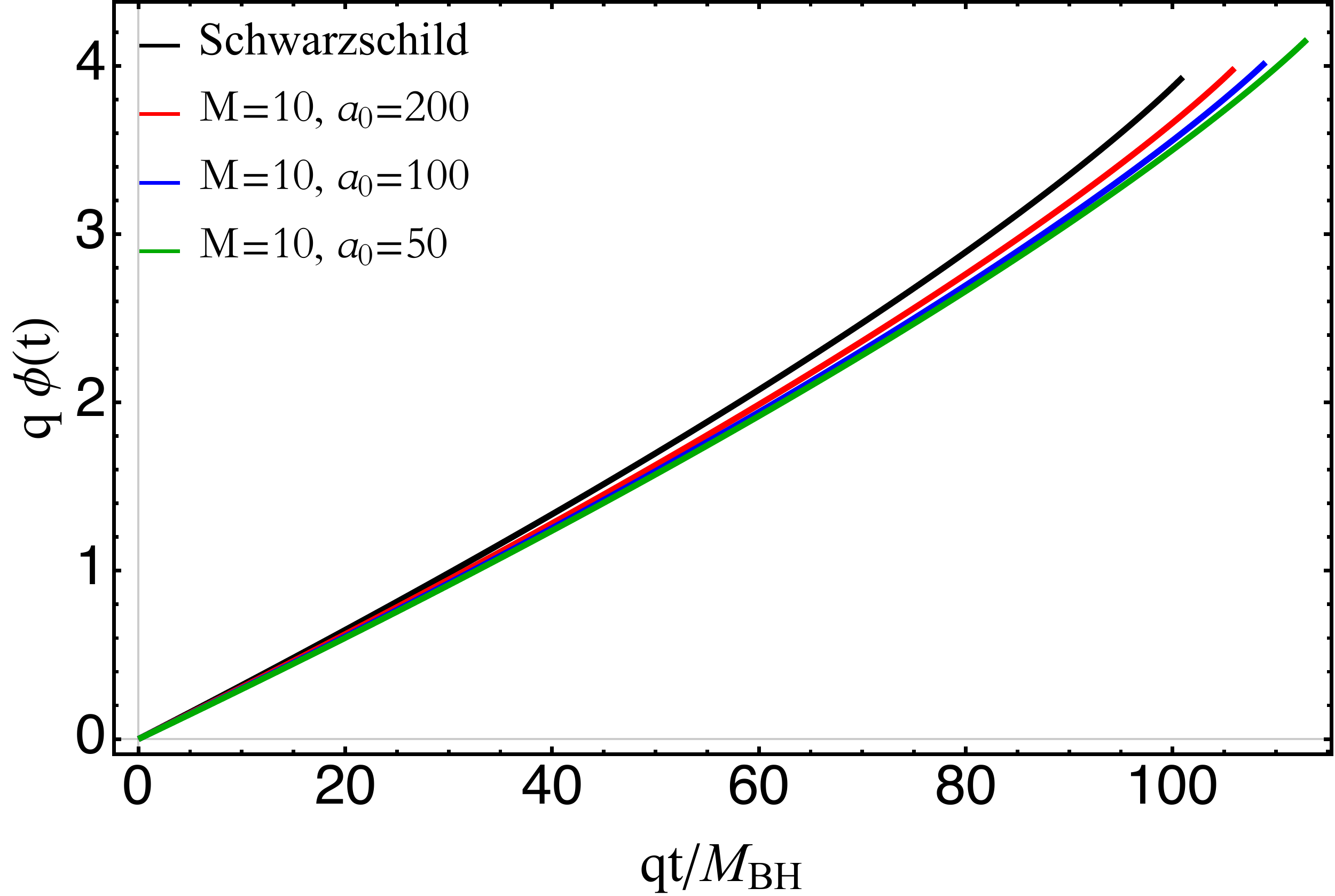}
% \caption{Wormholes for $\Lambda=0$}
	\endminipage\hfill
 \centering
	\minipage{0.48\textwidth}
	\includegraphics[width=\linewidth]{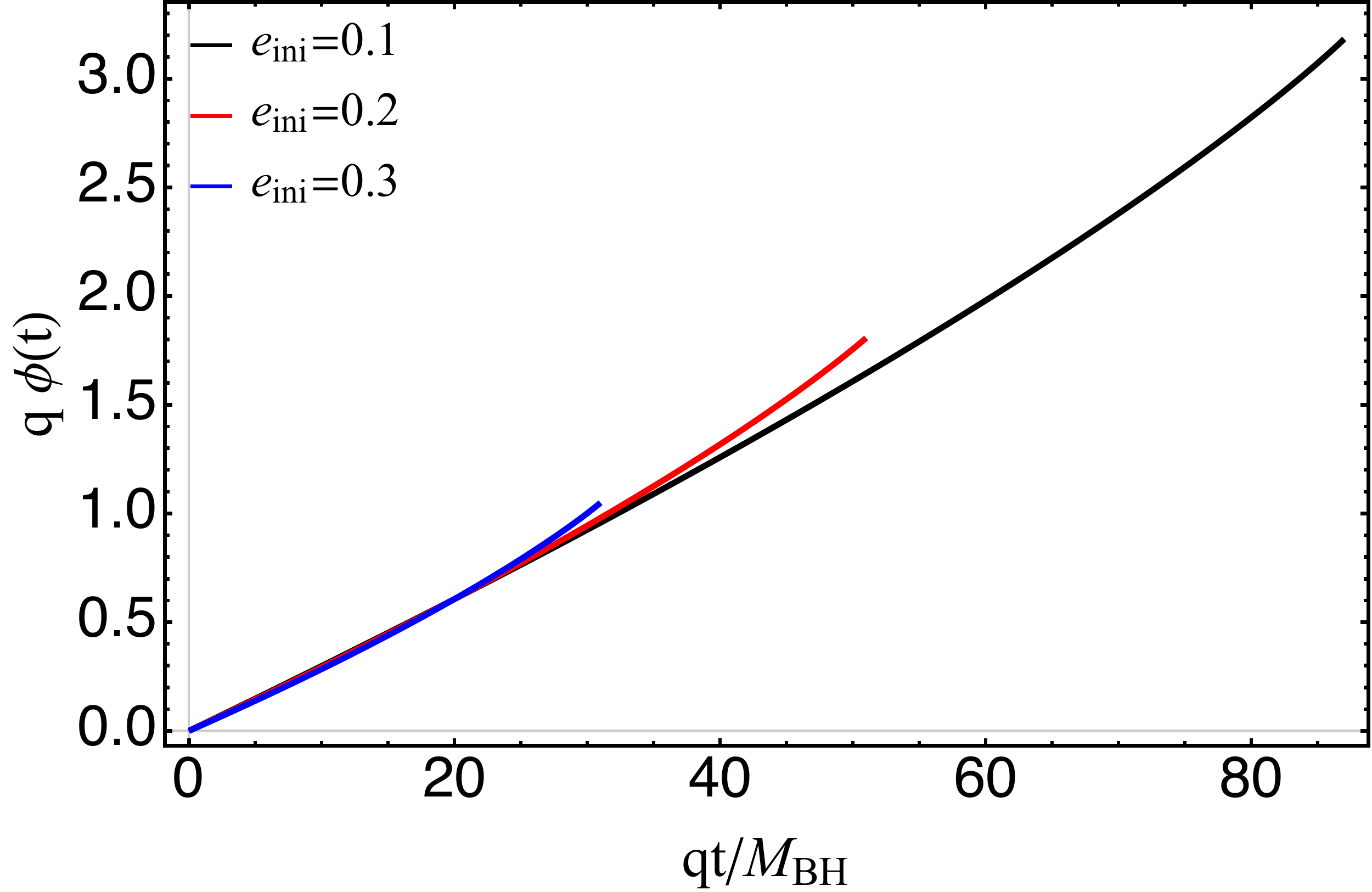}
% \caption{Wormholes for $\Lambda=0$}
	\endminipage
	\caption{The time evolution of the orbital phase is shown. The left panel shows the phase evolution  for different values of dark matter parameters with $e_{\textrm{ini}}=0.01$. The right panel shows the same for $M=10$ and $a_0=100$, considering different values of initial eccentricity. As before, we consider that the evolution starts at $p_{\textrm{ini}}=10$ and ends when the secondary object reaches the last stable orbit $p_{\textrm{end}}$. }\label{fig_phi_evolution}
\end{figure}

In this section, we investigate the possibility of detecting the signature of dark matter through observations of extreme mass-ratio inspirals (EMRIs). To do that, we first calculate the orbital phase through the following relation
%%%%%%%%%%%%%%%%%%%%%%%%%%%%%%%%%%%%%%%%%%%%%%%%%%%%%%%%%%%%%%%%
\begin{equation}
    \begin{aligned}
        \frac{d\varphi_i(t)}{dt}=\left\langle\Omega_i\left(p(t),e(t)\right)\right\rangle=\frac{1}{T_P}\int_{0}^{2\pi}d\chi \frac{dt}{d\chi}\Omega_i\left(p(t),e(t),\chi\right)\,,\qquad{i}\in \{\phi,r\}
    \end{aligned}
\end{equation}
%%%%%%%%%%%%%%%%%%%%%%%%%%%%%%%%%%%%%%%%%%%%%%%%%%%%%%%%%%%%%%%%
where $p(t)$ and $e(t)$ is the solution of Eq. (\ref{petoEJ}). Here, $\left\langle\Omega_i\right\rangle$ is the orbit-averaged frequency.
We consider the phase at the start of the inspiral to be $\varphi_i(0)=0$. Given that, $\varphi_r\ll\varphi_\phi$, orbital dephasing is primarily dictated by the azimuthal phase shift, i.e., $\varphi_\phi(t)\sim \phi(t)$ \cite{Barsanti:2022ana}.
We use explicit Euler's method to calculate $\phi(t)$. The result is presented in Fig. (\ref{fig_phi_evolution}). In the left panel, we show the phase shift $\phi(t)$ for different values of dark matter parameters and with $p_{\textrm{ini}}=10$, $e_{\textrm{ini}}=0.01$. In the right panel, we depict the same but for $M=10$, $a_0=100$ and $p_{\textrm{ini}}=10$ and with different values of initial eccentricity.  We notice that the phase shift is more when the halo compactness parameter is large. Furthermore, the phase shift is significantly larger when the initial eccentricity value is small.
%%%%%%%%%%%%%%%%%%%%%%%%%%%%%%%%%%%%%%%%%%%%%%%%%%%%%%%%%%%%%%%%
%%%%%%%%%%%%%%%%%%%%%%%%%%%%%%%%%
%%%%%%%%%%%%%%%%%%%%%%%%%%%%%%%%%%%%%%%%%%%%%%%%%%%%%%%%%%%%%%%%%

The gravitational wave phase of the dominant $\{2,2\}$ mode is related to the orbital phase by the relation $\Phi_{\textrm{GW}}^{\textrm{DM}}(t)\approx 2\phi(t)$. To examine the signature of dark matter, we consider the gravitational wave emitted from an EMRI system with a Schwarzschild black hole as the reference waveform. We calculate the GW dephasing with respect to this reference waveform. Specifically, we define the dephasing up to a certain time $t_{\textrm{obs}}$ between a black hole in the presence of dark matter, described by Eq. (\ref{met1}), and a Schwarzschild black hole as

%%%%%%%%%%%%%%%%%%%%%%%%%%%%%%%%%%%%%%%%%%%%%%%%%%%%%%%%%%%%%%%%%%%%%%%%%%%%%%%%%%%%%%%%%%%%%%%%%%%
\begin{equation}\label{dephasing}
\Delta\Phi(t_{\textrm{obs}})=\left|\Phi_{\textrm{GW}}^{\textrm{DM}}(t_{\textrm{obs}})-\Phi_{\textrm{GW}}^{\textrm{Schld}}(t_{\textrm{obs}})\right|\,.
\end{equation}
%%%%%%%%%%%%%%%%%%%%%%%%%%%%%%%%%%%%%%%%%%%%%%%%%%%%%%%%%%%%%%%%%%%%%%%%%%%%%%%%%%%%%%%%%%%%%%%%%%%

\begin{figure}[t!]
	%%%%%%%%%%%%%%%%%%%%%%%%
	\centering
	\minipage{0.48\textwidth}
	\includegraphics[width=\linewidth]{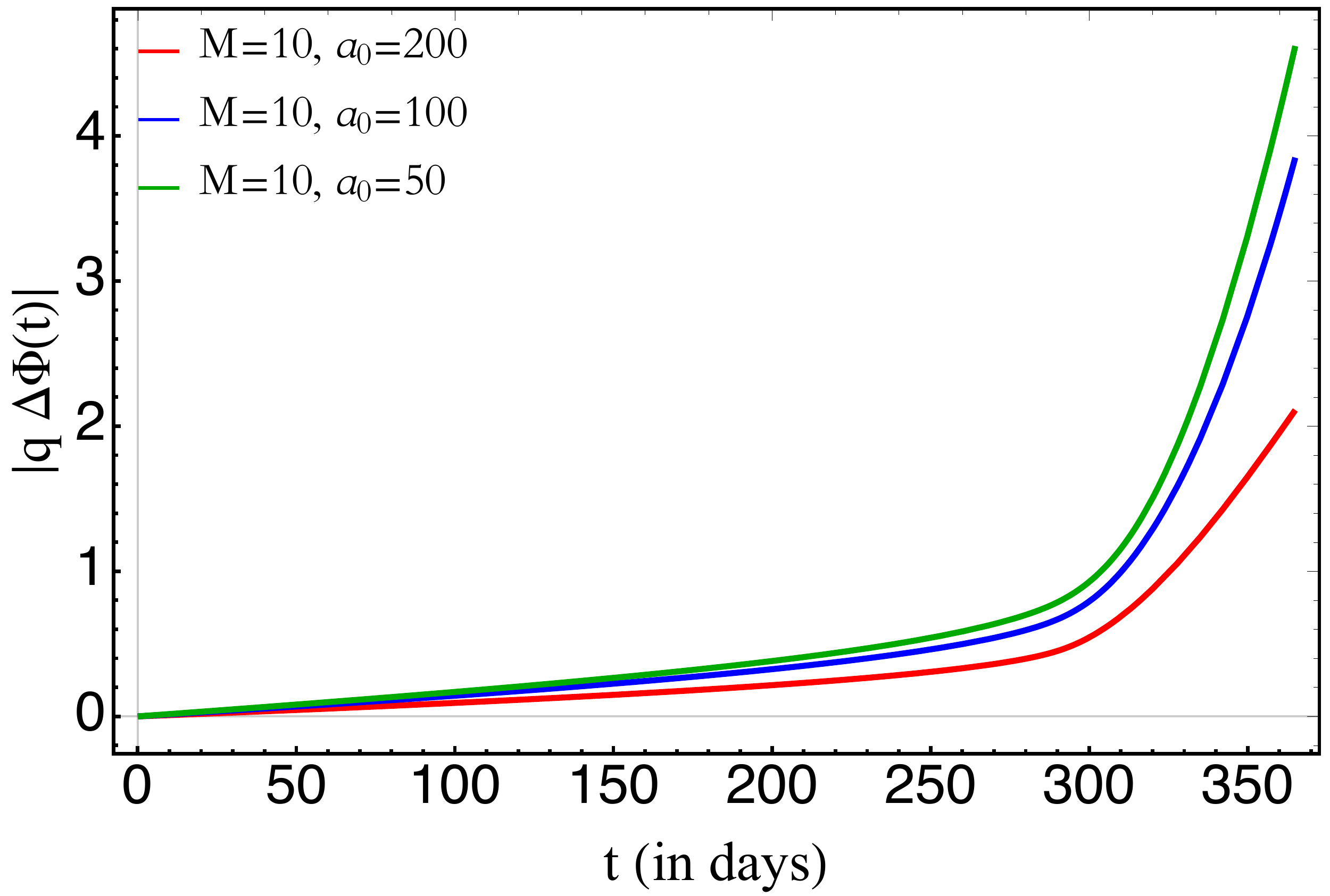}
	\endminipage\hfill
 \centering
	\minipage{0.48\textwidth}
	\includegraphics[width=\linewidth]{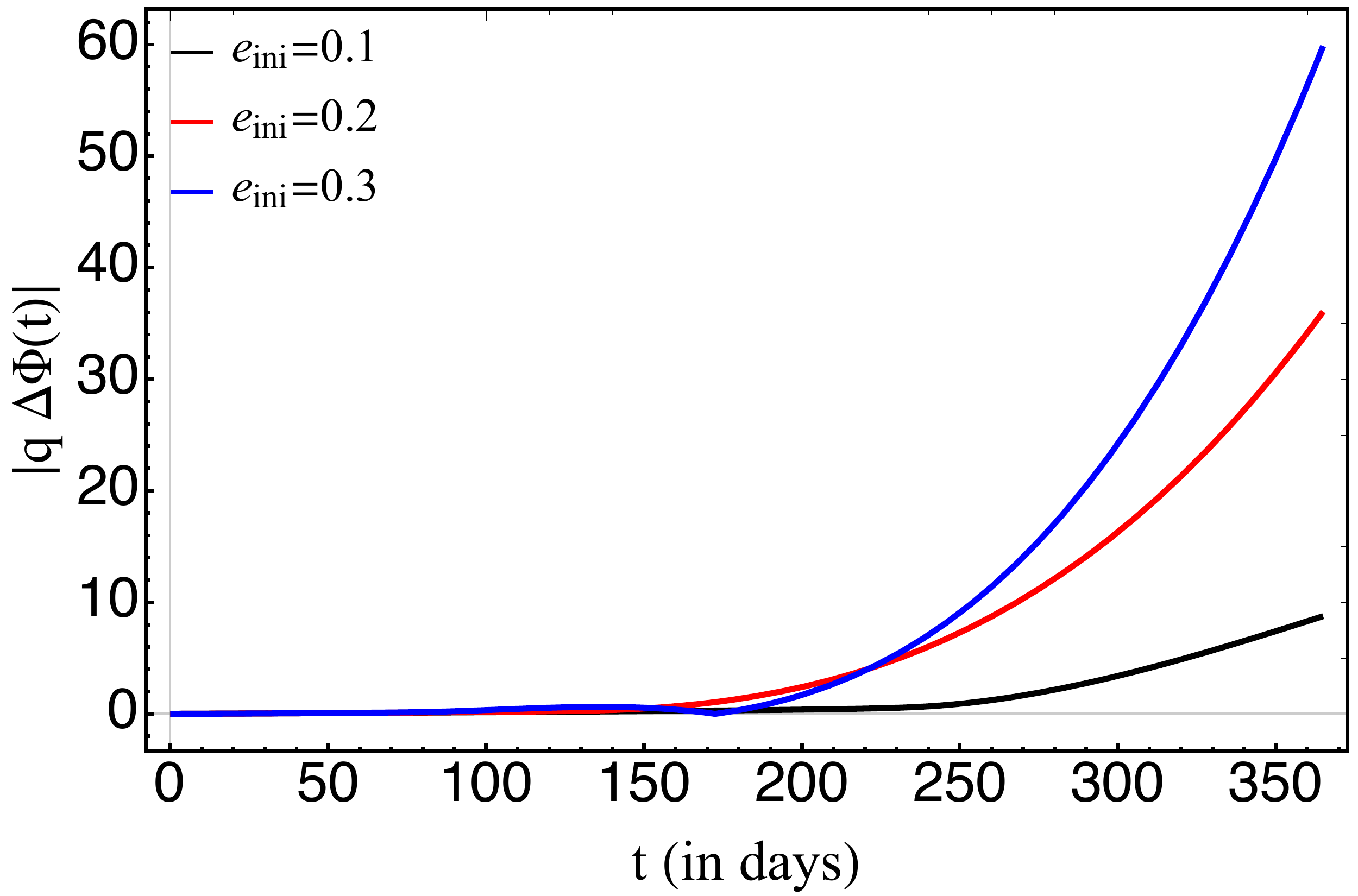}
% \caption{Wormholes for $\Lambda=0$}
	\endminipage
	\caption{The dephasing $\Delta \Phi$ (defined in Eq.~(\ref{dephasing})) in one year observation time   is plotted. Here, we consider an EMRI system with $M_{\textrm{BH}}=10^{6}~M_\odot$ and $\mu=20~M_\odot$. The left panel shows the dephasing for different values of dark matter parameters with $e_{\textrm{ini}}=0.01$. The right panel shows the same for $M=10$ and $a_0=100$, considering different values of initial eccentricity.  }\label{fig_deltaphi_evolution}
\end{figure}

Here, $\Phi_{\textrm{GW}}^{\textrm{Schld}}$ represents the phase of the gravitational wave emitted from a vacuum EMRI. We set the observation time to be $t_{\textrm{obs}}=1$ year. The results are presented in Fig.~(\ref{fig_deltaphi_evolution}) for an EMRI system with a black hole mass of $M_{\textrm{BH}}=10^6~M_{\odot}$ and $\mu=20~M_{\odot}$. The left panel of the figure illustrates the dephasing for different values of the dark matter parameters. We assume the inspiral begins with $p_{\textrm{ini}}=10$ and $e_{\textrm{ini}}=0.01$. As expected, we observe a greater dephasing when the halo compactness is large. In the right panel of Fig. (\ref{fig_deltaphi_evolution}), we show the dephasing for $M=10$, $a_0=100$, and different initial eccentricities ($e_{\textrm{ini}}$). \textit{It is noticeable that the dephasing is larger for highly eccentric orbits.}

%%%%%%%%%%%%%%%%%%%%%%%%%%%%%%%%%%%%%%%%%%%%%%%%%%%%%%%%%%%%%%%%

Considering that the average signal-to-noise ratio (SNR) for LISA observations is approximately $\textrm{SNR}\sim 30$, LISA is expected to detect the dephasing whenever $\Delta\Phi\gtrsim 0.1~\textrm{rad}$ \cite{PhysRevLett.123.101103}. Note that the presence of dark matter introduces a dephasing $\Delta\Phi\approx \mathcal{O}(1)/q$ for a halo compactness parameter $M/a_0\sim \mathcal{O}(10^{-1})$. Therefore, we anticipate that LISA will be able to detect the presence of dark matter with extremely high accuracy.
%%%
%%%%%%%%%%%%%%%%%%%%%%%%%%%%%%%%%%%%%%%%%%%%%%%%%%%%%%%%%%%%%%%%%%%%%%%%%%%%%%%%%%%%%%%%%%%%%%%%%%%

%%%%%%%%%%%%%%%%%%%%%%%%%%%%%%%%%%%%%%%%%%%%%%%%%%%%%%%%%%
%%%%%%%%%%%%%%%%%%%%%%%%%%%%%%%%%%%%%%%%%%%%%%%%%%%%%%%%%%%

\section{Discussion}\label{dscn}

%EMRI is a system that has gained noticeable attention with the potential detectability and impact of these sources on future detectors.
 Astrophysical systems like black hole binaries are seldom found in isolation. Instead, they often reside in complex environments that encompass various astrophysical components, including electromagnetic fields, accretion disks, dark matter, and other celestial objects. These surrounding elements significantly impact the dynamics of binaries, resulting in observable effects on the gravitational wave signals they emit \cite{Barausse:2014pra}. Therefore, studying the impact of the astrophysical environment on gravitational wave signals is crucial for gravitational wave astrophysics. It not only aids in detecting signals within noisy data but also provides insights into the surrounding environment of the binary system. This consideration becomes even more relevant for EMRIs where the systems consist of a stellar-mass black hole orbiting a supermassive black hole. The latter is often located at the centre of galaxies, one of the most environmentally rich regions in the universe.
In this paper, we have studied the dynamics of the eccentric EMRI system that is immersed in the dark matter halo, which follows the Hernquist density distribution. We have assumed that the primary object placed at the centre of the density distribution and the spacetime is described by the line element Eq.~(\ref{met1}). The secondary object revolves around the primary in an equatorial, eccentric orbit, characterized by the semi-latus rectum $p$ and eccentricity $e$. The study is crucial from an astrophysical point-of-view as the EMRIs can exhibit highly eccentric orbits.\par
\textit{By examining the orbital motion of a test particle in spacetime, we find that the location of the last stable orbit can significantly be influenced by the halo compactness parameter $M/a_0$. The last stable orbit refers to the smallest value of the semi-latus rectum $p_{\textrm{min}}$ that permits bound orbits for a given eccentricity. }
Specifically, as the halo compactness parameter increases, the position of the bound orbit decreases, consequently extending the parameter space in the $p$-$e$ plane that allows for bound orbits. We have adopted Regge-Wheeler-Zerilli formalism \cite{PhysRev.108.1063, PhysRevLett.24.737, PhysRevD.2.2141} to calculate the gravitational wave flux and study the system's evolution. When the halo compactness parameter is small, the value of the energy and angular momentum wave flux in the axial and polar sectors is almost identical to that of a Schwarzschild black hole. But as we increase the halo compactness parameter's value, the flux's value shifts away from the Schwarzschild value. Furthermore, the energy and momentum flux of $\{2,2\}$ mode is dominant over axial and higher-order polar modes. Thus, the orbital dynamics of the secondary is mainly governed by this leading order polar mode. \par
We have studied the secondary object's evolution within the adiabatic approximation framework. Since the adiabatic approximation breaks down beyond the last stable orbit $p_{\textrm{min}}(e)$, we confined our attention to the domain $p\in (p_{\textrm{ini}}=10, p_{\textrm{min}}(e))$. As a result of radiation backreaction, the value of the semi-latus rectum always decreases. The same can not be said about the eccentricity. In fact, near the separatrix $p_{\textrm{min}}(e)$, the eccentricity value can increase rapidly. Note that previous studies have shown that the same phenomenon is also observed in the Schwarzschild case \cite{PhysRevD.50.3816}. 
We presented the adiabatic evolution of $p$ and $e$ for various values of dark matter parameters (see Fig. (\ref{fig_orbital_evolution})) and different initial eccentricities (see Fig. (\ref{fig_orbital_evolution_diff_e})). \textit{Our observations indicate that the inspiral time lengthens in the presence of dark matter. However, highly eccentric orbits exhibit shorter inspiral times for a given set of dark matter parameters. Additionally, the total accumulated orbital phase during the inspiral rises with higher values of the halo compactness parameter or lower values of the initial eccentricity.}\par
To assess whether LISA can detect the presence of dark matter, we computed the phase shift between the black hole described in Eq. (\ref{met1}) and a Schwarzschild black hole. Our results reveal that dark matter introduces a dephasing of approximately $\Delta\Phi\approx \mathcal{O}(1)/q~\textrm{rad}$ when the halo compactness parameter satisfies $M/a_0\sim \mathcal{O}(10^{-1})$. Considering that LISA is anticipated to measure a phase shift on the order of $0.1~\textrm{rad}$ \cite{PhysRevLett.123.101103}, it follows that LISA will be capable of accurately detecting the presence of dark matter. \textit{Finally, our results are very encouraging from the point of view of detecting dark matter environments through eccentric EMRI systems. If the initial eccentricity is large, the dephasing becomes quite significant, helping us distinguish the effect of the dark matter environment.} \par
Our study focused on the orbital evolution considering the contributions from the leading order modes $\{2,2\}$ and $\{3,3\}$ only. However, for precise modelling of the gravitational wave signal and data analysis, it is also necessary to account for the contributions of higher modes. Nevertheless, incorporating these higher modes into the analysis poses significant computational challenges due to their computational expense.
Furthermore, to give a more accurate estimate on finding the signature of dark matter, it is essential to perform a Fisher-matrix analysis \cite{Vallisneri:2007ev}. However, in this paper, we have provided only an order-of-magnitude estimation on the detectability of dark matter and reserved the detailed Fisher matrix analysis as a future endeavour.
This study considers that the dark matter halo follows the Hernquist type density profile \cite{1990ApJ...356..359H}. It will be interesting to investigate the evolution of eccentric EMRI systems considering other types of dark matter distributions too, like NFW, Einsato density profile etc \cite{Navarro:1995iw,Graham:2005xx,DeLuca:2023laa,Figueiredo:2023gas}. However, this is beyond the scope of this paper. Last but not the least, dark matter environments can be modelled by certain ultra-light scalar and vector fields \cite{Baryakhtar:2022hbu, Bhattacharyya:2023kbh, Cardoso:2018tly}. It will be very interesting to extend our studies for some of those models, and we hope to investigate some of these issues in the near future.

%%%%%%%%%%%%%%%%%%%%%%%%%%%%%%%%%%%%%%%%%%%%%%%%%%%%%%%%%%%%%%%%%%%%%%%%%%%%%%%%%%%%%%%%%%%%%%%%%%%

\section*{Acknowledgements} 
A.B like to thank the participants of the (virtual) workshop ``Testing Aspects of  General Relativity-II" (11-13th April, 2023) and ``New insights into particle physics from quantum information and gravitational waves" (12-13th June, 2023) at Lethbridge University, Canada funded by McDonald Research Partnership-Building Workshop grant by McDonald Institute. A.B also like to thank the speakers of the online conference funded by Shastri Indo-Canadian Institute's Shastri Conference \& Lecture Series Grant (SCLSG) ``Testing Aspects of General Relativity," held between 11-14th March, 2022, as the idea was generated from some of the talks of the workshop. The Research of M.R. is funded by the National Post-Doctoral Fellowship (N-PDF) from SERB, DST, Government of India (Reg. No. PDF/2021/001234). S.K. is supported by the Post-Doctoral fellowship by the Indian Institute of Technology Gandhinagar and a consultancy project of A.B (CNS/ATPL/PH/P0300/2223/0031). A.B is supported by the Mathematical Research Impact Centric Support Grant (MTR/2021/000490) by the Department of Science and Technology Science and Engineering Research Board (India) and Relevant Research Project grant (202011BRE03RP06633-BRNS) by the Board Of Research In Nuclear Sciences (BRNS), Department of atomic Energy, India.
%%%%%%%%%%%%%%%%%%%%%%%%%%%%%%%%%%%%%%%%%%%%%%%%%%%%%%%%%%%%%%%%%%%%%%%%%%%%%%%%%%%%%%%%%%%%%%%%%%

\appendix

\section{Perturbation equations}\label{App_pert}
In this section and the subsequent section, we derive the governing equations for axial and polar perturbations. As mentioned earlier, the presence of the secondary object introduces perturbations to both the background metric and the energy-momentum tensor of dark matter (see Eq. (\ref{perturbation})). In this section, we focus on the discussion of metric perturbations. Subsequently, we discuss the perturbed energy-momentum tensor of dark matter and the energy-momentum tensor of the secondary object. We use \texttt{xACT} \cite{xACT}, a \texttt{Mathematica} package to obtain the equations for axial and polar perturbations.
%Owing to the small mass-ratio of the EMRI system, we consider that the secondary object perturbs the background spacetime, given in Eq.~(\ref{met1}). As a result, both the background metric and dark matter energy-momentum tensor get perturbed (see Eq.~())
\subsection{Metric Perturbation}
In the Regge-Wheeler-Zerilli gauge  \cite{PhysRevLett.24.737, PhysRevD.2.2141}, the metric perturbation can be decomposed into axial and polar sector as $g_{\mu\nu}^{(1)} = g_{\mu\nu}^{(1) \textrm{axial}}+g_{\mu\nu}^{(1) \textrm{polar}}$, where 
%Here, we provide expressions for axial and polar perturbations which can be written as follows
%using the first-order quantities, one can classify two perturbations-\textit{axial} and \textit{polar}, and further expand them into tensor harmonics. In the Regge-Wheeler gauge, these are defined as radial functions: $h_{0}^{lm}, h^{1}_{lm}$ for axial and $K^{lm}, H_{0}^{lm}, H_{1}^{lm}, H_{2}^{lm}$ for polar, together with angular basis functions \cite{Sarbach:2001qq, Martel:2005ir, PhysRev.108.1063, PhysRevLett.24.737, PhysRevD.2.2141}. 
%Let us now look at the perturbation equations in detail. As mentioned, we employ the Regge-Wheeler gauge in order for the metric components to follow
%\begin{align*}
  %  g_{\theta\phi} =& 0 \hspace{3mm} ; \hspace{3mm} g_{\phi\phi} = g_{\theta\theta}\sin^{2}\theta \nonumber \\
   % \partial_{\phi}\Big(\frac{g_{t\phi}}{\sin\theta}\Big)+\partial_{\theta}(g_{t\theta}\sin\theta) =& 0 \hspace{3mm} ; \hspace{3mm} \partial_{\phi}\Big(\frac{g_{r\phi}}{\sin\theta}\Big)+\partial_{\theta}(g_{r\theta}\sin\theta) = 0
%\end{align*}
%With this consideration, the metric perturbations 
%\begin{align*}
  %  g_{\mu\nu}^{(1)} = g_{\mu\nu}^{(1) \textrm{axial}}+g_{\mu\nu}^{(1) \textrm{polar}}
%\end{align*}
%take the following form
\begin{equation} \label{prt1}
    \begin{aligned}
    g_{\mu\nu}^{(1)axial}(t,r,\theta,\phi) =& \sum_{l=2}^{\infty}\sum_{m=-l}^{l} \frac{\sqrt{2l(l+1)}}{r}\Big(ih_{1}^{lm}c_{lm,\mu\nu}-h_{0}^{lm}c^{0}_{lm, \mu\nu}\Big)  \\
    g_{\mu\nu}^{(1)polar}(t,r,\theta,\phi) =& \sum_{l=2}^{\infty}\sum_{m=-l}^{l}\Big(aH_{0}^{lm}a^{0}_{lm,\mu\nu}-i\sqrt{2}H_{1}^{lm}a^{1}_{lm,\mu\nu}+\frac{H_{2}^{lm}}{b}a_{lm,\mu\nu}+\sqrt{2}K^{lm}g_{lm,\mu\nu}\Big).
\end{aligned}
\end{equation}
The perturbation in the axial sector is described in terms of functions: $h_{0}^{lm}, h^{1}_{lm}$ while the perturbation in the polar sector is defined in terms of $K^{lm}, H_{0}^{lm}, H_{1}^{lm}, H_{2}^{lm}$,%We have omitted writing the above relations in the form of coordinate functions in order to make it reader-friendly. However, 
where $h_{0}^{lm}, h_{1}^{lm}, H_{0}^{lm}, H_{1}^{lm}, H_{2}^{lm}$ and $K^{lm}$ are functions of ($t, r$). $c_{lm,\mu\nu}, c^{0}_{lm,\mu\nu}$ and $g_{lm,\mu\nu}$ are functions of ($r, \theta, \phi$). Further, $a^{0}_{lm,\mu\nu}, a^{1}_{lm,\mu\nu}$ and $a_{lm,\mu\nu}$ are functions of ($\theta, \phi$). $\{c_{lm,\mu\nu}, c^{0}_{lm,\mu\nu}, a^{0}_{lm,\mu\nu}, a_{lm,\mu\nu}, g_{lm,\mu\nu}\}$ are six tensor spherical harmonics which, including four more, can be found in \cite{PhysRevD.67.104017}. Next, we describe the source terms of the astrophysical environment and the secondary object.

\subsection{Perturbed energy-momentum tensor of the dark matter halo}
The perturbed density, pressure of the dark matter halo can also be decomposed into tensor spherical harmonics in the following manner 
\begin{align}
\rho^{(1)}(t,r,\theta,\phi)&=\sum_{l=2}^\infty\sum_{m=-l}^{l}\delta \rho_{l m}(t,r)Y_{l m}(\theta,\phi) \nonumber \\
p^{(1)}_t(t,r,\theta,\phi)&=\sum_{l=2}^\infty\sum_{m=-l}^{l}\delta p_{t}^{l m}(t,r)Y_{l m}(\theta,\phi) \nonumber \\
p^{(1)}_r(t,r,\theta,\phi)&=\sum_{l=2}^\infty\sum_{m=-l}^{l}\delta p_{r}^{l m}(t,r)Y_{l m}(\theta,\phi),
\label{math:prepertrad}
\end{align} 
where $Y_{lm}(\theta, \phi)$ denotes the spherical harmonics on 2-sphere.
In order to construct the perturbed energy-momentum tensor of the environment, we first need to perturb the four-velocity of the fluid and the normal vector as $u^{\mu}=u^{\mu}_{(0)}+u^{\mu}_{(1)}$ and  $k^{\mu}=k^{\mu}_{(0)}+k^{\mu}_{(1)}$. Note that, $u^{\mu}_{(1)}$ and $k^{\mu}_{(1)}$ can be described by functions $\{U_{l m}(t,r), V_{l m}(t,r), W_{l m}(t,r)\}$ \cite{PhysRevD.67.104017}. The normalization condition mentioned in the main text should hold up to the first order (see Section \ref{BH}). With this, the perturbed 4-velocity is given by

\begin{equation}\label{fluid_4_vel}
\begin{aligned}
u^{t}_{(1)} &= \frac{1}{2f^{1/2}}\sum_{lm} H^{l m}_0 Y_{l m} \hspace{3mm} ; \hspace{3mm}
u^{r}_{(1)} = \frac{f^{1/2}}{b}\frac{1}{4\pi\kappa}\sum_{lm} W_{l m}Y_{l m}\,, \\
u^{\theta}_{(1)}&=\frac{f^{1/2}}{4\pi\kappa r^2}\sum_{lm}\bigg[V_{l m}\partial_\theta -\frac{U_{l m}}{\sin\theta}\partial_\phi\bigg]Y_{l m}\,, \\
u^{\phi}_{(1)}&=\frac{f^{1/2}}{4\pi\kappa r^2\sin^2\theta}\sum_{lm}\bigg[V_{l m}\partial_\phi +\frac{U_{l m}}{\sin\theta}\partial_\theta\bigg]Y_{l m} ,
\end{aligned}
\end{equation}
where $\kappa = \rho_{\textrm{DM}}(r)+p_{t}(r)$ as we are considering only anisotropic background with vanishing radial pressure. The perturbed normal vector is given by 
\begin{equation}\label{fluid_normal_vec}
\begin{aligned}
k_{t}^{(1)} &= \frac{i\omega}{b^{1/2}}\sum_{lm} Z_{l m} Y_{l m} \hspace{3mm} ; \hspace{3mm}
k_{r}^{(1)} = \frac{1}{2b^{1/2}}\sum_{lm} H_2^{l m}Y_{l m}\,, \\
k_{\theta}^{(1)}&=\frac{1}{b^{1/2}}\sum_{lm} Z_{l m} \partial_{\theta}Y_{l m} \,,\qquad
k_{\phi}^{(1)}=0 ,
\end{aligned}
\end{equation}
where $Z_{lm}=i f W_{lm}/(4 \pi  \omega  b \kappa) $. Replacing the values of the perturbed 4-velocity $u^{\mu}$ and the normal vector $k^{\mu}$ in Eq.~(\ref{stt1}), we can find the expression for perturbed energy-momentum tensor $T_{\mu\nu}^{DM (1)}$. The non-vanishing components of which are given by 
\begin{align}\label{fluid_pert_em_tensor}
T_{tt}^{\textrm{DM}(1)} =& f(r)\sum_{l=2}^\infty\sum_{m=-l}^{l}(\delta\rho_{l m}-H^0_{l m}\rho_{\textrm{DM}})Y_{l m} \hspace{5mm} ; \hspace{5mm} T_{\phi\phi}^{\textrm{DM}(1)} =  T_{\theta\theta}^{\textrm{DM}(1)}\sin^2\theta  \nonumber\\
T_{tr}^{\textrm{DM}(1)} =& -\sum_{l=2}^\infty\sum_{m=-l}^{l}\left[\frac{f(r)\left(\rho_{\textrm{DM}} + p_r\right)}{4\pi\kappa  b^2}W_{l m}+H^{1}_{l m} \rho_{\textrm{DM}}\right] Y_{l m} \nonumber\\
T_{t\theta}^{\textrm{DM}(1)} =&
\frac{f(r)\left(\rho_{\textrm{DM}} + p_t\right)}{4\pi\kappa}\sum_{l=2}^\infty\sum_{m=-l}^{l}\left[\csc\theta  U_{l m} \partial_{\phi}-V_{l m}\partial_{\theta}\right]Y_{l m} \nonumber\\
T_{t\phi}^{\textrm{DM}(1)} =&
-\frac{f(r)\left(\rho_{\textrm{DM}} + p_t\right)}{4\pi \kappa}\sum_{l=2}^\infty\sum_{m=-l}^{l}\left[V_{l m} \partial_{\phi}+U_{l m}\sin\theta\partial_{\theta}\right]Y_{l m} \nonumber\\
T_{rr}^{\textrm{DM}(1)} =& \frac{1}{b}\sum_{l=2}^\infty\sum_{m=-l}^{l}  \delta p_{r}^{l m} Y_{l m} \hspace{3mm} ; \hspace{3mm}
T_{\theta\theta}^{\textrm{DM}(1)} = r^2 \sum_{l=2}^\infty\sum_{m=-l}^\ell(p_t K_{l m}+
\delta p_{t}^{l m})Y_{l m} .
\end{align}
where, $\delta p^{lm}_{t} (t, r) = c^{2}_{st}(r) \delta\rho^{lm}(t, r)$ and $\delta p^{lm}_{r} (t, r) = c^{2}_{sr}(r) \delta\rho^{lm}(t, r)$, defined with the radial and transverse sound speeds ($c_{sr}(r), c_{st}(r)$). %It is to add that we ($l, m$) indices in the suffix now for improving the readability. 
%Further, the perturbations to the fluid environment are given by
%
Let us next consider the details of the secondary source.
% Collectively, in general, a perturbed quantity $\mathcal{C}^{(1)}$ can be written in the following form
%  \begin{align}
%      \mathcal{C}^{(1)} = \sum_{l=2}^{\infty}\sum_{m=-l}^{l} \delta \mathcal{C}^{lm}(t,r) Y_{lm}(\theta, \phi),
%  \end{align}
%  The perturbations generated by the secondary are known, once the fluctuation $\delta\mathcal{C}^{lm}$ is calculated where the $\mathcal{C}$ quantity is $\rho_{\textrm{DM}}$, $p_{t}$ and $p_{r}$.

\subsection{Energy-momentum tensor of the secondary object}\label{App_pont_particle}
The stress tensor for the pointlike particle given by Eq.~(\ref{src}) can also be decomposed in angular basis \cite{PhysRevD.2.2141, PhysRevD.67.104017}. % that further allows the separation of the equations of motion. The secondary source Eq. (\ref{src}) in the EMRI system generates the metric and fluid perturbations which can be written in the form of tensor harmonics \cite{PhysRevD.67.104017}. 
The resultant energy-momentum tensor for the secondary object takes the following form 
\begin{align}
T^{P}_{\mu\nu}(t, r, \theta, \phi)=\sum_{l=2}^{\infty}\sum_{m =-l}^{l}
\bigg[&{A}^{0}_{l m }a^{0}_{l m,\mu\nu}(\theta,\phi)+{A}^{1}_{l m}a^{1}_{l m,\mu\nu}(\theta,\phi) +{A}_{l m }a_{l m,\mu\nu}(\theta,\phi)+{B}^{0}_{l m }b^{0}_{l m,\mu\nu}(r,\theta,\phi)\nonumber\\
&+{B}_{l m }b_{l m,\mu\nu}(r,\theta,\phi)+{Q}^{0}_{l m }c^{0}_{l m,\mu\nu}(r,\theta,\phi)+{Q}_{l m }c_{l m,\mu\nu}(r,\theta,\phi)\nonumber\\
&+{D}_{l m }d_{l m,\mu\nu}(r,\theta,\phi)+{G}_{l m}g_{l m,\mu\nu}(r,\theta,\phi)+{F}_{l m }f_{l m,\mu\nu}(r,\theta,\phi)\bigg]. \label{harmonicexp}
\end{align}
Here, as mentioned earlier, the letters in small Latin represent the tensor harmonics quantities which can be found in the appendix (A) of \cite{PhysRevD.67.104017}.

Now we consider the equatorial eccentric orbits at radius $r$ and $\theta_{P}=\frac{\pi}{2}$. The four-velocity of the secondary object, exhibiting eccentric motion, can be written as
\begin{align}
    U^{\mu}_{P} = \Big(U^{t}, U^{r}, 0, U^{\phi}\Big),
\end{align}
where the expressions for $U^{t},~ U^{r}$ and $U^{\phi}$ are given in Eq.~(\ref{gd1}). With this construction of the 4-velocity, the energy-momentum tensor of the secondary has the following non-vanishing coefficients
\begin{align} \label{dfvcxv}
A_{l m }^{0}=& \frac{\mu}{r^{2}} U^{t} \sqrt{f^{3}b} Y^{*}_{lm} \delta(r-r_{P}) \hspace{15mm} ; \hspace{3mm}
A^{1}_{lm} = -\frac{i\sqrt{2}\mu}{r^{2}\sqrt{fb}}fU^{r}Y^{*}_{lm}\delta(r-r_{P}) \nonumber\\
A_{lm} =& \frac{\mu (U^{r})^{2}}{r^{2}U^{t}\sqrt{fb^{3}}}Y^{*}_{lm}\delta(r-r_{P}) \hspace{15mm} ; \hspace{3mm}
B^{0}_{lm} = \frac{\mu\sqrt{2}}{r} U^{\phi} \sqrt{\frac{fb}{l(l+1)}}\partial_{\phi}Y^{*}_{lm}\delta(r-r_{P}) \nonumber \\
B_{lm} =& -\frac{\mu\sqrt{2}U^{r}U^{\phi}}{rU^{t}\sqrt{l(l+1)bf}} Y^{*}_{lm}\delta(r-r_{P}) \hspace{2mm} ; \hspace{3mm} 
Q_{lm}^{0} = -\frac{\mu\sqrt{2}}{r}\sqrt{\frac{bf}{l(l+1)}}U^{\phi}\delta(r-r_{P})\partial_{\theta}Y^{*}_{lm} \nonumber\\
Q_{lm} =& \frac{i\sqrt{2}\mu U^{r}U^{\phi}}{U^{t}r\sqrt{l(l+1)bf}} \delta(r-r_{P})\partial_{\theta}Y^{*}_{lm} \hspace{3mm} ; \hspace{3mm}
D_{lm} = \frac{\mu\sqrt{2}}{U^{t}} \sqrt{\frac{b}{\Tilde{n}f}} (U^{\phi})^{2}\delta(r-r_{P})\partial_{\theta\phi}Y^{*}_{lm} \nonumber\\
G_{lm} =& \frac{\mu}{U^{t}\sqrt{2}}(U^{\phi})^{2}\sqrt{\frac{b}{f}}Y^{*}_{lm} \delta(r-r_{P}) \hspace{7mm} ; \hspace{3mm}
F_{lm} = \frac{\mu (U^{\phi})^{2}}{U^{t}\sqrt{2}}\sqrt{\frac{b}{\Tilde{n} f}} (l+l^{2}-2m^{2})Y^{*}_{lm}\delta(r-r_{P}),
\end{align}
where $\Tilde{n} = l(l-1)(l+1)(l+2)$ and $Y^{*}_{lm} = Y^{*}_{lm}(\theta_{P}, \phi_{P})$. This sums up the construction of the source term of the secondary object. Next, we investigate the Master perturbation equations.

%%%%%%%%%%%%%%%%%%%%%%%%%%%%%%%%%%%%%%%%%%%%%%%%%%%%%%%%%%%%%%%

\section{Axial and Polar perturbations}\label{App_axial_polar}
In this section, we elaborate on the details of the perturbations in the axial and polar sectors. Let us begin with revisiting the axial perturbations.

\subsection{Axial sector}\label{App_axial}
 In this section, we supply relevant expressions for deriving the master equation for axial perturbation, i.e., Eq. (\ref{eq:master_axial}). We begin with the  combination $\mathcal{E_{\theta\theta}}-\mathcal{E_{\phi\phi}}/\sin^{2}\theta$ which gives the following, 
\begin{align}
    h_{0}^{lm} = \frac{i h_{1}^{lm}(r) (f(r) b'(r)+b(r) f'(r))}{2 \omega }+\frac{i b(r) f(r) h_{1}^{lm'}(r)}{\omega }+\frac{8 \pi  \sqrt{2} r^2 D_{lm}(r) f(r)}{\sqrt{(l-1) l (l+1) (l+2)} \omega }.
\end{align}
% \begin{align}\color{blue}
% \frac{\partial h^{lm}_0}{\partial t}=f b \frac{dh^{lm}_1}{dr}
% 	+\frac{f(1-b+r b')}{2r}h^{lm}_{1}-
% 	i \frac{8\sqrt{2}\pi r^2 f}{\sqrt{\Tilde{n}}}{\cal D}^{lm} .
% \end{align}
Then using the $\mathcal{E}_{r\phi}$ component we get, 
\begin{equation}
\begin{aligned}
 \partial^{2}_{r}h_{1}^{lm}(r) = \partial_{r}h_{1}^{lm}(r) (-\frac{3 b'}{2 b}-\frac{3 f'}{2 f}+\frac{2}{r})+\frac{h_{1}^{lm} \tilde{h}_{w}^{lm}(r)}{2 r^{2} b f^{2}}+\frac{8 i \pi  \sqrt{2} r \tilde{h}^{lm}_{z}}{l (l+1) \sqrt{l^{2}+l-2} b} +\frac{8 i \pi  \sqrt{2} r^{2} D_{lm} f}{\sqrt{\tilde{n}} b f},
\end{aligned}
\end{equation}
with $$\tilde{h}_{w}^{lm}(r)=r f \left(\left(4 b-r b'\right) f'+r b f''-2 r \omega ^2\right)+f^2 \left(2 \left(2 r \left(b'-8 \pi  r p_{t}\right)+l^2+l-2\right)-r^2 b''\right)-r^2 b f'^2$$, and $$\tilde{h}^{lm}_{z}(r)=\sqrt{l (l+1)} r D_{lm}'+\sqrt{(l-1) l (l+1) (l+2)} Q^{0}_{lm}.$$
Here, prime denotes the derivative of a function with respect to $r$. We further define  $h_{1}^{lm}(r)=\frac{r}{\sqrt{fb}}\mathcal{R}_{lm\omega}(r)$ and in terms of this new variable we obtain a second-order non-homogeneous differential equation. Finally, the master equation in the frequency domain can be written in the following form, which we also consider mentioning in the main text of the draft (Eq.(\ref{eq:master_axial})),
% \begin{align}
% \Big( - \frac{\partial^2 }{\partial t^2} + \dfrac{\partial^2 }{\partial r_*^2} - V^{ax}\Big) \mathcal{R}_{lm\omega}(r)=  \mathcal{S}^{ax}_{lm\omega} ,\label{eq:master_axial} 
% \end{align}
\begin{align}
    \Big(\frac{d^{2}}{dr_{*}^{2}}+\omega^{2}-V^{ax}(r)\Big)\mathcal{R}_{lm\omega} = S^{ax}_{lm\omega}, 
\end{align}
where the tortoise coordinate follows the relation $dr_{*}=dr/\sqrt{fb}$, and the potential is given by
\begin{equation}
V^{ax}(r) = \dfrac{f}{r^2}\Big(l(l+1) - \dfrac{6m(r)}{r} + m'(r)  \Big)\ ,
\end{equation}
together with the source term
\begin{align}
\label{eq:Source}
\mathcal{S}^{ax}_{lm\omega}(r) = \frac{8 i \sqrt{2} \pi  b(r) f(r)}{\sqrt{l (l+1) b(r)f(r)}} \left(\frac{r f(r) D_{lm}'(r)}{\sqrt{(l-1)(l+2)}}+\frac{r D_{lm}(r) f'(r)}{\sqrt{(l-1) (l+2)}}+f(r) Q^{0}_{lm}(r)\right).
\end{align}
%with
%\begin{align}
%\mathcal{G} = & \\
%\mathcal{F} =& \\
%\mathcal{R}_P =&  
%\end{align}

Let us now turn our discussion to boundary conditions. We need to put appropriate boundary conditions in order to obtain the solution of the perturbation equation. We follow \cite{Pani} for deriving the boundary conditions of the homogeneous perturbation equation. Let us recast the Eq. (\ref{eq:master_axial}) in the following form
% \begin{align}
%     \Big(\frac{d^{2}}{dr_{*}^{2}}+\omega^{2}-V^{ax}\Big)\mathcal{R}_{lm\omega} = S^{ax}_{lm\omega}
% \end{align}
% where $\frac{dr_{*}}{dr} = \frac{1}{\sqrt{fb}}$. 
%We can now recast the above equation in the following form
\begin{equation}
\Delta^{2}\mathcal{R}''_{lm\omega}+r^{2}\Delta \frac{d\tilde f}{dr} \mathcal{R}_{lm\omega}+(\omega^{2}-V^{ax})r^{4}\mathcal{R}_{lm\omega} = 0,\label{sn1}
\end{equation}
where $\Delta=r^{2}\sqrt{f(r)b(r)}$ and $\tilde{f}(r)=\frac{\Delta}{r^{2}}$. Let us first consider the boundary condition at the horizon. For this, it is convenient to write the above equation as
\begin{equation}
(r-r_+)^2\frac{d^2\mathcal{R}_{lm\omega}}{dr^2} + (r-r_+) p_H(r)\frac{d\mathcal{R}_{lm\omega}}{dr} + q_H(r)\mathcal{R}_{lm\omega}=0,
\end{equation}
where 
\begin{align}
p_H(r) &=(r-r_{+}) \frac{h_{s}'}{h_{s}}+1 \hspace{3mm} ; \hspace{3mm}
q_H(\hat{r}) = \frac{1}{h_{s}^{2}}(\omega^{2}-V^{ax}) .
\end{align}
Here, we note that $\sqrt{f(r)b(r)} = (r-r_{+})h_{s}(r)$ and $h_{s}=e^{\frac{\Gamma}{2}}\Big(1-\frac{2(r-r_{+})M}{(a_{0}+r)^{2}}\Big)$. We use the Frobenius method for constructing the power series solution
\begin{equation}\label{bc_axial_horizon}
\mathcal{R}_{lm\omega} = (r-r_+)^d \displaystyle\sum_{n=0}^{\infty} a_n (r-r_+)^n,
\end{equation}
where $d$ is one of the solutions of the indicial equation

\begin{equation}
I(d) = d(d-1) + p_H(r_+) d + q_H(r_+) =0 \Longrightarrow d = \pm\frac{i\omega}{\sqrt{h_{s}(r_{+})}}.
\end{equation}
Thus the solution is given by
\begin{equation}
\mathcal{R}_{lm\omega}=\exp\bigg\{\pm \frac{i\omega}{\sqrt{h_{s}(r_{+})}} \log(\hat{r} - \hat{r}_+)\bigg\} \displaystyle \sum_{n=0}^{\infty} 
a_n (r-r_+)^n \label{eq:BCHor} \ .
\end{equation}
The recursion relation for $a_n$, with setting $a_{0}=1$ is
\begin{equation}
a_n = - \frac{1}{I(d+n)}  \displaystyle\sum_{k=0}^{n-1}\frac{(k+d)p_H^{(n-k)}(\hat{r}_+) +q_H^{(n-k)}(r_+)}{(n-k)!} a_k .
\end{equation}
Here, $p_H^{(k)}(r_+) $ and $q_H^{(k)}(r_+) $ are the $k$-th derivatives of $p_H(r)$ and $q_H(r)$, further evaluated at $r_+$. This sets up the boundary condition at the horizon. Let us now turn to the boundary condition at infinity.

We rewrite the Eq. (\ref{sn1}) in the following form in order to derive the boundary conditions at infinity
\begin{equation}
\mathcal{R}''_{lm\omega}+p_{\infty}\mathcal{R}'_{lm\omega}+q_{\infty}\mathcal{R}_{lm\omega} = 0 , 
\end{equation}
where 
\begin{align}
p_{\infty} = \frac{r^{2}}{\Delta}\frac{d\tilde{f}}{dr} \hspace{3mm} ; \hspace{3mm} q_{\infty} = (\omega^{2}-V^{ax})\frac{r^{4}}{\Delta^{2}}\,.
\end{align}
Following \cite{Pani}, one can write down the series solution of $\mathcal{R}$ as
%$p_{\infty}$ and $q_{\infty}$ are analytic functions and their series solutions are written in the following form
%\begin{alignat*}{2}
%p_\infty(r) = \displaystyle \sum_{n=0}^{\infty} \frac{1}{n!}  \frac{p_\infty^{(n)}}{r^n} \hspace{3mm} ; \hspace{3mm} 
%q_\infty(r) = 
%\displaystyle \sum_{n=0}^{\infty} \frac{1}{n!}  \frac{q_\infty^{(n)}}{r^n} ,
%\end{alignat*}
%where $p_\infty^{(n)}$ and $q_\infty^{(n)}$ are $n$-th derivatives of the coefficients $p_\infty$ and $q_\infty$  with respect to $\hat{r}$. One can write down the series solution if at least one of $p_\infty^{(0)}$, $q_\infty^{(0)}$ or $q_\infty^{(1)}$ is nonzero.
\begin{equation}\label{bc_axial_inf}
\mathcal{R}_{lm\omega} = e^{\gamma r} r^{\tilde{\xi}} \displaystyle\sum_{n=0}^{\infty} \frac{b_n}{r^n} ,
\end{equation}
where $\gamma = \pm  i\omega$ and  $\tilde{\xi} = \pm 2i (M+M_{\textrm{BH}}) \omega $.
%where $\gamma$ is one of the solutions of the characteristic equation
%\begin{equation}
%\gamma^2 + p^{(0)}_\infty \gamma + q^{(0)}_\infty =0 \hspace{3mm} ; \hspace{3mm} \textup{and} \hspace{3mm} ; \hspace{3mm} \tilde{\xi}= -\frac{p_\infty^{(1)}\gamma +q_\infty^{(1)}}{p_\infty^{(0)}+2 \gamma} \ .
%\end{equation}
%We further have,
%\begin{align}
%p_\infty^{(0)} &= 0= p_\infty^{(1)} \hspace{2mm} ; \hspace{2mm} q_\infty^{(0)} = \omega^2 \hspace{2mm} ; \hspace{2mm}  q_\infty^{(1)} = 4(M+M_{\textrm{BH}}) \omega^2 \hspace{2mm} ; \hspace{2mm}
%\gamma^2 + \omega^2 =0  \hspace{2mm} ; \hspace{2mm} \tilde{\xi} = \pm 2i (M+M_{\textrm{BH}}) \omega .
%\end{align}
Therefore, we have two series solutions
\begin{equation}
\mathcal{R}_{lm\omega}=\exp \{\pm i \omega [r +2(M+M_{\textrm{BH}}) \log(r)]\} \displaystyle 
\sum_{n=0}^{\infty}\frac{b_n}{r^n} \,.
\label{eq:BCInf}
\end{equation}
The general recursion relation for the coefficients $b_n$ with $b_0=1$ is
\begin{align}
(p_\infty^{(0)} + 2 \gamma) n b_n = (n - \tilde{\xi})(n-1 -\tilde{\xi})b_{n-1} 
+\displaystyle \sum_{k=1}^{n}\Big[\gamma p_\infty^{(k+1)}+q_\infty^{(k+1)} - (n-k-\tilde{\xi})p_\infty^{(k)}\Big]b_{n-k}\ .
\end{align}
This concludes the boundary conditions at infinity. We use these boundary conditions, at the horizon and at infinity, to solve the axial perturbation equation numerically.

\subsection{Polar sector}

In this section, we derive a set of first-order differential equations for the metric polar perturbations using components of $\mathcal{E}_{\mu\nu}$. The components $\mathcal{E}_{tr}$, $\mathcal{E}_{t\theta}$ and $\mathcal{E}_{r\theta}$ give three inhomogeneous differential equations for $\frac{dK^{lm}}{dr}$, $\frac{dH_{1}^{lm}}{dr}$ and $\frac{dH_{0}^{lm}}{dr}$ respectively. We further use the conservation of energy-momentum tensor ($\nabla_{\mu}T^{\mu\nu}=0$) in order to find two other differential equations for variables $\partial_{r}W^{lm}(r)$ and $\partial_{r}\delta\rho^{lm}(r)$ together with an algebraic relation for fluid velocity component $V$. We simplify the equations by replacing $H_{2}$ and its derivatives with the help of $\mathcal{E}_{\theta\phi}$. With this setup, we get five coupled inhomogeneous first-order ordinary differential equations (ODEs) for $\vec{\psi}=(H^{lm}_1, H^{lm}_0, K^{lm}, W^{lm},\delta\rho^{lm})$ that can be written in the matrix form in the following way (also mentioned in the main text Eq. (\ref{ppte1})),
\begin{align}
  \frac{d\vec{\psi}_{lm\omega}}{dr}-\boldsymbol{\alpha}\vec{\psi}_{lm\omega}=\vec S^{pol}_{lm\omega} ,
\end{align}
where $\vec S^{pol}_{lm\omega}$ denotes the source term and $\boldsymbol{\alpha}$ is a matrix whose non-zero coefficients are given by
\begin{align} \label{alpha_vals}
\alpha_{11}&=\frac{r m'-2 m}{r (r-2 m)} \hspace{3mm} ; \hspace{3mm}
\alpha_{12}=-\frac{i \left(2 f m'+r^2 \omega ^2\right)}{r \omega  (r-2 m)} \hspace{3mm} ; \hspace{3mm}
\alpha_{13}=-\frac{i r \omega }{r-2 m} \hspace{3mm} ; \hspace{3mm}
\alpha_{15}= \frac{16 i \pi  c_{st}^2 r f}{\omega(r -2 m)}\,, \nonumber\\
\alpha_{21}&=\frac{i l (l+1)}{2 r^2 \omega }-\frac{i \omega }{f} \hspace{3mm} ; \hspace{3mm}
\alpha_{22}=\frac{1}{2 m-r}+\frac{2}{r} \hspace{3mm} ; \hspace{3mm}
\alpha_{23}=-\frac{r-3 m}{r(r-2 m)} \hspace{3mm} ; \hspace{3mm} 
\alpha_{24}=\frac{4 i r^2 f (r-m)}{\omega  (2 r-3 m) (r-2 m)^2}\,, \nonumber\\
\alpha_{31}&=\frac{i l (l+1)}{2 r^2 \omega } \hspace{3mm} ; \hspace{3mm} \alpha_{32}=\frac{1}{r} \hspace{3mm} ; \hspace{3mm}
\alpha_{33}=-\frac{r-3 m}{r(r-2 m)} \hspace{3mm} ; \hspace{3mm}
\alpha_{34}=\frac{4 i r^2 f}{-7 r \omega  m+6 \omega  m^2+2 r^2 \omega } \nonumber\\
\alpha_{42}&=\frac{i (2 r-3 m) m'\left(l (l+1) f+r^2 \omega ^2\right)}{4 r^5 \omega  f} \hspace{3mm} ; \hspace{3mm}
\alpha_{43}=\frac{i \omega  (2 r-3 m)^2 m'}{4 r^3 f (r-2 m)}\,, \nonumber \\
\alpha_{44}&=\frac{-r^2 \left(5 m'+4\right)+r m \left(9 m'+13\right)-12 m^2}{r (2 r-3 m) (r-2 m)} \hspace{3mm} ; \hspace{3mm}
\alpha_{45}=\frac{2 i \pi  (2 r-3 m) \left(r^2 \omega ^2-c_{st}^2 l (l+1) f\right)}{r^3 \omega  f}\,, \nonumber \\ 
\alpha_{51}&=\frac{i m' \left(l (l+1) f (r-m)+2 r^2 \omega ^2 (r-2 m)\right)}{16 \pi  c_{sr}^2 r^4 \omega  f (r-2 m)} \hspace{3mm} ; \hspace{3mm} \alpha_{52}=\frac{(r-3 m) m'}{8 \pi  c_{sr}^2 r^3 (r-2 m)}\,, \nonumber\\
\alpha_{53}&=-\frac{(r-3 m) (r-m) m'}{8 \pi  c_{sr}^2 r^3 (r-2 m)^2} \hspace{3mm} ; \hspace{3mm}
\alpha_{54}=\frac{i \left(m \left(f \left(l^2+l-m^2\right)-4 r^2 \omega ^2\right)+2 r f m'+2 r^3 \omega ^2\right)}{4 \pi  c_{sr}^2 \omega  (2 r-3 m) (r-2 m)^2}\,, \nonumber\\
\alpha_{55}&=\frac{\left(3 c_{sr}^2-4 c_{st}^2-1\right) m+2 r \left(c_{st}^2-c_{sr}^2\right)}{c_{sr}^2 r (r-2 m)}, 
\end{align}
and we recall that $m$ and $f$ are functions of $r$.

%\section{Results}
The source term can be written as $\vec S^{pol}_{lm\omega}=(S_1,S_2,S_3,S_4,S_5)$ where

\begin{equation}\label{source_polar}
    \begin{aligned}
        S_1&=\frac{8 i \pi  \sqrt{2} B_{lm}^{0} r^2}{\sqrt{l (l+1)} (r-2 m)}+\frac{16 i \pi  \sqrt{2} F_{lm} r \left(r^2 \omega ^2 (r-2 m)-m f(r) m'\right)}{\sqrt{(l-1) l (l+1) (l+2)} \omega  (r-2 m)^2}\,,\\
        S_2&=\frac{4 \pi  \sqrt{2} A_{lm}^{1}}{\omega }+\frac{8 \pi  \sqrt{2} B_{lm} r}{\sqrt{l (l+1)}}+\frac{16 \pi  \sqrt{2} F_{lm} m r}{\sqrt{(l-1) l (l+1) (l+2)} (r-2 m)}\,,\\
        S_3&=\frac{4 \sqrt{2} \pi  A_{lm}^{1}}{\omega }-\frac{16 \sqrt{2} \pi  F_{lm} r}{\sqrt{(l-1) l (l+1) (l+2)}}\,,\\
        S_4&=\frac{2 i \sqrt{2} \pi  F_{lm} (2 r-3 m) m' \left(l (l+1) m f(r)+2 r^2 \omega ^2 (2 m-r)\right)}{\sqrt{(l-1) l (l+1) (l+2)} r^3 \omega  f(r) (r-2 m)}\,,\\
        S_5&=\frac{A_{lm}^{1} (r-m) m'}{\sqrt{2} c_{sr}^2 r^2 \omega  (r-2 m)}+\frac{\sqrt{2} B_{lm} m'}{c_{sr}^2 \sqrt{l (l+1)} r}\,.
    \end{aligned}
\end{equation}
\newpage
%Given that all the eigenvalues $\lambda_j,~(j=1,2,..,5)$ of $\tilde{\boldsymbol{\alpha}}(r=2)$ are distinct, the fundamental matrix solution of the above equation has the following form
%%%%%%%%%%%%%%%%%%%%%%%%%%%%%%%%%%%%%%%%%%%%%%%%%%%%%%%%%%%%%%%%
%\begin{equation}\label{fundamental_matrix_regular}
 %   \begin{aligned}
%        \boldsymbol{\Psi}(r)=\boldsymbol{Q}(r) (r-2)^{\textrm{diag}(\lambda_1,~\lambda_2,...,~\lambda_5)}
%    \end{aligned}
%\end{equation}
%%%%%%%%%%%%%%%%%%%%%%%%%%%%%%%%%%%%%%%%%%%%%%%%%%%%%%%%%%%%%%%%
%where, $\boldsymbol{Q}(r)$ is holomorphic at $r=2$ and $\det \boldsymbol{Q}\neq 0$.
%This equation has a fundamental matrix solution of the form 

\section{Fluxes: for different modes, halo parameters and eccentricities}\label{apenteu1}

\begin{figure}[htb!]
	%%%%%%%%%%%%%%%%%%%%%%%%
	\centering
	\minipage{0.48\textwidth}
	\includegraphics[width=\linewidth]{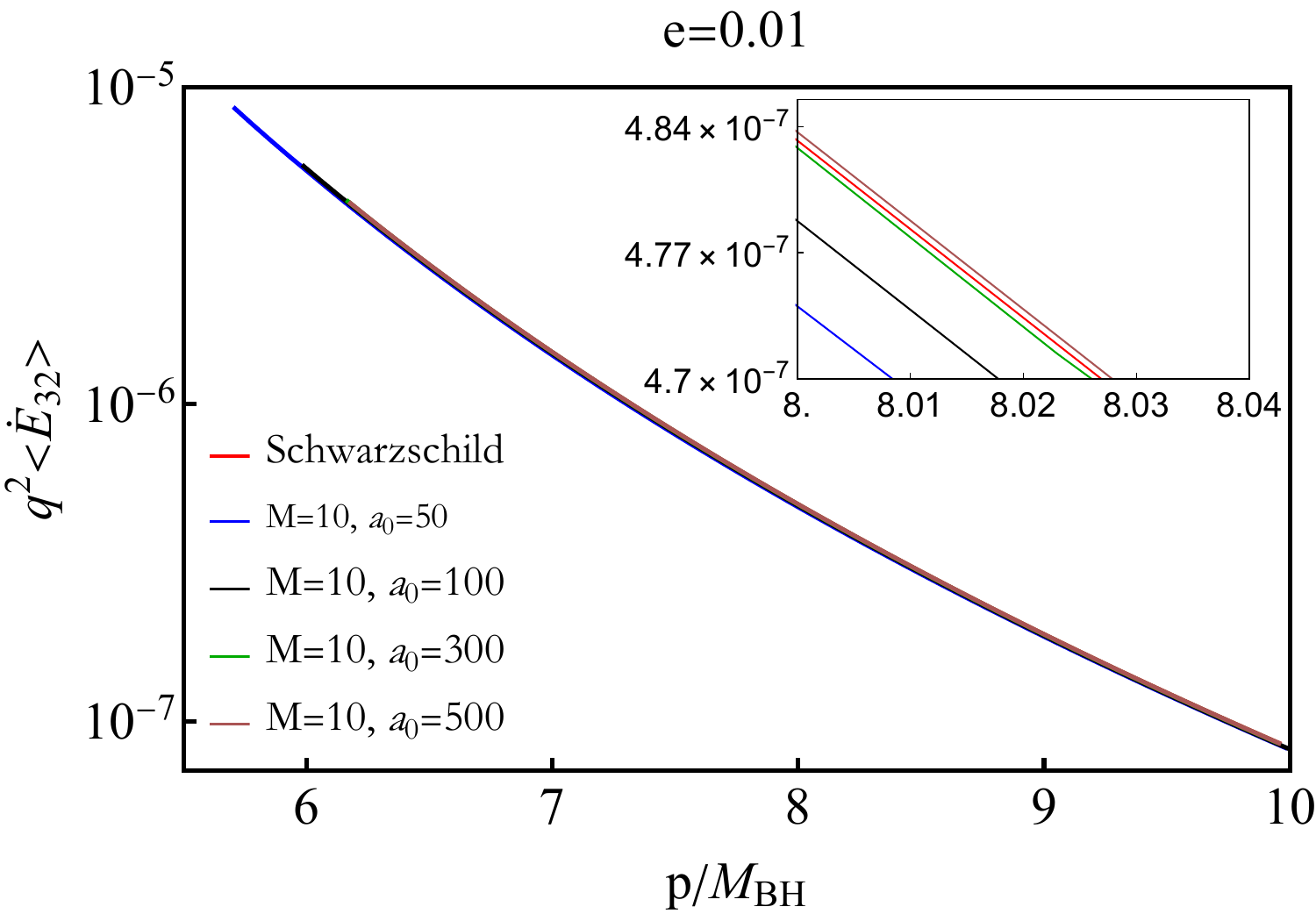}
	\endminipage\hfill
	%%%%%%%%%%%%%%%%%%%%%%%%
	\minipage{0.48\textwidth}
	\includegraphics[width=\linewidth]{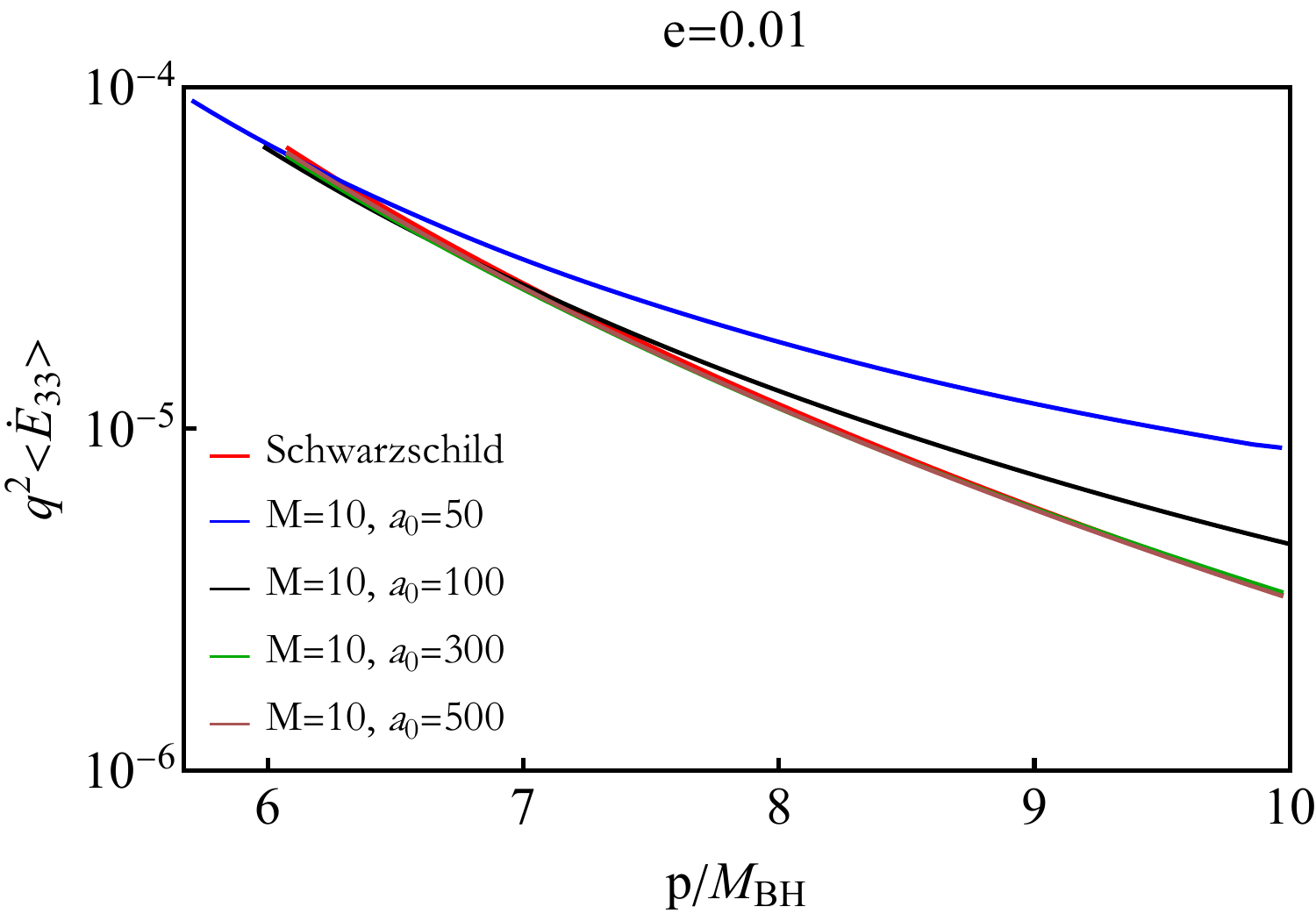}
	\endminipage\hfill
 \minipage{0.48\textwidth}
	\includegraphics[width=\linewidth]{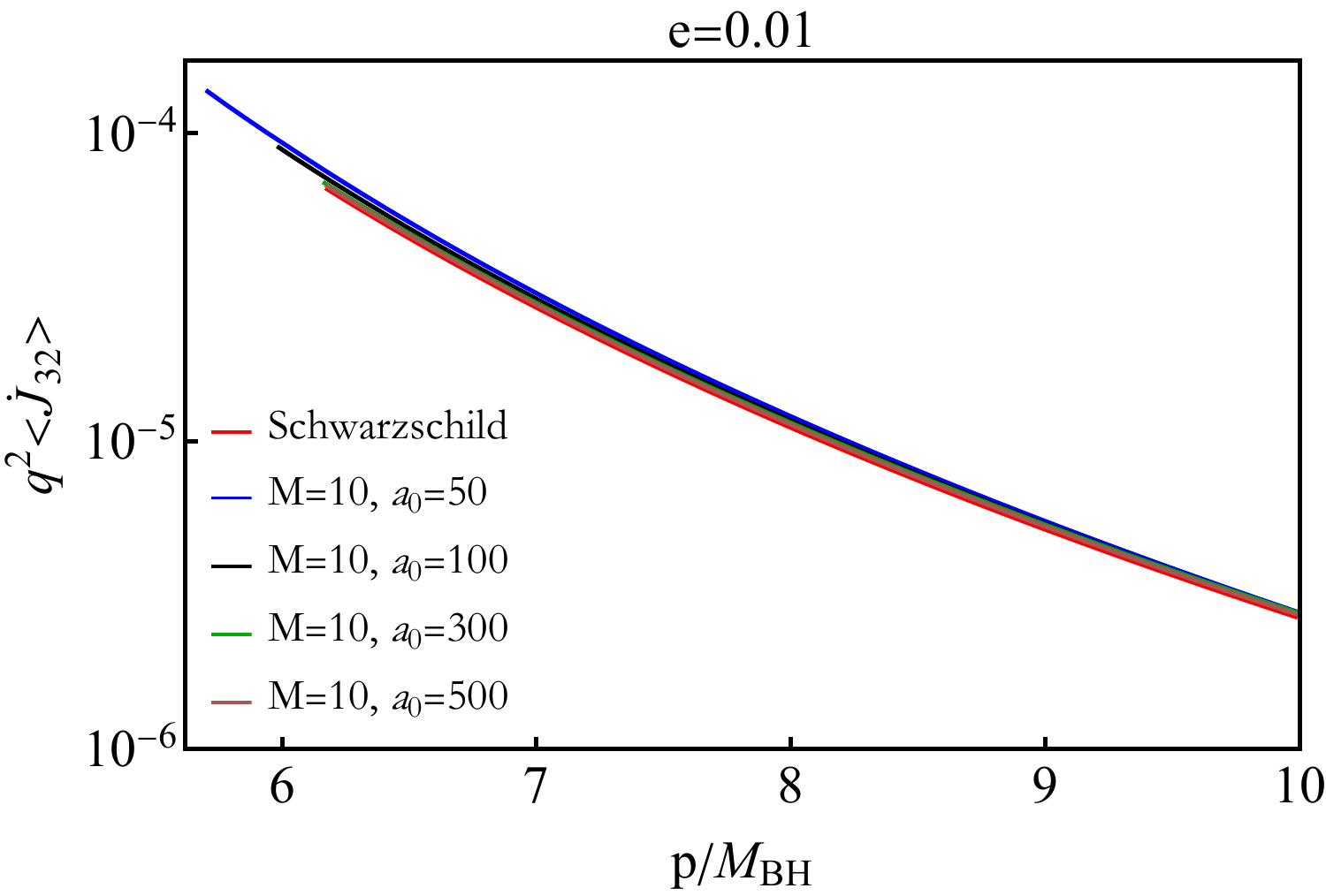}
	\endminipage\hfill
 \centering
	\minipage{0.48\textwidth}
	\includegraphics[width=\linewidth]{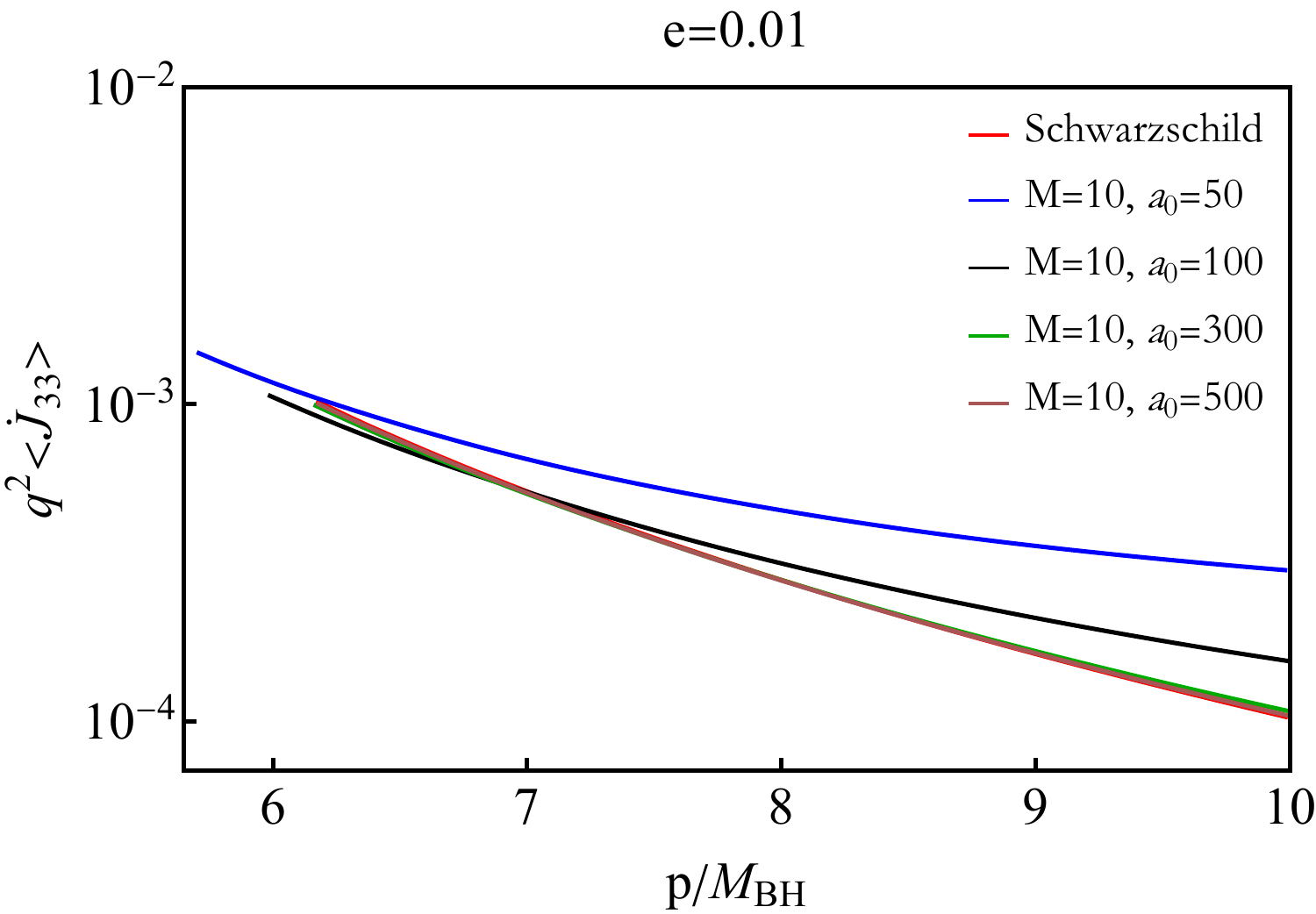}
 \endminipage
	\caption{In the upper panel, we show the average energy flux for $\{3,2\}$ (left panel) and $\{3,3\}$ (right panel) modes in the range $p\in(p_{\textrm{min}}(e),~p_{\textrm{ini}})$ for a fixed value of orbital eccentricity and different values of dark matter parameters $M$ and $a_0$. The lower panel depicts the same for average angular momentum flux. We set $p_{\textrm{ini}}=10$. We take the eccentricity value $e=0.01$. The red line in each of these plots represents the flux for the Schwarzschild black hole.
 }\label{fig_flux_higher}
\end{figure}
%%%%%%%%%%%%%%%%%%%%%%%%%%%%%%%%%%%%%%%%%%%%%%%
\begin{figure}[htb!]
	%%%%%%%%%%%%%%%%%%%%%%%%
	\centering
	\minipage{0.48\textwidth}
	\includegraphics[width=\linewidth]{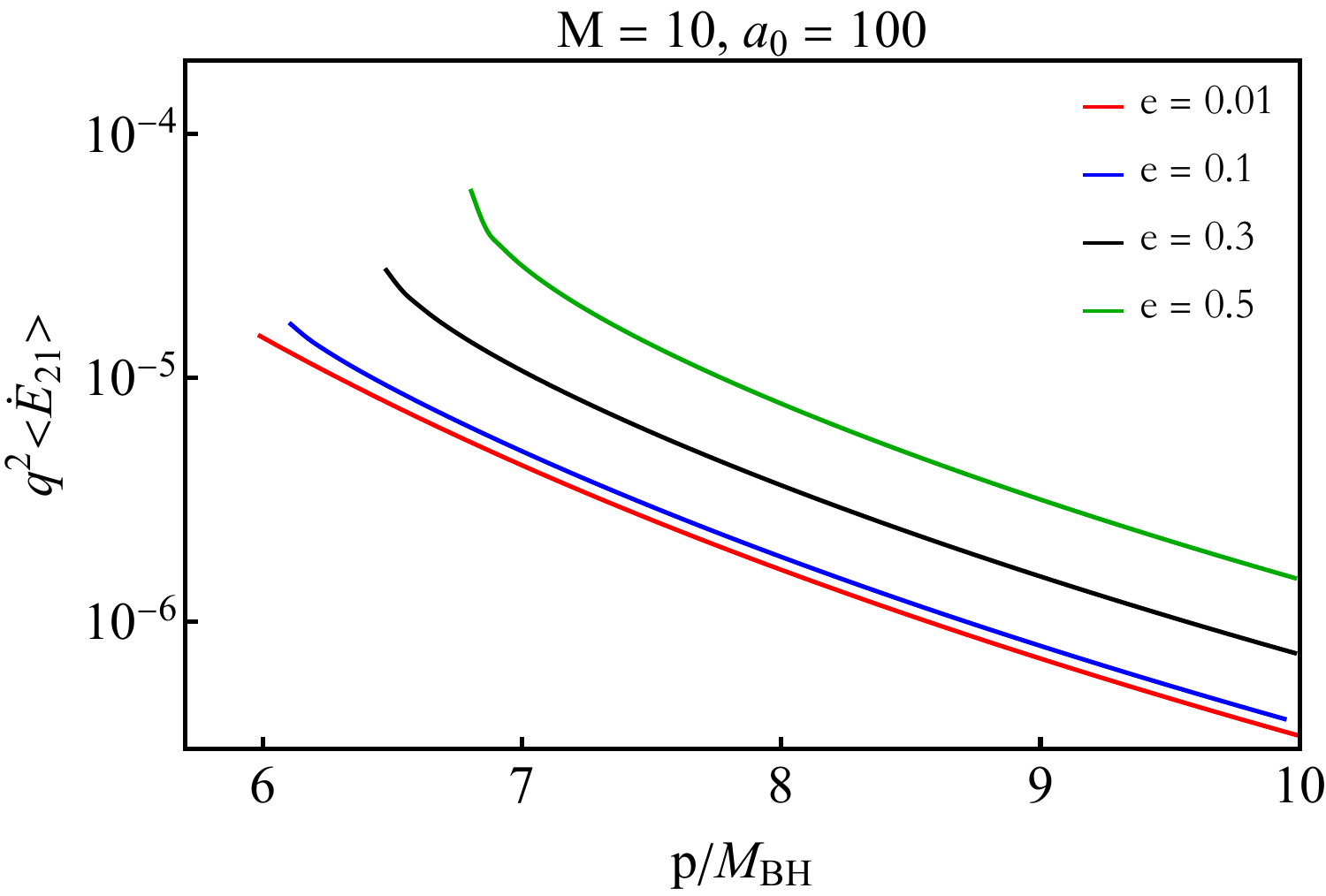}
 % \caption{Both left and right panels describe the behaviour of eccentricity for fixed halo parameters (with $M=10, a_{0}=100$) on average energy flux for modes $\lbrace 2,1 \rbrace$ and $\lbrace 2,2 \rbrace$}
	\endminipage
	%%%%%%%%%%%%%%%%%%%%%%%%
	\minipage{0.48\textwidth}
	\includegraphics[width=\linewidth]{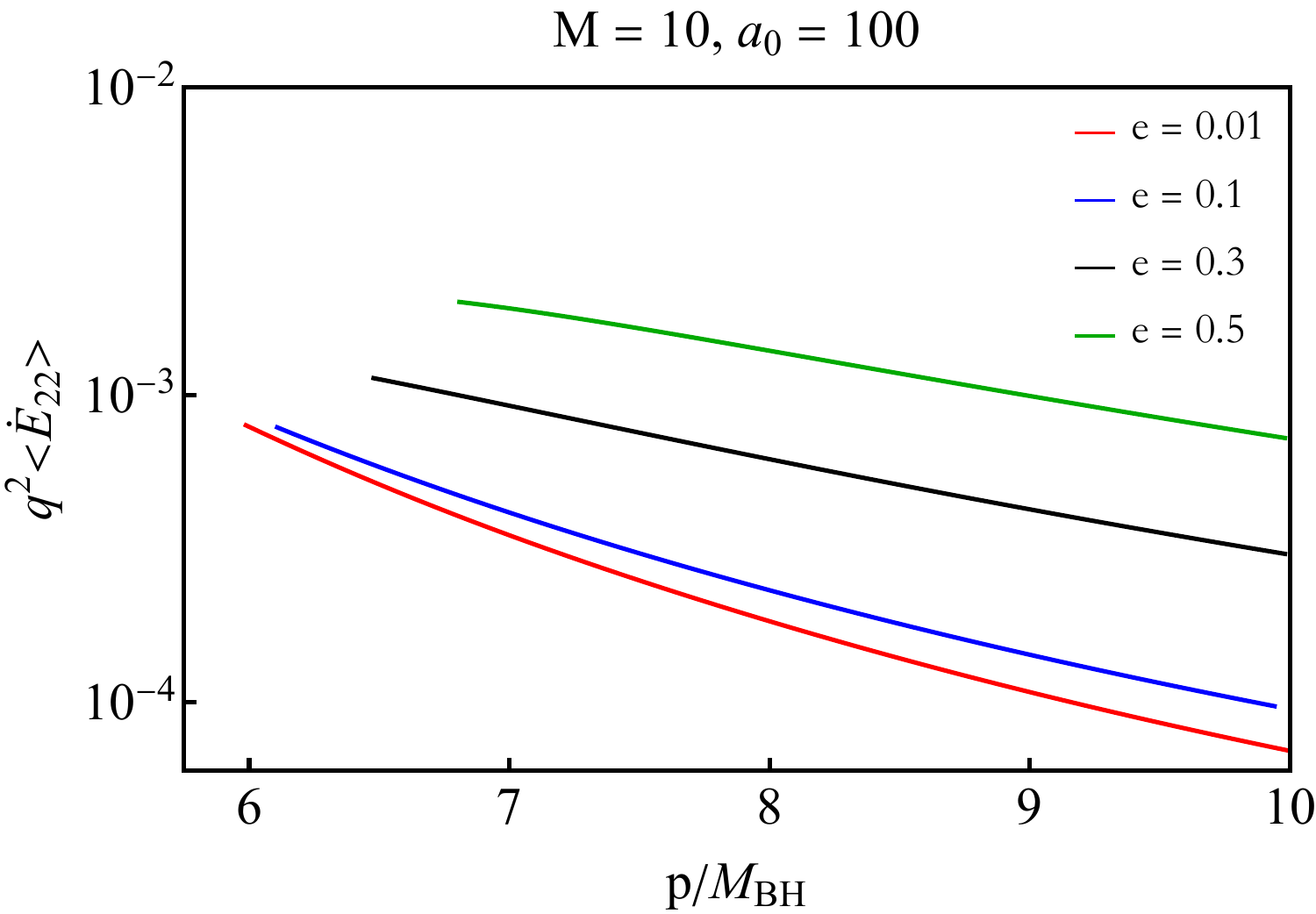}
	\endminipage\hfill
 \minipage{0.48\textwidth}
	\includegraphics[width=\linewidth]{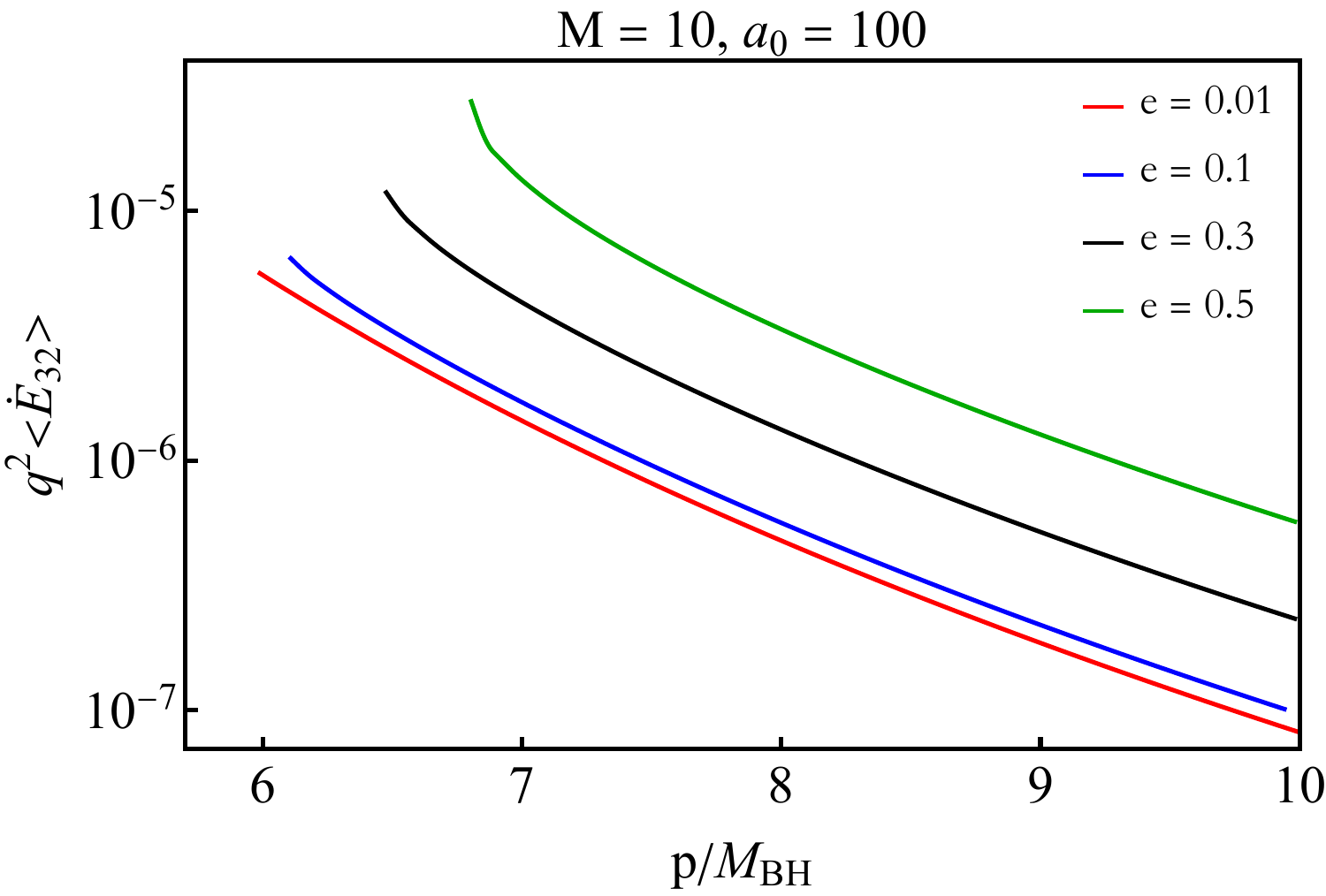}
	\endminipage\hfill
 \centering
	\minipage{0.48\textwidth}
	\includegraphics[width=\linewidth]{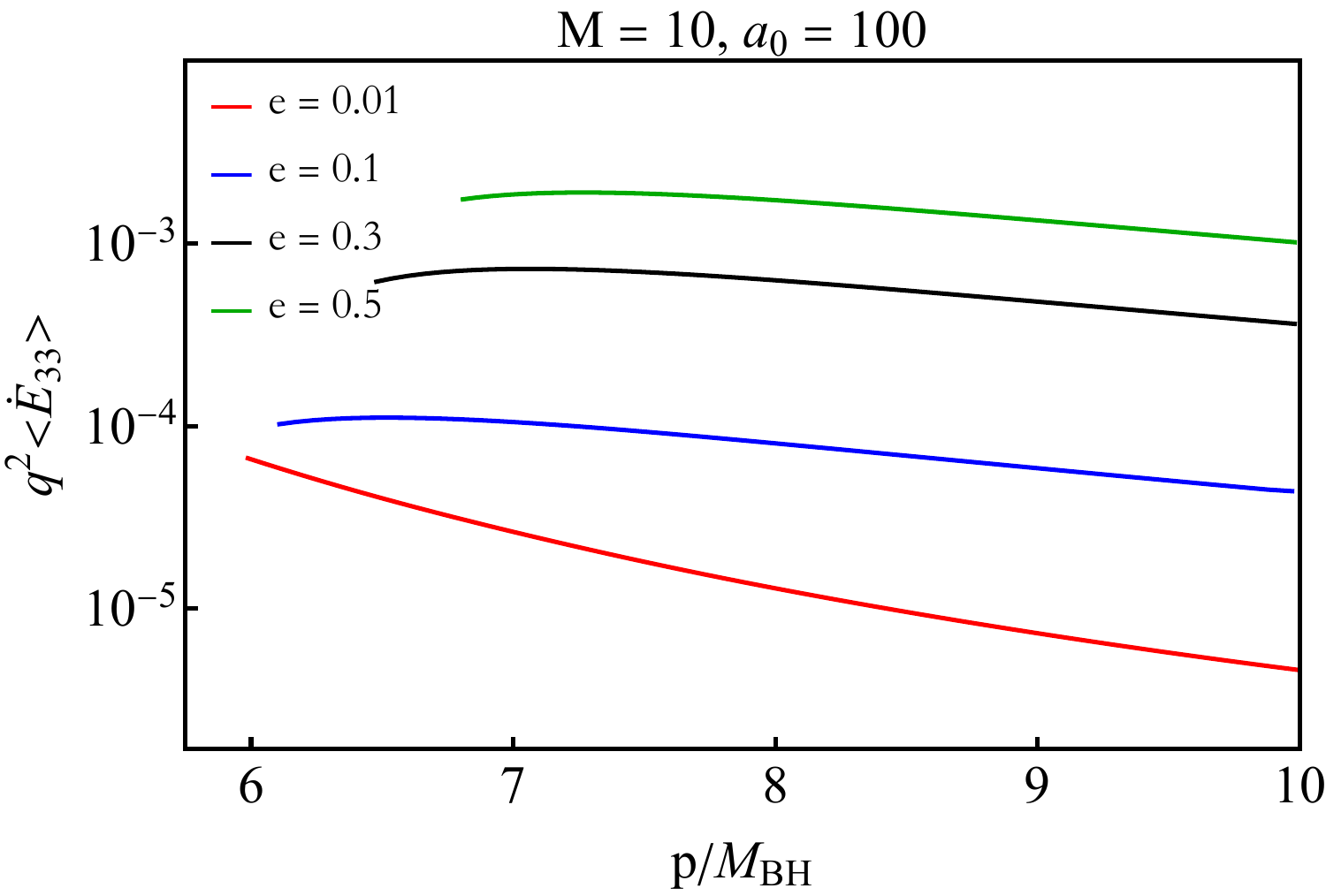}
 \endminipage
	\caption{In the upper panel, we show the average energy flux for $\{2,1\}$ (left panel) and $\{2,2\}$ (right panel) modes in the range $p\in(p_{\textrm{min}}(e),~p_{\textrm{ini}})$ for the fixed values of dark matter parameters $(M, a_{0})=(10, 100)$ and different values of orbital eccentricity ($e$). The lower panel depicts the same for $\{3,2\}$ (left panel) and $\{3,3\}$ (right panel) modes. We set $p_{\textrm{ini}}=10$.
 }\label{fig_enr_flux_same_DM}
\end{figure}
%%%%%%%%%%%%%%%%%%%%%%%%%%%%%%%%%%%%%%%%%%%%%%%%%%%%%%%%%
%%%%%%%%%%%%%%%%%%%%%%%%%%%%%%%%%%%%%%%%%%%%%%%%%%%%%%%%%%%%%%%%
%\newpage
Here we provide average energy and angular momentum fluxes for different higher-order modes. The Fig. (\ref{fig_flux_higher}) depicts the corresponding quantities for different modes with the fixed value of eccentricity and various dark matter parameters. Further, the Fig. (\ref{fig_enr_flux_same_DM}) depicts the same for $(M,a_{0})=(10,100)$ and different values of eccentricity. %The various plots show the behaviour of dark matter mass, halo radius and eccentricity in the estimated quantities. 

\newpage

\bibliography{DM}

\providecommand{\href}[2]{#2}\begingroup\raggedright\begin{thebibliography}{10}

\bibitem{1990ApJ...356..359H}
L.~{Hernquist}, \emph{{An Analytical Model for Spherical Galaxies and Bulges}},
  \href{https://doi.org/10.1086/168845}{\emph{Astrophys. J.} {\bfseries 356}
  (1990) 359}.

\bibitem{Navarro:1995iw}
J.~F. Navarro, C.~S. Frenk and S.~D.~M. White, \emph{{The Structure of cold
  dark matter halos}}, \href{https://doi.org/10.1086/177173}{\emph{Astrophys.
  J.} {\bfseries 462} (1996) 563}
  [\href{https://arxiv.org/abs/astro-ph/9508025}{{\ttfamily
  astro-ph/9508025}}].

\bibitem{Freese:2008cz}
K.~Freese, \emph{{Review of Observational Evidence for Dark Matter in the
  Universe and in upcoming searches for Dark Stars}},
  \href{https://doi.org/10.1051/eas/0936016}{\emph{EAS Publ. Ser.} {\bfseries
  36} (2009) 113} [\href{https://arxiv.org/abs/0812.4005}{{\ttfamily
  0812.4005}}].

\bibitem{Clowe:2006eq}
D.~Clowe, M.~Bradac, A.~H. Gonzalez, M.~Markevitch, S.~W. Randall, C.~Jones
  et~al., \emph{{A direct empirical proof of the existence of dark matter}},
  \href{https://doi.org/10.1086/508162}{\emph{Astrophys. J. Lett.} {\bfseries
  648} (2006) L109} [\href{https://arxiv.org/abs/astro-ph/0608407}{{\ttfamily
  astro-ph/0608407}}].

\bibitem{Bertone:2004pz}
G.~Bertone, D.~Hooper and J.~Silk, \emph{{Particle dark matter: Evidence,
  candidates and constraints}},
  \href{https://doi.org/10.1016/j.physrep.2004.08.031}{\emph{Phys. Rept.}
  {\bfseries 405} (2005) 279}
  [\href{https://arxiv.org/abs/hep-ph/0404175}{{\ttfamily hep-ph/0404175}}].

\bibitem{vandenBergh:1999sa}
S.~van~den Bergh, \emph{{The Early history of dark matter}},
  \href{https://doi.org/10.1086/316369}{\emph{Publ. Astron. Soc. Pac.}
  {\bfseries 111} (1999) 657}
  [\href{https://arxiv.org/abs/astro-ph/9904251}{{\ttfamily
  astro-ph/9904251}}].

\bibitem{Persic:1995ru}
M.~Persic, P.~Salucci and F.~Stel, \emph{{The Universal rotation curve of
  spiral galaxies: 1. The Dark matter connection}},
  \href{https://doi.org/10.1093/mnras/278.1.27}{\emph{Mon. Not. Roy. Astron.
  Soc.} {\bfseries 281} (1996) 27}
  [\href{https://arxiv.org/abs/astro-ph/9506004}{{\ttfamily
  astro-ph/9506004}}].

\bibitem{Corbelli:1999af}
E.~Corbelli and P.~Salucci, \emph{{The Extended Rotation Curve and the Dark
  Matter Halo of M33}},
  \href{https://doi.org/10.1046/j.1365-8711.2000.03075.x}{\emph{Mon. Not. Roy.
  Astron. Soc.} {\bfseries 311} (2000) 441}
  [\href{https://arxiv.org/abs/astro-ph/9909252}{{\ttfamily
  astro-ph/9909252}}].

\bibitem{Massey:2010hh}
R.~Massey, T.~Kitching and J.~Richard, \emph{{The dark matter of gravitational
  lensing}}, \href{https://doi.org/10.1088/0034-4885/73/8/086901}{\emph{Rept.
  Prog. Phys.} {\bfseries 73} (2010) 086901}
  [\href{https://arxiv.org/abs/1001.1739}{{\ttfamily 1001.1739}}].

\bibitem{Ellis:2010kf}
J.~Ellis and K.~A. Olive, \emph{{Supersymmetric Dark Matter Candidates}},
  \href{https://arxiv.org/abs/1001.3651}{{\ttfamily 1001.3651}}.

\bibitem{Kormendy:2013dxa}
J.~Kormendy and L.~C. Ho, \emph{{Coevolution (Or Not) of Supermassive Black
  Holes and Host Galaxies}},
  \href{https://doi.org/10.1146/annurev-astro-082708-101811}{\emph{Ann. Rev.
  Astron. Astrophys.} {\bfseries 51} (2013) 511}
  [\href{https://arxiv.org/abs/1304.7762}{{\ttfamily 1304.7762}}].

\bibitem{Harris:2015vxa}
W.~Harris, G.~Harris and M.~Hudson, \emph{{Dark Matter Halos in Galaxies and
  Globular Cluster Populations. II: Metallicity and Morphology}},
  \href{https://doi.org/10.1088/0004-637X/806/1/36}{\emph{Astrophys. J.}
  {\bfseries 806} (2015) 36}
  [\href{https://arxiv.org/abs/1504.03199}{{\ttfamily 1504.03199}}].

\bibitem{Macedo:2013qea}
C.~F.~B. Macedo, P.~Pani, V.~Cardoso and L.~C.~B. Crispino, \emph{{Into the
  lair: gravitational-wave signatures of dark matter}},
  \href{https://doi.org/10.1088/0004-637X/774/1/48}{\emph{Astrophys. J.}
  {\bfseries 774} (2013) 48} [\href{https://arxiv.org/abs/1302.2646}{{\ttfamily
  1302.2646}}].

\bibitem{Coogan:2021uqv}
A.~Coogan, G.~Bertone, D.~Gaggero, B.~J. Kavanagh and D.~A. Nichols,
  \emph{{Measuring the dark matter environments of black hole binaries with
  gravitational waves}},
  \href{https://doi.org/10.1103/PhysRevD.105.043009}{\emph{Phys. Rev. D}
  {\bfseries 105} (2022) 043009}
  [\href{https://arxiv.org/abs/2108.04154}{{\ttfamily 2108.04154}}].

\bibitem{Baryakhtar:2022hbu}
M.~Baryakhtar et~al., \emph{{Dark Matter In Extreme Astrophysical
  Environments}},  in \emph{{Snowmass 2021}}, 3, 2022,
  \href{https://arxiv.org/abs/2203.07984}{{\ttfamily 2203.07984}}.

\bibitem{Singh:2022wvw}
D.~Singh, A.~Gupta, E.~Berti, S.~Reddy and B.~S. Sathyaprakash,
  \emph{{Constraining properties of asymmetric dark matter candidates from
  gravitational-wave observations}},
  \href{https://doi.org/10.1103/PhysRevD.107.083037}{\emph{Phys. Rev. D}
  {\bfseries 107} (2023) 083037}
  [\href{https://arxiv.org/abs/2210.15739}{{\ttfamily 2210.15739}}].

\bibitem{Bhattacharya:2023stq}
S.~Bhattacharya, B.~Dasgupta, R.~Laha and A.~Ray, \emph{{Can LIGO Detect
  Asymmetric Dark Matter?}},
  \href{https://arxiv.org/abs/2302.07898}{{\ttfamily 2302.07898}}.

\bibitem{Bhattacharyya:2023kbh}
A.~Bhattacharyya, S.~Ghosh and S.~Pal, \emph{{Worldline effective field theory
  of inspiralling black hole binaries in presence of dark photon and axionic
  dark matter}},  \href{https://arxiv.org/abs/2305.15473}{{\ttfamily
  2305.15473}}.

\bibitem{AbhishekChowdhuri:2023cle}
A.~Chowdhuri, R.~K. Singh, K.~Kangsabanik and A.~Bhattacharyya,
  \emph{{Gravitational Radiation from hyperbolic encounters in the presence of
  dark matter}},  \href{https://arxiv.org/abs/2306.11787}{{\ttfamily
  2306.11787}}.

\bibitem{BENSON201033}
A.~J. Benson, \emph{Galaxy formation theory},
  \href{https://doi.org/https://doi.org/10.1016/j.physrep.2010.06.001}{\emph{Physics
  Reports} {\bfseries 495} (2010) 33}.

\bibitem{Xu_2018}
Z.~Xu, X.~Hou, X.~Gong and J.~Wang, \emph{Black hole space-time in dark matter
  halo}, \href{https://doi.org/10.1088/1475-7516/2018/09/038}{\emph{Journal of
  Cosmology and Astroparticle Physics} {\bfseries 2018} (2018) 038}.

\bibitem{PhysRevD.104.124082}
C.~Zhang, T.~Zhu and A.~Wang, \emph{Gravitational axial perturbations of
  schwarzschild-like black holes in dark matter halos},
  \href{https://doi.org/10.1103/PhysRevD.104.124082}{\emph{Phys. Rev. D}
  {\bfseries 104} (2021) 124082}.

\bibitem{PhysRevD.104.104042}
D.~Liu, Y.~Yang, S.~Wu, Y.~Xing, Z.~Xu and Z.-W. Long, \emph{Ringing of a black
  hole in a dark matter halo},
  \href{https://doi.org/10.1103/PhysRevD.104.104042}{\emph{Phys. Rev. D}
  {\bfseries 104} (2021) 104042}.

\bibitem{Jusufi2020}
K.~Jusufi, M.~Jamil and T.~Zhu, \emph{Shadows of sgr $a^{*}$ black hole
  surrounded by superfluid dark matter halo},
  \href{https://doi.org/10.1140/epjc/s10052-020-7899-5}{\emph{The European
  Physical Journal C} {\bfseries 80} (2020) 354}.

\bibitem{Hou_2018}
X.~Hou, Z.~Xu, M.~Zhou and J.~Wang, \emph{Black hole shadow of sgr a* in dark
  matter halo},
  \href{https://doi.org/10.1088/1475-7516/2018/07/015}{\emph{Journal of
  Cosmology and Astroparticle Physics} {\bfseries 2018} (2018) 015}.

\bibitem{KONOPLYA20191}
R.~Konoplya, \emph{Shadow of a black hole surrounded by dark matter},
  \href{https://doi.org/https://doi.org/10.1016/j.physletb.2019.05.043}{\emph{Physics
  Letters B} {\bfseries 795} (2019) 1}.

\bibitem{PhysRevD.105.L061501}
V.~Cardoso, K.~Destounis, F.~Duque, R.~P. Macedo and A.~Maselli, \emph{Black
  holes in galaxies: Environmental impact on gravitational-wave generation and
  propagation}, \href{https://doi.org/10.1103/PhysRevD.105.L061501}{\emph{Phys.
  Rev. D} {\bfseries 105} (2022) L061501}.

\bibitem{Konoplya:2022hbl}
R.~A. Konoplya and A.~Zhidenko, \emph{{Solutions of the Einstein Equations for
  a Black Hole Surrounded by a Galactic Halo}},
  \href{https://doi.org/10.3847/1538-4357/ac76bc}{\emph{Astrophys. J.}
  {\bfseries 933} (2022) 166}
  [\href{https://arxiv.org/abs/2202.02205}{{\ttfamily 2202.02205}}].

\bibitem{Dai:2023cft}
N.~Dai, Y.~Gong, Y.~Zhao and T.~Jiang, \emph{{Extreme mass ratio inspirals in
  galaxies with dark matter halos}},
  \href{https://arxiv.org/abs/2301.05088}{{\ttfamily 2301.05088}}.

\bibitem{Xavier:2023exm}
S.~V. M. C.~B. Xavier, H.~C.~D. Lima, Junior. and L.~C.~B. Crispino,
  \emph{{Shadows of black holes with dark matter halo}},
  \href{https://doi.org/10.1103/PhysRevD.107.064040}{\emph{Phys. Rev. D}
  {\bfseries 107} (2023) 064040}
  [\href{https://arxiv.org/abs/2303.17666}{{\ttfamily 2303.17666}}].

\bibitem{Stuchlik:2021gwg}
Z.~Stuchl\'\i{}k and J.~Vrba, \emph{{Supermassive black holes surrounded by
  dark matter modeled as anisotropic fluid: epicyclic oscillations and their
  fitting to observed QPOs}},
  \href{https://doi.org/10.1088/1475-7516/2021/11/059}{\emph{JCAP} {\bfseries
  11} (2021) 059} [\href{https://arxiv.org/abs/2110.07411}{{\ttfamily
  2110.07411}}].

\bibitem{Figueiredo:2023gas}
E.~Figueiredo, A.~Maselli and V.~Cardoso, \emph{{Black holes surrounded by
  generic dark matter profiles: Appearance and gravitational-wave emission}},
  \href{https://doi.org/10.1103/PhysRevD.107.104033}{\emph{Phys. Rev. D}
  {\bfseries 107} (2023) 104033}
  [\href{https://arxiv.org/abs/2303.08183}{{\ttfamily 2303.08183}}].

\bibitem{Cardoso:2022whc}
V.~Cardoso, K.~Destounis, F.~Duque, R.~Panosso~Macedo and A.~Maselli,
  \emph{{Gravitational Waves from Extreme-Mass-Ratio Systems in Astrophysical
  Environments}},
  \href{https://doi.org/10.1103/PhysRevLett.129.241103}{\emph{Phys. Rev. Lett.}
  {\bfseries 129} (2022) 241103}
  [\href{https://arxiv.org/abs/2210.01133}{{\ttfamily 2210.01133}}].

\bibitem{Destounis:2022obl}
K.~Destounis, A.~Kulathingal, K.~D. Kokkotas and G.~O. Papadopoulos,
  \emph{{Gravitational-wave imprints of compact and galactic-scale environments
  in extreme-mass-ratio binaries}},
  \href{https://doi.org/10.1103/PhysRevD.107.084027}{\emph{Phys. Rev. D}
  {\bfseries 107} (2023) 084027}
  [\href{https://arxiv.org/abs/2210.09357}{{\ttfamily 2210.09357}}].

\bibitem{LIGOScientific:2016aoc}
{\scshape LIGO Scientific, Virgo} collaboration, \emph{{Observation of
  Gravitational Waves from a Binary Black Hole Merger}},
  \href{https://doi.org/10.1103/PhysRevLett.116.061102}{\emph{Phys. Rev. Lett.}
  {\bfseries 116} (2016) 061102}
  [\href{https://arxiv.org/abs/1602.03837}{{\ttfamily 1602.03837}}].

\bibitem{LIGOScientific:2017vox}
{\scshape LIGO Scientific, Virgo} collaboration, \emph{{GW170608: Observation
  of a 19-solar-mass Binary Black Hole Coalescence}},
  \href{https://doi.org/10.3847/2041-8213/aa9f0c}{\emph{Astrophys. J. Lett.}
  {\bfseries 851} (2017) L35}
  [\href{https://arxiv.org/abs/1711.05578}{{\ttfamily 1711.05578}}].

\bibitem{Babak:2017tow}
S.~Babak, J.~Gair, A.~Sesana, E.~Barausse, C.~F. Sopuerta, C.~P.~L. Berry
  et~al., \emph{{Science with the space-based interferometer LISA. V: Extreme
  mass-ratio inspirals}},
  \href{https://doi.org/10.1103/PhysRevD.95.103012}{\emph{Phys. Rev. D}
  {\bfseries 95} (2017) 103012}
  [\href{https://arxiv.org/abs/1703.09722}{{\ttfamily 1703.09722}}].

\bibitem{Amaro-Seoane:2007osp}
P.~Amaro-Seoane, J.~R. Gair, M.~Freitag, M.~Coleman~Miller, I.~Mandel, C.~J.
  Cutler et~al., \emph{{Astrophysics, detection and science applications of
  intermediate- and extreme mass-ratio inspirals}},
  \href{https://doi.org/10.1088/0264-9381/24/17/R01}{\emph{Class. Quant. Grav.}
  {\bfseries 24} (2007) R113}
  [\href{https://arxiv.org/abs/astro-ph/0703495}{{\ttfamily
  astro-ph/0703495}}].

\bibitem{Barack:2009ux}
L.~Barack, \emph{{Gravitational self force in extreme mass-ratio inspirals}},
  \href{https://doi.org/10.1088/0264-9381/26/21/213001}{\emph{Class. Quant.
  Grav.} {\bfseries 26} (2009) 213001}
  [\href{https://arxiv.org/abs/0908.1664}{{\ttfamily 0908.1664}}].

\bibitem{Hinderer:2008dm}
T.~Hinderer and E.~E. Flanagan, \emph{{Two timescale analysis of extreme mass
  ratio inspirals in Kerr. I. Orbital Motion}},
  \href{https://doi.org/10.1103/PhysRevD.78.064028}{\emph{Phys. Rev. D}
  {\bfseries 78} (2008) 064028}
  [\href{https://arxiv.org/abs/0805.3337}{{\ttfamily 0805.3337}}].

\bibitem{Drasco:2005kz}
S.~Drasco and S.~A. Hughes, \emph{{Gravitational wave snapshots of generic
  extreme mass ratio inspirals}},
  \href{https://doi.org/10.1103/PhysRevD.73.024027}{\emph{Phys. Rev. D}
  {\bfseries 73} (2006) 024027}
  [\href{https://arxiv.org/abs/gr-qc/0509101}{{\ttfamily gr-qc/0509101}}].

\bibitem{Gair:2004iv}
J.~R. Gair, L.~Barack, T.~Creighton, C.~Cutler, S.~L. Larson, E.~S. Phinney
  et~al., \emph{{Event rate estimates for LISA extreme mass ratio capture
  sources}}, \href{https://doi.org/10.1088/0264-9381/21/20/003}{\emph{Class.
  Quant. Grav.} {\bfseries 21} (2004) S1595}
  [\href{https://arxiv.org/abs/gr-qc/0405137}{{\ttfamily gr-qc/0405137}}].

\bibitem{Sopuerta:2009iy}
C.~F. Sopuerta and N.~Yunes, \emph{{Extreme and Intermediate-Mass Ratio
  Inspirals in Dynamical Chern-Simons Modified Gravity}},
  \href{https://doi.org/10.1103/PhysRevD.80.064006}{\emph{Phys. Rev. D}
  {\bfseries 80} (2009) 064006}
  [\href{https://arxiv.org/abs/0904.4501}{{\ttfamily 0904.4501}}].

\bibitem{Rahman:2021eay}
M.~Rahman and A.~Bhattacharyya, \emph{{Prospects for determining the nature of
  the secondaries of extreme mass-ratio inspirals using the spin-induced
  quadrupole deformation}},
  \href{https://doi.org/10.1103/PhysRevD.107.024006}{\emph{Phys. Rev. D}
  {\bfseries 107} (2023) 024006}
  [\href{https://arxiv.org/abs/2112.13869}{{\ttfamily 2112.13869}}].

\bibitem{Rahman:2022fay}
M.~Rahman, S.~Kumar and A.~Bhattacharyya, \emph{{Gravitational wave from
  extreme mass-ratio inspirals as a probe of extra dimensions}},
  \href{https://doi.org/10.1088/1475-7516/2023/01/046}{\emph{JCAP} {\bfseries
  01} (2023) 046} [\href{https://arxiv.org/abs/2212.01404}{{\ttfamily
  2212.01404}}].

\bibitem{Fan:2022wio}
H.-M. Fan, S.~Zhong, Z.-C. Liang, Z.~Wu, J.-d. Zhang and Y.-M. Hu,
  \emph{{Extreme-mass-ratio burst detection with TianQin}},
  \href{https://doi.org/10.1103/PhysRevD.106.124028}{\emph{Phys. Rev. D}
  {\bfseries 106} (2022) 124028}
  [\href{https://arxiv.org/abs/2209.13387}{{\ttfamily 2209.13387}}].

\bibitem{Liang:2022gdk}
D.~Liang, R.~Xu, Z.-F. Mai and L.~Shao, \emph{{Probing vector hair of black
  holes with extreme-mass-ratio inspirals}},
  \href{https://doi.org/10.1103/PhysRevD.107.044053}{\emph{Phys. Rev. D}
  {\bfseries 107} (2023) 044053}
  [\href{https://arxiv.org/abs/2212.09346}{{\ttfamily 2212.09346}}].

\bibitem{Maselli:2021men}
A.~Maselli, N.~Franchini, L.~Gualtieri, T.~P. Sotiriou, S.~Barsanti and
  P.~Pani, \emph{{Detecting fundamental fields with LISA observations of
  gravitational waves from extreme mass-ratio inspirals}},
  \href{https://doi.org/10.1038/s41550-021-01589-5}{\emph{Nature Astron.}
  {\bfseries 6} (2022) 464} [\href{https://arxiv.org/abs/2106.11325}{{\ttfamily
  2106.11325}}].

\bibitem{Drummond:2023loz}
L.~V. Drummond, A.~G. Hanselman, D.~R. Becker and S.~A. Hughes, \emph{{Extreme
  mass-ratio inspiral of a spinning body into a Kerr black hole I: Evolution
  along generic trajectories}},
  \href{https://arxiv.org/abs/2305.08919}{{\ttfamily 2305.08919}}.

\bibitem{PhysRevD.102.024041}
G.~A. Piovano, A.~Maselli and P.~Pani, \emph{Extreme mass ratio inspirals with
  spinning secondary: A detailed study of equatorial circular motion},
  \href{https://doi.org/10.1103/PhysRevD.102.024041}{\emph{Phys. Rev. D}
  {\bfseries 102} (2020) 024041}.

\bibitem{Gair:2017ynp}
J.~R. Gair, S.~Babak, A.~Sesana, P.~Amaro-Seoane, E.~Barausse, C.~P.~L. Berry
  et~al., \emph{{Prospects for observing extreme-mass-ratio inspirals with
  LISA}}, \href{https://doi.org/10.1088/1742-6596/840/1/012021}{\emph{J. Phys.
  Conf. Ser.} {\bfseries 840} (2017) 012021}
  [\href{https://arxiv.org/abs/1704.00009}{{\ttfamily 1704.00009}}].

\bibitem{LISA:2022kgy}
{\scshape LISA} collaboration, \emph{{New horizons for fundamental physics with
  LISA}}, \href{https://doi.org/10.1007/s41114-022-00036-9}{\emph{Living Rev.
  Rel.} {\bfseries 25} (2022) 4}
  [\href{https://arxiv.org/abs/2205.01597}{{\ttfamily 2205.01597}}].

\bibitem{Cardoso:2020iji}
V.~Cardoso, C.~F.~B. Macedo and R.~Vicente, \emph{{Eccentricity evolution of
  compact binaries and applications to gravitational-wave physics}},
  \href{https://doi.org/10.1103/PhysRevD.103.023015}{\emph{Phys. Rev. D}
  {\bfseries 103} (2021) 023015}
  [\href{https://arxiv.org/abs/2010.15151}{{\ttfamily 2010.15151}}].

\bibitem{Barsanti:2022ana}
S.~Barsanti, N.~Franchini, L.~Gualtieri, A.~Maselli and T.~P. Sotiriou,
  \emph{{Extreme mass-ratio inspirals as probes of scalar fields: Eccentric
  equatorial orbits around Kerr black holes}},
  \href{https://doi.org/10.1103/PhysRevD.106.044029}{\emph{Phys. Rev. D}
  {\bfseries 106} (2022) 044029}
  [\href{https://arxiv.org/abs/2203.05003}{{\ttfamily 2203.05003}}].

\bibitem{PhysRevD.103.104045}
V.~Skoup\'y and G.~Lukes-Gerakopoulos, \emph{Spinning test body orbiting around
  a kerr black hole: Eccentric equatorial orbits and their asymptotic
  gravitational-wave fluxes},
  \href{https://doi.org/10.1103/PhysRevD.103.104045}{\emph{Phys. Rev. D}
  {\bfseries 103} (2021) 104045}.

\bibitem{PhysRevD.50.3816}
C.~Cutler, D.~Kennefick and E.~Poisson, \emph{Gravitational radiation reaction
  for bound motion around a schwarzschild black hole},
  \href{https://doi.org/10.1103/PhysRevD.50.3816}{\emph{Phys. Rev. D}
  {\bfseries 50} (1994) 3816}.

\bibitem{PhysRevD.77.124050}
P.~A. Sundararajan, \emph{Transition from adiabatic inspiral to geodesic plunge
  for a compact object around a massive kerr black hole: Generic orbits},
  \href{https://doi.org/10.1103/PhysRevD.77.124050}{\emph{Phys. Rev. D}
  {\bfseries 77} (2008) 124050}.

\bibitem{PhysRevD.67.104017}
N.~Sago, H.~Nakano and M.~Sasaki, \emph{Gauge problem in the gravitational
  self-force: Harmonic gauge approach in the schwarzschild background},
  \href{https://doi.org/10.1103/PhysRevD.67.104017}{\emph{Phys. Rev. D}
  {\bfseries 67} (2003) 104017}.

\bibitem{PhysRevLett.24.737}
F.~J. Zerilli, \emph{Effective potential for even-parity regge-wheeler
  gravitational perturbation equations},
  \href{https://doi.org/10.1103/PhysRevLett.24.737}{\emph{Phys. Rev. Lett.}
  {\bfseries 24} (1970) 737}.

\bibitem{PhysRevD.2.2141}
F.~J. Zerilli, \emph{Gravitational field of a particle falling in a
  schwarzschild geometry analyzed in tensor harmonics},
  \href{https://doi.org/10.1103/PhysRevD.2.2141}{\emph{Phys. Rev. D} {\bfseries
  2} (1970) 2141}.

\bibitem{Poisson:2011nh}
E.~Poisson, A.~Pound and I.~Vega, \emph{{The Motion of point particles in
  curved spacetime}}, \href{https://doi.org/10.12942/lrr-2011-7}{\emph{Living
  Rev. Rel.} {\bfseries 14} (2011) 7}
  [\href{https://arxiv.org/abs/1102.0529}{{\ttfamily 1102.0529}}].

\bibitem{wasow1965asymptotic}
W.~Wasow, \emph{Asymptotic expansions for ordinary differential equations},
  Pure and Applied Mathematics, Vol. XIV. Interscience Publishers John Wiley \&
  Sons, Inc., New York-London-Sydney, 1965.

\bibitem{Mathematica}
W.~R. Inc., ``Mathematica, {V}ersion 12.0.''

\bibitem{cardoso}
{\relax grit repo,}.
  \href{https://centra.tecnico.ulisboa.pt/network/grit/files/}{https://centra.tecnico.ulisboa.pt/network/grit/files/}.

\bibitem{maseli}
{\relax sgrep repo,}.
  \href{https://github.com/masellia/SGREP}{https://github.com/masellia/SGREP}.

\bibitem{Berti:2009kk}
E.~Berti, V.~Cardoso and A.~O. Starinets, \emph{{Quasinormal modes of black
  holes and black branes}},
  \href{https://doi.org/10.1088/0264-9381/26/16/163001}{\emph{Class. Quant.
  Grav.} {\bfseries 26} (2009) 163001}
  [\href{https://arxiv.org/abs/0905.2975}{{\ttfamily 0905.2975}}].

\bibitem{Cardoso:2017njb}
V.~Cardoso and P.~Pani, \emph{{The observational evidence for horizons: from
  echoes to precision gravitational-wave physics}},
  \href{https://arxiv.org/abs/1707.03021}{{\ttfamily 1707.03021}}.

\bibitem{Cardoso:2008bp}
V.~Cardoso, A.~S. Miranda, E.~Berti, H.~Witek and V.~T. Zanchin,
  \emph{{Geodesic stability, Lyapunov exponents and quasinormal modes}},
  \href{https://doi.org/10.1103/PhysRevD.79.064016}{\emph{Phys. Rev. D}
  {\bfseries 79} (2009) 064016}
  [\href{https://arxiv.org/abs/0812.1806}{{\ttfamily 0812.1806}}].

\bibitem{RevModPhys.52.299}
K.~S. Thorne, \emph{Multipole expansions of gravitational radiation},
  \href{https://doi.org/10.1103/RevModPhys.52.299}{\emph{Rev. Mod. Phys.}
  {\bfseries 52} (1980) 299}.

\bibitem{PhysRevD.70.084044}
E.~Poisson, \emph{Absorption of mass and angular momentum by a black hole:
  Time-domain formalisms for gravitational perturbations, and the small-hole or
  slow-motion approximation},
  \href{https://doi.org/10.1103/PhysRevD.70.084044}{\emph{Phys. Rev. D}
  {\bfseries 70} (2004) 084044}.

\bibitem{PhysRevD.69.044025}
K.~Martel, \emph{Gravitational waveforms from a point particle orbiting a
  schwarzschild black hole},
  \href{https://doi.org/10.1103/PhysRevD.69.044025}{\emph{Phys. Rev. D}
  {\bfseries 69} (2004) 044025}.

\bibitem{PhysRevD.53.3064}
F.~D. Ryan, \emph{Effect of gravitational radiation reaction on nonequatorial
  orbits around a kerr black hole},
  \href{https://doi.org/10.1103/PhysRevD.53.3064}{\emph{Phys. Rev. D}
  {\bfseries 53} (1996) 3064}.

\bibitem{PhysRevD.78.064028}
T.~Hinderer and E.~E. Flanagan, \emph{Two-timescale analysis of extreme mass
  ratio inspirals in kerr spacetime: Orbital motion},
  \href{https://doi.org/10.1103/PhysRevD.78.064028}{\emph{Phys. Rev. D}
  {\bfseries 78} (2008) 064028}.

\bibitem{PhysRevD.103.104014}
S.~A. Hughes, N.~Warburton, G.~Khanna, A.~J.~K. Chua and M.~L. Katz,
  \emph{Adiabatic waveforms for extreme mass-ratio inspirals via multivoice
  decomposition in time and frequency},
  \href{https://doi.org/10.1103/PhysRevD.103.104014}{\emph{Phys. Rev. D}
  {\bfseries 103} (2021) 104014}.

\bibitem{Isoyama:2021jjd}
S.~Isoyama, R.~Fujita, A.~J.~K. Chua, H.~Nakano, A.~Pound and N.~Sago,
  \emph{{Adiabatic Waveforms from Extreme-Mass-Ratio Inspirals: An Analytical
  Approach}}, \href{https://doi.org/10.1103/PhysRevLett.128.231101}{\emph{Phys.
  Rev. Lett.} {\bfseries 128} (2022) 231101}
  [\href{https://arxiv.org/abs/2111.05288}{{\ttfamily 2111.05288}}].

\bibitem{holmes2006introduction}
M.~Holmes, \emph{Introduction to Numerical Methods in Differential Equations},
  Texts in Applied Mathematics. Springer New York, 2006.

\bibitem{PhysRevLett.123.101103}
B.~Bonga, H.~Yang and S.~A. Hughes, \emph{Tidal resonance in extreme mass-ratio
  inspirals}, \href{https://doi.org/10.1103/PhysRevLett.123.101103}{\emph{Phys.
  Rev. Lett.} {\bfseries 123} (2019) 101103}.

\bibitem{Barausse:2014pra}
E.~Barausse, V.~Cardoso and P.~Pani, \emph{{Environmental Effects for
  Gravitational-wave Astrophysics}},
  \href{https://doi.org/10.1088/1742-6596/610/1/012044}{\emph{J. Phys. Conf.
  Ser.} {\bfseries 610} (2015) 012044}
  [\href{https://arxiv.org/abs/1404.7140}{{\ttfamily 1404.7140}}].

\bibitem{PhysRev.108.1063}
T.~Regge and J.~A. Wheeler, \emph{Stability of a schwarzschild singularity},
  \href{https://doi.org/10.1103/PhysRev.108.1063}{\emph{Phys. Rev.} {\bfseries
  108} (1957) 1063}.

\bibitem{Vallisneri:2007ev}
M.~Vallisneri, \emph{{Use and abuse of the Fisher information matrix in the
  assessment of gravitational-wave parameter-estimation prospects}},
  \href{https://doi.org/10.1103/PhysRevD.77.042001}{\emph{Phys. Rev. D}
  {\bfseries 77} (2008) 042001}
  [\href{https://arxiv.org/abs/gr-qc/0703086}{{\ttfamily gr-qc/0703086}}].

\bibitem{Graham:2005xx}
A.~W. Graham, D.~Merritt, B.~Moore, J.~Diemand and B.~Terzic, \emph{{Empirical
  models for Dark Matter Halos. I. Nonparametric Construction of Density
  Profiles and Comparison with Parametric Models}},
  \href{https://doi.org/10.1086/508988}{\emph{Astron. J.} {\bfseries 132}
  (2006) 2685} [\href{https://arxiv.org/abs/astro-ph/0509417}{{\ttfamily
  astro-ph/0509417}}].

\bibitem{DeLuca:2023laa}
V.~De~Luca and J.~Khoury, \emph{{Superfluid dark matter around black holes}},
  \href{https://doi.org/10.1088/1475-7516/2023/04/048}{\emph{JCAP} {\bfseries
  04} (2023) 048} [\href{https://arxiv.org/abs/2302.10286}{{\ttfamily
  2302.10286}}].

\bibitem{Cardoso:2018tly}
V.~Cardoso, O.~J.~C. Dias, G.~S. Hartnett, M.~Middleton, P.~Pani and J.~E.
  Santos, \emph{{Constraining the mass of dark photons and axion-like particles
  through black-hole superradiance}},
  \href{https://doi.org/10.1088/1475-7516/2018/03/043}{\emph{JCAP} {\bfseries
  03} (2018) 043} [\href{https://arxiv.org/abs/1801.01420}{{\ttfamily
  1801.01420}}].

\bibitem{xACT}
{ Jos\`{e} M. Mart\`{i}n-Garc\`{i}a et. al.,}. \href{http://www.xact.es/}{xACT,
  {V}ersion 1.2.0}.

\bibitem{Pani}
G.~A. Piovano, A.~Maselli and P.~Pani, \emph{Extreme mass ratio inspirals with
  spinning secondary: A detailed study of equatorial circular motion},
  \href{https://doi.org/10.1103/PhysRevD.102.024041}{\emph{Phys. Rev. D}
  {\bfseries 102} (2020) 024041}.

\end{thebibliography}\endgroup


\begin{thebibliography}{10}

\bibitem{PhysRevLett.116.061102}
{\scshape LIGO Scientific Collaboration and Virgo Collaboration} collaboration,
  \emph{Observation of gravitational waves from a binary black hole merger},
  \href{https://doi.org/10.1103/PhysRevLett.116.061102}{\emph{Phys. Rev. Lett.}
  {\bfseries 116} (2016) 061102}.

\bibitem{PhysRevLett.125.101102}
{\scshape LIGO Scientific Collaboration and Virgo Collaboration} collaboration,
  \emph{Gw190521: A binary black hole merger with a total mass of $150\text{
  }\text{ }{M}_{\ensuremath{\bigodot}}$},
  \href{https://doi.org/10.1103/PhysRevLett.125.101102}{\emph{Phys. Rev. Lett.}
  {\bfseries 125} (2020) 101102}.

\end{thebibliography}
\bibliographystyle{JHEP}
\end{document}